\documentclass[12pt]{article}

\usepackage[utf8]{inputenc}   
\usepackage[T1]{fontenc}      

\usepackage{amsmath}
\usepackage{graphicx}
\usepackage{enumerate}
\usepackage[round]{natbib}
\usepackage{url}
\usepackage{hyperref}
\usepackage{bm}
\usepackage{accents}
\usepackage{booktabs}
\usepackage[most]{tcolorbox}
\usepackage{multirow}
\usepackage{pdfpages}

\newcommand\munderbar[1]{%
  \underaccent{\bar}{#1}}

\newcommand{\bzero}{\textbf{0}}
\newcommand{\bone}{\textbf{1}}

\newcommand{\bz}{\textbf{z}}

\newcommand{\bA}{\textbf{A}}

\newcommand{\bB}{\textbf{B}}

\newcommand{\bI}{\textbf{I}}

\newcommand{\bp}{\textbf{p}}

\newcommand{\bs}{\textbf{s}}

\newcommand{\bT}{\textbf{T}}

\newcommand{\bv}{\textbf{v}}

\newcommand{\bw}{\textbf{w}}

\newcommand{\bx}{\textbf{x}}

\newcommand{\bY}{\textbf{Y}}

\DeclareMathOperator*{\Prob}{Pr}

\DeclareMathOperator*{\T}{T}

\DeclareMathOperator*{\cov}{cov}

\DeclareMathOperator*{\var}{var}
\DeclareMathOperator{\dist}{dist}

\begin{document}

\title{\bf Joint space-time modelling for upper daily maximum and minimum temperature record-breaking}

\author{Jorge Castillo-Mateo\footnote{Corresponding author. \href{email:jorgecm@unizar.es}{jorgecm@unizar.es}} \\ Department of Statistical Methods, University of Zaragoza \\
  Zeus Gracia-Tabuenca \\ Department of Statistical Methods, University of Zaragoza \\
  Jesús Asín \\ Department of Statistical Methods, University of Zaragoza \\
  Ana C. Cebrián \\ Department of Statistical Methods, University of Zaragoza \\
  Alan E. Gelfand \\ Department of Statistical Science, Duke University \\}

\maketitle

\begin{abstract}
Record-breaking temperature events are now frequently in the news, proffered as evidence of climate change, and often bring significant economic and human impacts. Our previous work undertook the first substantial spatial modelling investigation of temperature record-breaking across years for any given day within the year, employing a dataset consisting of over sixty years of daily maximum temperatures across peninsular Spain.  That dataset also supplies daily minimum temperatures (which, in fact, are now available through 2023).  Here, the dataset is converted into a daily pair of binary events, indicators, for that day, of whether a yearly record was broken for the daily maximum temperature and/or for the daily minimum temperature.  Joint modelling addresses several inference issues: (i) defining/modelling record-breaking with bivariate time series of yearly indicators, (ii) strength of relationship between record-breaking events, (iii) prediction of joint, conditional and marginal record-breaking, (iv) persistence in record-breaking across days, (v) spatial interpolation across peninsular Spain.  We substantially expand our previous work to enable investigation of these issues.  We observe strong correlation between both processes but a growing trend of climate change that is well differentiated between them both spatially and temporally as well as different strengths of persistence and spatial dependence.
\end{abstract}

\noindent%

\section{Introduction}\label{sec:intro}

Most studies seeking to quantify the effects of climate change on daily temperatures focus on the analysis of daily mean or maximum temperatures \citep[see, e.g.,][]{krock2022,castillo2022b,healy2021}. However, many other studies related to the consequences of high temperatures highlight the importance of considering daily minimum temperatures \citep{correa2024}. For example, it was found that both maximum and minimum temperatures, especially the combination of extreme values in both, have an impact on human health and mortality \citep{hajat2006}. The effects of increasing minimum temperatures are also relevant in agriculture and crop production \citep{Battisti2009}. \cite{kaur2010} suggested that increase in both maximum and minimum temperatures yields a reduction in wheat and gram production and changes in phenological development. As a consequence, an increasing number of definitions of \emph{heat wave} incorporate information from both temperatures; see, e.g., \cite{plummer1999}, \cite{tryhorn2006}, or the US National Weather Service definition. Changes in the diurnal temperature range (DTR) are closely associated with both daily maximum and minimum temperatures. Studies by \cite{huang2023}, \cite{daramola2024} and \cite{liu2024} indicate that, since around the 1980s, daily maximum temperatures have risen more rapidly than minimum temperatures, leading to an overall increase in DTR, although with regional variations.

To learn about maximum and minimum temperatures, it is appropriate to develop a joint model that takes into account the dependence between both temperatures in order to share the strength of both processes. For example, \cite{saez2012} modelled both daily maximum and minimum temperature to estimate long-term trends and assess the spatial and temporal variability. \cite{Kleiber2013} proposed a bivariate stochastic model for maximum and minimum temperatures for simulating series to be used as input for climate studies and hydrological and agricultural models. \cite{North2021} proposed a multivariate spatio-temporal model to quantify the spatial and temporal changes in minimum and maximum temperature seasonal cycles. There is particular interest in temperature extremes. \cite{abaurrea2018} modelled the occurrence of extreme heat events in daily maximum and minimum temperatures using a non-homogeneous common Poisson shock process. \cite{lewis2015} found that the number of upper maximum and minimum temperature records increased dramatically in recent decades.

In the context of impact studies, the interest in modelling the extremes of both temperature signals is clear. Although the analysis of extreme values is often studied in terms of block maxima and exceedances over high thresholds \citep{davison2015}, the frequency of calendar day records in a period of time and area is a useful dimensionless metric used to measure the effect of climate change on the upper tail of temperature; see \cite{castillo2024} and  references therein. The analysis of records does not directly correspond to the analysis of excessive heat events and may present a noisier signal as record values depend on past observations. Nonetheless, this approach has several important advantages.
First, it enables fair comparisons across time series with differing distributions (e.g., from different locations) without requiring standardization or the identification of local thresholds. Second, under a stationary climate, the occurrence of records is governed by well-established distribution-free probabilistic properties, described in the next paragraph. This provides a robust theoretical framework to detect and quantify deviations from climate stationarity.
Further, as far as we know, no work has addressed the joint modelling of records in maximum and minimum temperatures.

Evidently, records would occur in what is referred to as a stationary climate, since the probability of a record at time $t$ is $1/t$ for any series of continuous i.i.d. variables \citep[][Chapter~2]{arnold1998}. Thus, we need to address the question of whether the rate of observed records could arise without the influence of climate change, to quantify any increase in the rate if it exists, and to study whether the occurrence pattern varies across space or across days within the years. Adding to the challenges of modelling records in a univariate framework, we must consider that there is no clear definition of a record in a bivariate vector \citep[][Chapter~8]{arnold1998}.  Modelling the dependence between both types of records, which need not be simultaneous, adds further complexity.
 
Most of the literature on univariate records is related to the probabilistic properties of the occurrence and value under simple stochastic models, e.g., under the classical record model that assumes series of continuous i.i.d. variables. These properties have been used for studying deviations from stationarity; see, e.g., \cite{benestad2003}, \cite{naveau2018}, \cite{cebrian2022a}, or \cite{castillo2023}. This is even more evident in the study of records in a bivariate context, where some probabilistic properties have been obtained assuming simple models \citep{kemalbay2019,balakrishnan2020,fill2023}. The analysis of records is of interest in many fields, but only a few attempts at statistical modelling of their occurrence or value have been made; a few examples appear in sports \citep{noubary2004}, finance \citep{wergen2014b}, hydrology \citep{serinaldi2018} or climate \citep{wergen2014a}.  

The work of \cite{castillo2024} undertook the first substantial spatial modelling investigation of temperature record-breaking events.  They employed a dataset consisting of daily maximum temperatures  from 40 locations across peninsular Spain spanning the years 1960--2021. Records are considered in a given location for individual calendar days across years. The responses are binary time series that describe, for a given day, whether or not a record was broken in a given year for that location.  In the work here, we jointly model the occurrence of records defined marginally, i.e., on a given calendar day a new \emph{partial} bivariate record occurs at time $t$ for the maxima when the maximum for that day within year $t$ exceeds all preceding maxima for that calendar day and location, and similarly for the minima. We define as a \emph{joint} bivariate record when both maximum and minimum temperatures are records for a given calendar day and location within a given year.

The peninsular Spain dataset considered in \cite{castillo2024} now supplies daily minimum temperatures which extend to include 2022 and 2023, two extremely hot summers in Spain \citep{serrrano2023}.  This updated dataset motivates our work here.  Only the summer season is considered, as the most relevant impacts correspond to extreme warm temperatures.  We convert the dataset into a daily pair of binary events, i.e., an indicator of whether a yearly record was broken for the daily maximum temperature on that day and also an indicator of whether a record was broken for the daily minimum temperature for that day.  Then, we consider the joint daily record-breaking behaviour. The interest in joint modelling enables consideration of several inference issues.  How do we define joint record-breaking with bivariate time series of yearly indicators? Could the actual record rates be observed without the influence of climate change? How strong is the relationship between record-breaking events? Is the occurrence of maximum and minimum temperature records similar over time and across  space? Can we find persistence in record-breaking, e.g., if today's daily minimum temperature set a record, what is the chance that today's daily maximum will set a record?  Can we learn about persistence in record-breaking from the previous day's events?  Can we illuminate these issues spatially across peninsular Spain, i.e., can we predict record-breaking behaviour at unobserved locations?  Can we develop spatial record-breaking surfaces to reveal the extent of joint or marginal record-breaking over the study region or subregions? We substantially expand the work in \cite{castillo2024} in order to address these issues.

We explore these questions through hierarchical modelling implemented in a Bayesian framework.  The modelling novelties include: (i) a bivariate binary spatio-temporal model that captures persistence between the binary events through a vector autoregression (VAR) structure with linear predictors; (ii) linear predictors including both fixed effects (long-term trends, spatial covariates, and interactions) and necessary random effects; (iii) bivariate spatial dependence modelled through Gaussian processes with spatially varying coregionalisation, and correlation structure dependent on covariates. Note that these novelties relax the assumptions of stationarity and isotropy in the spatial Gaussian processes. We observe that the bivariate setting offers greater flexibility than individual univariate models. In particular, it captures persistence not only in the occurrence of each marginal record but also in the occurrence of records for both temperature signals, which identifies the most dangerous situations in terms of impact. We also provide space-time interpolation and model-based tools for joint, conditional, and marginal record-breaking behaviour.

As noted above, our aim is to develop a modelling framework focused specifically on the occurrence of record-breaking events, using annual time series of binary indicators denoting whether a record was set. An alternative approach, which utilises the raw temperature data fully, would be to model daily temperatures directly using a spatio-temporal mean model, and then derive properties related to records based on the fitted temperature process. However, such models are driven by the bulk of the distribution and often fail to capture tail behaviour adequately, limiting their utility for studying extremes. Selecting an appropriate temperature model would require substantial additional effort and would depend heavily on the specific inferential objective. Modelling the binary indicators directly avoids these issues, provides a more targeted approach to studying record-breaking events, and does not require specifying the full temperature distribution, simplifying most parts of the modelling process.

The outline of the paper is as follows. Section~\ref{sec:data} presents the daily temperature data and exploratory analysis of the record indicators in both temperature series. Section~\ref{sec:model} introduces the modelling framework for the bivariate spatial binary model together with Bayesian model-based inference tools. It offers a model fitting implementation using data augmentation Markov chain Monte Carlo (MCMC), and offers a performance metric to validate and compare models in terms of how well they capture records. Section~\ref{sec:results} shows the model comparison and the results from the selected model, while Section~\ref{sec:end} presents conclusions and future work.

\section{Data and exploratory analysis}\label{sec:data}

\subsection{Data}

\begin{figure}[tb]
    \centering
    \includegraphics[width=0.8\textwidth]{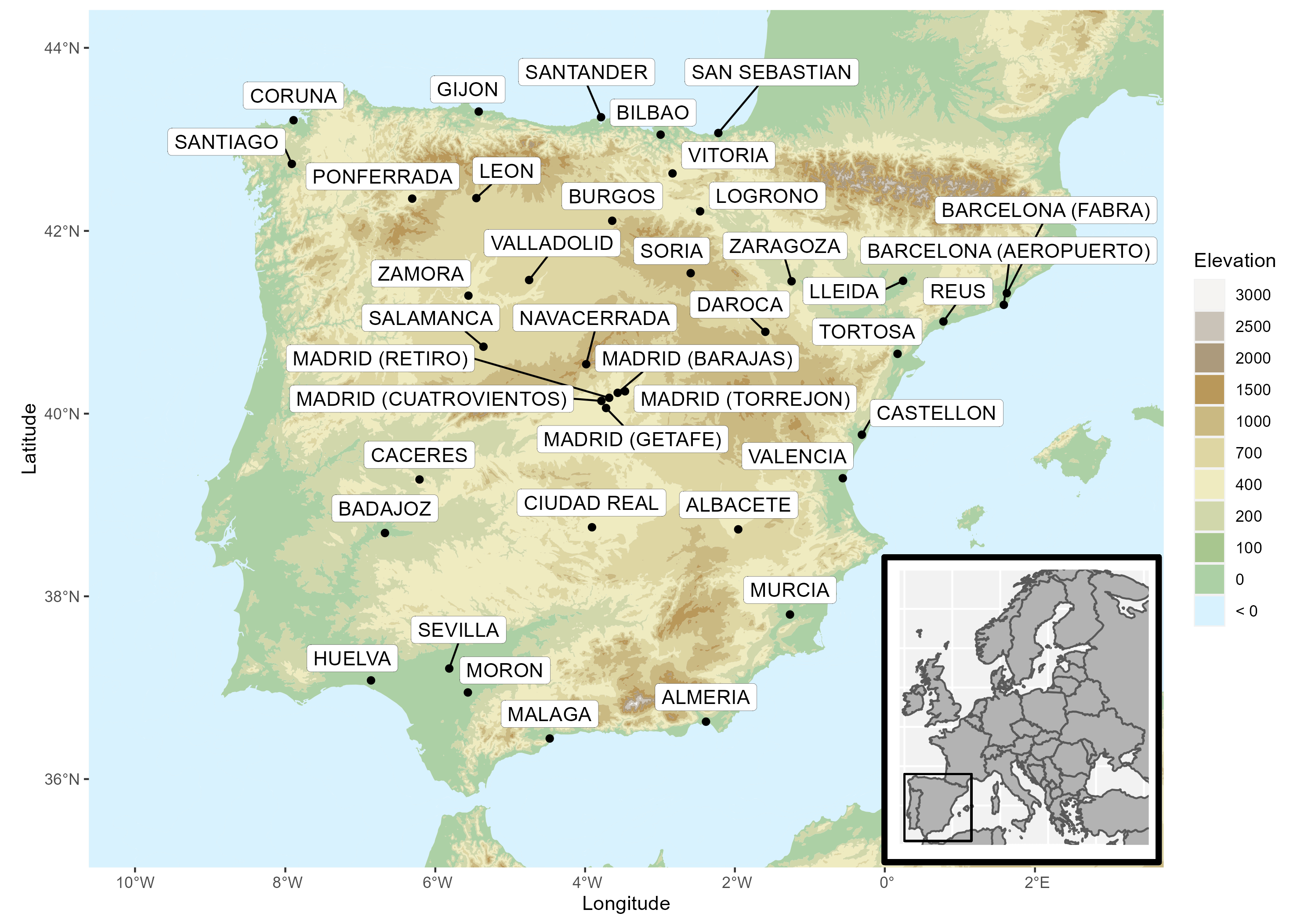}
    \caption{Map of the 40 Spanish stations in southwestern Europe.}
    \label{fig:map}
\end{figure}

Our aim is to study the occurrence of calendar day records for both daily maximum ($\T_{x}$) and minimum ($\T_{n}$) temperatures throughout the summer season, i.e., June, July, and August (henceforth JJA), over peninsular Spain ($492{,}175$ km$^{2}$). The point-referenced dataset contains the observed $\T_{x}$ and $\T_{n}$ series at 40 weather stations, from the 1st of January 1960 to the 31st of December 2023, obtained from the European Climate Assessment \& Dataset \citep[ECA\&D;][]{tank2002}. Given that the analysis focuses on the temperature record indicators defined within a summer day across years, the dataset is organised as two sets with $92$ binary series of length $64$ for each site, a total of $471{,}040$ observations. The locations of the weather stations are shown in Figure~\ref{fig:map}. The stations are irregularly distributed across the region, capturing the diverse climatic zones in Spain, with several mountain ranges and a long coastline.

The dataset may be incomplete on two accounts; we have addressed these as in \cite{castillo2024}. First, there are missing values and, second, the data is rounded. For the first, the retained 40 stations have a small amount of missing values, on average, less than $0.07\%$ in the period of interest and equivalent between both types of temperature. We assigned a value of $-\infty$ (or equivalently any large negative value) to missing temperature data, whether in $\T_{x}$ or $\T_{n}$. The idea is to impose that missing values are not considered records, except when they occur at $t = 1$. As shown by \cite{castillo2024} in a simulation study, the impact on the results of a small amount of missing data is negligible.  For the second, temperature series are measured to the nearest $1/10$th of a $^{\circ}$C.  This discretisation results in some ties when records are identified.  An observation that is at least as large as any previous observation is called a \emph{weak} record.  To deal with ties, \cite{castillo2024} defined a \emph{tied} record in terms of \emph{equal} rather than \emph{higher} or \emph{equal}, i.e., an $r$-tied record ($r \ge 2$) arises when an observation shares the same value with $r - 1$ preceding weak records. Among all non-trivial and weak records, the proportion of tied records across stations ranges from $3.2\%$ to $12.2\%$ for the maxima and from $5.5\%$ to $20.6\%$ for the minima. Tied records are handled by sampling their indicators as $\text{Bernoulli}(1 / r)$ within the Bayesian model fitting.

\subsection{Exploring the occurrence of bivariate records} \label{sec:EDA}

We implemented an extensive exploratory analysis focused on the occurrence of bivariate records with two main objectives.  We want to reveal the evolution of the occurrence of records in peninsular Spain, and we want to identify temporal and spatial patterns that should be captured by fixed and random effects in the model.  The main results, linked to long-term trends, persistence, and regional variability, are summarised in this section. Additional results are shown in Section~1 of the Supplementary material.  The exploratory analysis considers that indicators associated with tied records are zero, and it only includes indicators from 36 stations; only one of the five Madrid stations is used to avoid its over-representation. 

We denote by $I_{t\ell}(\bs)$ the record indicator at location $\bs$ for whether a record on calendar day $\ell$ was broken on year $t$. More formally,
\begin{equation}
    I_{t\ell}(\bs) =
    \begin{cases}
        1 & \text{if } X_{t\ell}(\bs) > \max\{X_{1\ell}(\bs),X_{2\ell}(\bs),\ldots,X_{t-1,\ell}(\bs)\}, \\ 
        0 & \text{otherwise.}
    \end{cases}
\end{equation}
By definition $I_{1\ell}(\bs) = 1$, and $X_{t\ell}(\bs)$ denotes a daily temperature for day $\ell$ within year $t$ at location $\bs$. In particular, $\bar{I}_{t\ell}(\bs)$ refers to records in $\T_{x}$ and $\munderbar{I}_{t\ell}(\bs)$ refers to records in $\T_{n}$. A joint bivariate record is defined when both maximum and minimum temperatures are
records for a given calendar day within a given year, i.e., the joint record indicator is $\bar{\munderbar{I}}_{t\ell}(\bs) = \bar{I}_{t\ell}(\bs) \munderbar{I}_{t\ell}(\bs)$. The period of interest includes the days $\ell \in JJA$ within years $t = 1,\ldots,T$, where $1$ is year 1960 and $T = 64$ is year 2023.  The location of the $i$th weather station in the exploratory analysis is denoted by $\bs_{i} \in \{\bs_{1},\ldots,\bs_{36}\}$.

\subsubsection{Identifying potential explanatory variables} \label{sec:EDA_fx}

Here, we explore long-term trends, bivariate relationships between indicators, persistence terms, and spatial patterns that should be captured by fixed effects in the model.

\begin{figure}[tb]
    \centering
    \includegraphics[width=.45\textwidth]{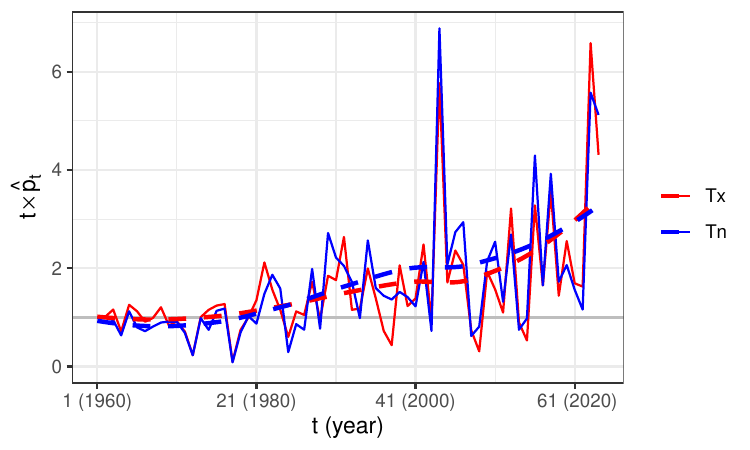}
    \includegraphics[width=.45\textwidth]{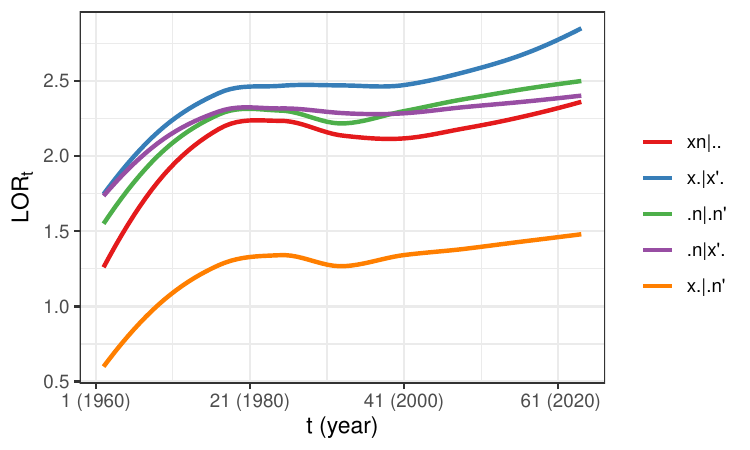}
    \caption{Left: time series plot and LOESS curve of $t \times \hat{p}_{t}$ against $t$ with reference value $1$. Right: LOESS curve of the $LOR_t$'s for the $\bar{I}_{t\ell}(\bs)$ and $\protect\munderbar{I}_{t\ell}(\bs)$ concurrence and persistence against $t$. In the legend, $x$ and $n$ refer to the occurrence of $\T_{x}$ and $\T_{n}$ records, $x^{\prime}$ and $n^{\prime}$ refer to the occurrence of $\T_{x}$ and $\T_{n}$ records the previous day.}
    \label{fig:eda:fixedeffects}
\end{figure}

\paragraph{\textbf{Long-term trend}} We focus on exploring the long-term trend of the probabilities of record. An empirical estimate of the probability of a record in year $t$ can be obtained by averaging in space and days within year, i.e., $\hat{p}_{t} = \sum_{i=1}^{36} \sum_{\ell \in JJA} I_{t\ell}(\bs_i) / (36 \times 92)$.  The left plot in Figure~\ref{fig:eda:fixedeffects} shows $t \times \hat{p}_{t}$ against $t$ for $\bar{I}_{t\ell}(\bs)$ and $\munderbar{I}_{t\ell}(\bs)$, both series evolve away from $1$, the expected value under stationary conditions.\footnote{Under stationarity, the variance of $I_{t}$ is $\var(I_{t}) = (1/t) (1 - 1/t)$, so proportionally the variance of $t \times \hat{p}_t$ starts at $1/2$, grows with $t$ and approaches $1$. In the non-stationary case with space-time dependence, we expect a similar pattern. Indeed, except for some peaks, the variance seems to stabilise around $t \approx 20$.} Figure~1 of the Supplementary material shows analogous results distinguishing climatic regions; all regions show departures from stationarity and the spatial behaviour of the long-term trend differs between $\bar{I}_{t\ell}(\bs)$ and $\munderbar{I}_{t\ell}(\bs)$.

\begin{figure}[tb]
    \centering
    \includegraphics[width=.45\textwidth]{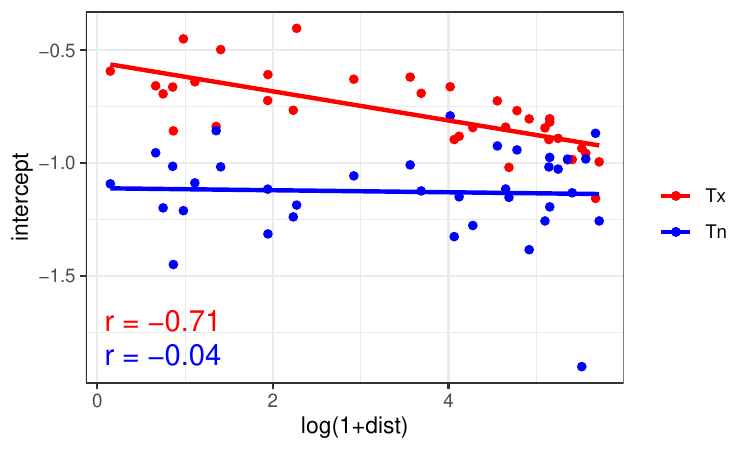}
    \includegraphics[width=.45\textwidth]{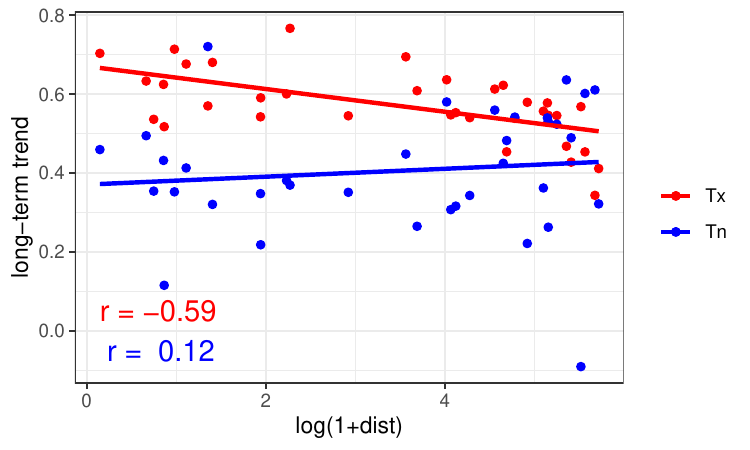}
    \caption{Scatterplots of the maximum likelihood intercept (left) and long-term trend (right) coefficients in the univariate local probit models against the logarithm of one plus distance to the coast.}
    \label{fig:eda:fx_dist}
\end{figure}

\paragraph{\textbf{Bivariate relationship}} The strength of dependence between the variables of the bivariate distribution $[\bar{I}_{t\ell}(\bs), \munderbar{I}_{t\ell}(\bs)]$ can be expressed for each year $t$ by aggregating in space and days within year and using a log odds ratio (LOR).  In particular, $LOR_{t,xn}$, where $xn$ denotes the concurrence of records between $\T_{x}$ and $\T_{n}$, is defined by
\begin{equation}
    LOR_{t,xn} = \log\frac{(n_{t,00}+0.5)(n_{t,11}+0.5)}{(n_{t,10}+0.5)(n_{t,01}+0.5)}.
\end{equation}
Here $n_{t,jk} = \sum_{i=1}^{36} \sum_{\ell \in JJA} \bone(\bar{I}_{t\ell}(\bs_i) = j, \munderbar{I}_{t\ell}(\bs_i) = k)$ for $j,k \in \{0,1\}$.  Positive, zero, and negative values reflect positive, neutral, or negative concurrence of records between $\T_x$ and $\T_n$, respectively.  The evolution of $LOR_{t,xn}$ is represented by the red line in the right plot of Figure~\ref{fig:eda:fixedeffects}; it reaches values above two, indicating strong dependence in the concurrence of both variables.

\paragraph{\textbf{Persistence terms}} The persistence of the phenomenon can be summarised using the joint distribution conditioned on the state of the previous day, $[\bar{I}_{t\ell}(\bs), \munderbar{I}_{t\ell}(\bs) \mid \bar{I}_{t,\ell-1}(\bs), \munderbar{I}_{t,\ell-1}(\bs)]$. Generalising the previous LOR notation, given $j,k,j^{\prime},k^{\prime} \in \{0,1\}$, we can define $n_{t,jk \mid j^{\prime}k^{\prime}} = \sum_{i=1}^{36} \sum_{\ell \in JJA} \bone(\bar{I}_{t\ell}(\bs_i) = j, \munderbar{I}_{t\ell}(\bs_i) = k, \bar{I}_{t,\ell-1}(\bs_i) = j^{\prime}, \munderbar{I}_{t,\ell-1}(\bs_i) = k^{\prime})$; or equivalently reduced expressions like $n_{t,j. \mid j^{\prime}.} = \sum_{i=1}^{36} \sum_{\ell \in JJA} \bone(\bar{I}_{t\ell}(\bs_i) = j, \bar{I}_{t,\ell-1}(\bs_i) = j^{\prime})$.  We explored persistence in different situations, e.g., for the occurrence of records in $\T_{x}$ conditional on the occurrence of a record in the previous day also in $\T_{x}$,
\begin{equation}
    LOR_{t,x. \mid x^{\prime}.} = \log\frac{(n_{t,0. \mid 0.}+0.5)(n_{t,1. \mid 1.}+0.5)}{(n_{t,1. \mid 0.}+0.5)(n_{t,0. \mid 1.}+0.5)}.
\end{equation}
In this notation, the subscript to the left of the vertical bar indicates the temperature considered and the subscript to the right of the bar indicates the temperature on the previous day.

The right plot in Figure~\ref{fig:eda:fixedeffects} shows $LOR_{t,x. \mid x^{\prime}.}$; the blue line predominates over all the others, indicating that the strongest dependence is associated with the occurrence of records in $\T_{x}$ for consecutive days. Close but smaller values appear for the green and purple lines, i.e., for $LOR_{t,.n \mid .n^{\prime}}$ and $LOR_{t,.n \mid x^{\prime}.}$, which denotes the occurrence of consecutive $\T_{n}$ records and the occurrence of $\T_{x}$ records given a $\T_{n}$ record the previous day, respectively. Finally, the orange line, i.e., the $LOR_{t,x. \mid .n^{\prime}}$ which accounts for the occurrence of $\T_{x}$ records given a $\T_{n}$ record the previous day, is clearly positive but much weaker than the others. Note that the particular case of $LOR_{t,xn \mid ..}$ equals $LOR_{t,xn}$, which captures the dependence of the $\T_{x}$ and $\T_{n}$ records and is shown alongside the persistence combinations for reference.

The relationships observed thus far justify including fixed effect terms for long-term trend and persistence, along with their interactions, in the model.

\paragraph{\textbf{Spatial terms}} To explore the spatial variability of these previous effects, for each location $\bs_{i}$, two local univariate probit models were fitted for $\bar{I}_{t\ell}(\bs_{i})$ and $\munderbar{I}_{t\ell}(\bs_{i})$, respectively. With the probit link, these models include the $\Phi^{-1}(1 / t)$ covariate in the linear predictor, which, on the probability scale, is $1/t$, the probability of record in the stationary record model (we give additional details in Section~\ref{sec:univariate}). The models consider an intercept, the long-term trend $\Phi^{-1}(1 / t)$, persistence terms $\bar{I}_{t,\ell-1}(\bs_{i})$ and $\munderbar{I}_{t,\ell-1}(\bs_{i})$, and the interaction between the long-term trend and the persistence terms, i.e., a total of six regression coefficients.

Plots in Figure~\ref{fig:eda:fx_dist} show the coefficients obtained through maximum likelihood in the $2 \times 36$ local models against the natural logarithm of one plus the minimum distance in km between the location and the coast, on the left for the intercept and on the right for the long-term trend coefficient. The observed relationship suggests that the distance to the coast should be included in the fixed effects, especially for $\bar{I}_{t\ell}(\bs)$, both independently and in interaction with the long-term trend.

\subsubsection{Analysis of residual variability} \label{sec:EDA_re}

We use the deviance residuals from each of the univariate local probit models as a tool to study deviations from stationarity and isotropy in the spatial dependence remaining in the residuals.

\begin{figure}[tb]
    \centering
    \includegraphics[width=.45\textwidth]{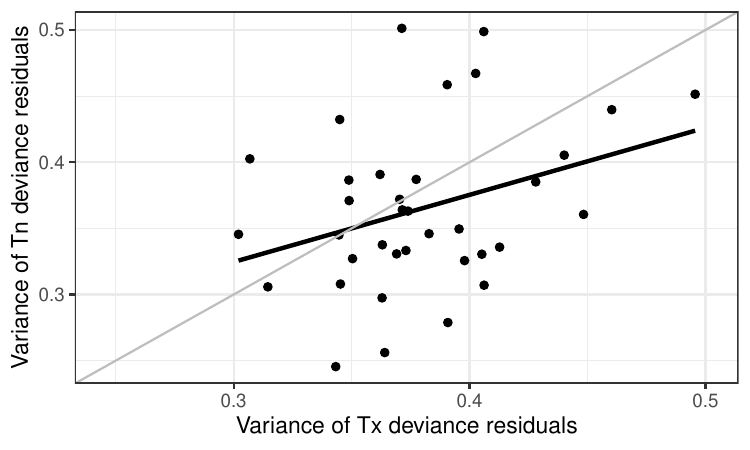}
    \includegraphics[width=0.45\linewidth]{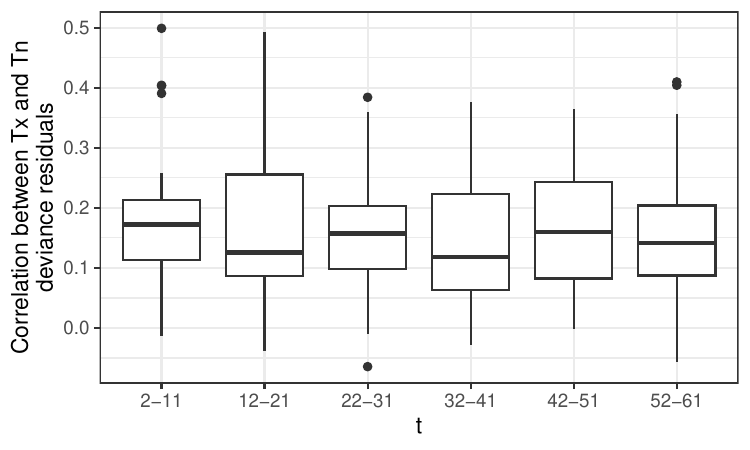}
    \caption{Left: scatterplot of the variance of the deviance residuals in the univariate local probit models of $\protect\munderbar{I}_{t\ell}(\bs_{i})$ against those of $\bar{I}_{t\ell}(\bs_{i})$. 
    Right: boxplots of the correlations between the deviance residuals for $\bar{I}_{t\ell}(\bs_{i})$ and those of $\protect\munderbar{I}_{t\ell}(\bs_{i})$ for six disjoint decades.}
    \label{fig:GLMs:var_res_xn}
\end{figure}

\paragraph{\textbf{Spatio-temporal variability in variance and covariance}} First, we examine deviations from stationarity by assessing spatial differences in the residual variance across stations. The left plot in Figure~\ref{fig:GLMs:var_res_xn} shows the variance of the residuals of $\munderbar{I}_{t\ell}(\bs_{i})$ against those of $\bar{I}_{t\ell}(\bs_{i})$ for each station. The range of the residual variances of $\bar{I}_{t\ell}(\bs_{i})$ is similar to that of $\munderbar{I}_{t\ell}(\bs_{i})$, and these variances could vary spatially.

We can also look for bivariate dependence through the deviance residuals. The top right plot shows boxplots for the correlation between the residual series of $\bar{I}_{t\ell}(\bs_{i})$ and $\munderbar{I}_{t\ell}(\bs_{i})$, i.e., the boxplots are made with 36 points, one point for the correlation between the residuals of $\bar{I}_{t\ell}(\bs_{i})$ and $\munderbar{I}_{t\ell}(\bs_{i})$ at each station. The correlations vary up to $0.5$, which should be captured by the model. Correlation between $\bar{I}_{t\ell}(\bs_{i})$ and $\munderbar{I}_{t\ell}(\bs_{i})$ could be considered constant in time but they show a spatial relationship with distance to the coast; see Figure~2 of the Supplementary material.

\begin{figure}[tb]
    \centering
    \includegraphics[width=0.45\linewidth]{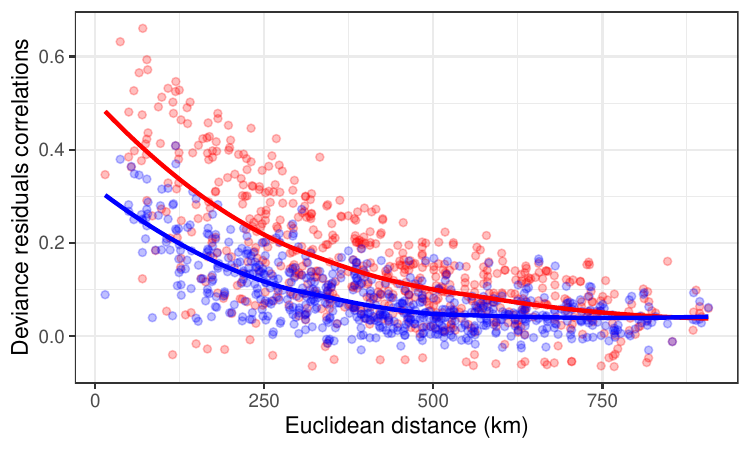}
    \includegraphics[width=0.45\linewidth]{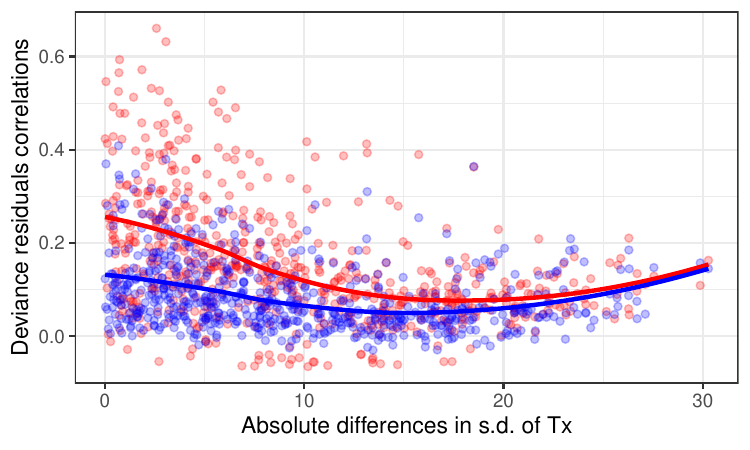}
    \caption{Scatterplot and LOESS curve of the pairwise site correlation of deviance residuals for $\bar{I}_{t\ell}(\bs_{i})$ (red) and $\protect\munderbar{I}_{t\ell}(\bs_{i})$ (blue) against Euclidean distance (left) and the differences on the logarithm of the standard deviation of $\T_{x}$ (right).}
    \label{fig:GLMs:cor_res_xn}
\end{figure}

\paragraph{\textbf{Anisotropic spatial patterns}} Finally, additional exploratory linear models were fitted to explain the spatial residual correlation through pairwise Euclidean distances between stations and differences in readily available climatic variables. That is, in addition to the usual assumption where the dependence between observations decreases with physical distance, we explore possible signs of anisotropy caused, e.g., by the orography of the region. First, we calculated the logarithm of the correlation between the deviance residuals for each pair of stations in $\bar{I}_{t\ell}(\bs_{i})$. Then, we fitted several linear models including different covariates: Euclidean distance, and differences in the logarithm of one plus distance to the coast or elevation, or in variables that express climatic conditions, like the average and standard deviation during JJA in the 30-year reference period of 1991--2020, which are themselves interpolated data products. Similar analyses were implemented for $\munderbar{I}_{t\ell}(\bs_{i})$.  The bottom plots in Figure~\ref{fig:GLMs:cor_res_xn} show the usual relationship between the residual correlation and Euclidean distance but also some relationship with the difference on the logarithm of the standard deviation of $\T_{x}$, i.e., a signal of anisotropic spatial dependence that should be considered in the modelling. Table~2 of the Supplementary material reports a comparison between these fitted linear models.

In summary, the foregoing exploratory analysis suggests that a model for the record indicators should include a long-term trend, autoregressive terms, and their interaction. Additionally, distance to the coast appears to influence both the location and the long-term trend, warranting the inclusion of an interaction term. Random effects should also be incorporated in order to account for unexplained bivariate dependence as well as the non-stationary and anisotropic structure of the spatial dependence.

\section{Modelling framework}\label{sec:model}

This section presents the modelling framework underlying our bivariate record-breaking specification. We first briefly review the univariate record-breaking model presented in \cite{castillo2024}. Then, we extend it to the bivariate setting, with more flexible modelling, resulting in the first general bivariate record-breaking model. We offer model fitting using MCMC with data augmentation, as well as subsequent prediction and validation. Further, we propose model-based inference tools for assessing joint record-breaking incidence.

\subsection{Record models}

\subsubsection{The univariate record model} \label{sec:univariate}

We describe the univariate record model \citep{castillo2024} in terms of any time series of observations indexed by year, calendar day, and location; the model can be applied to any suitable point-referenced time series. Let $I_{t\ell}(\bs)$ indicate whether, for year $t$, an observation is a record on calendar day $\ell$ and location $\bs$. The record indicators for $t > 1$ are modelled through a spatial generalised linear mixed model \citep{diggle1998} as follows:
\begin{equation} \label{eq:model}
    I_{t\ell}(\bs) \sim \text{ind. } \text{Bernoulli}(g^{-1}(\eta_{t\ell}(\bs))), \quad \text{with } \eta_{t\ell}(\bs) = \mu_{t\ell}(\bs) + v_{t\ell}(\bs),
\end{equation}
where $g$ is a link function, $\mu_{t\ell}(\bs)$ denotes fixed effects, and $v_{t\ell}(\bs)$ denotes random effects.  

We adopt the probit link $\Phi^{-1}$ in our model. An alternative would be the logit link, typically implemented via a Kolmogorov-Smirnov \citep{held2006} or Pólya-Gamma \citep{polson2013} representation. However, the probit link is more convenient computationally, as it allows for a simpler data augmentation representation with latent Gaussian variables \citep{albert1993}, i.e.,
\begin{equation}
    Y_{t\ell}(\bs) = \mu_{t\ell}(\bs) + v_{t\ell}(\bs) + \epsilon_{t\ell}(\bs), \quad \text{with } \epsilon_{t\ell}(\bs) \sim \text{i.i.d. } N(0, 1),
\end{equation}
where $I_{t\ell}(\bs) = 1$ if $Y_{t\ell}(\bs) > 0$ and $I_{t\ell}(\bs) = 0$ otherwise.\footnote{Note that $Y_{t\ell}(\bs)$ has nothing to do with a daily temperature model specification.} Note that the record indicators are conditionally independent given the linear term $\eta_{t\ell}(\bs)$, but become marginally dependent when integrating over the random effects $v_{t\ell}(\bs)$.

Anticipating the stationary record model, i.e., $I_{t\ell}(\bs) \sim \text{ind. } \text{Bernoulli}(1/t)$, as a special case of \eqref{eq:model} the fixed effects can be split in two parts, 
\begin{equation} \label{eq:trend}
    \mu_{t\ell}(\bs) = \mu_{0,t\ell}(\bs) + \mu_{1,t\ell}(\bs) g(1 / t).
\end{equation}
For example, $g(1 / t) = \Phi^{-1}(1 / t)$ with a probit link function. Note that if $v_{t\ell}(\bs)$ is a daily local adjustment modelled as a spatial Gaussian process with a spatial covariance function and the magnitude of this spatial effect is large with respect to that of the fixed effects, then the regression coefficients do not have a marginal interpretation, instead they have a conditional interpretation \citep{deoliveira2020}. Evidently, a fixed effects record model is a special case of \eqref{eq:model} obtained when $v_{t\ell}(\bs) = 0$, and the stationary record model is obtained when $\mu_{0,t\ell}(\bs) = 0$ and $\mu_{1,t\ell}(\bs) = 1$.

\subsubsection{The bivariate record model}

Turning to the bivariate case, we present the modelling in terms of daily maximum and minimum temperatures. Again, our  period of interest includes the days $\ell \in JJA$ within years $t = 1,\ldots,T$, where $1$ is year 1960 and $T = 64$ is year 2023.  The locations of interest comprise $\bs \in D$ for the study region $D$ being peninsular Spain. Again, $\bar{I}_{t\ell}(\bs)$ ($\munderbar{I}_{t\ell}(\bs)$) denotes whether, for year $t$, the maximum (minimum) temperature is a record for the daily maxima (minima) on day $\ell$ at location $\bs$. Let $\bI_{t\ell}(\bs) = (\bar{I}_{t\ell}(\bs), \munderbar{I}_{t\ell}(\bs))^{\top}$ specify a realisation of a bivariate binary process driven by $\bm{\eta}_{t\ell}(\bs) = (\bar{\eta}_{t\ell}(\bs), \munderbar{\eta}_{t\ell}(\bs))^{\top}$, where each element is analogous to \eqref{eq:model} with obvious notation for the bivariate vectors. 

In this case, $\bI_{t\ell}(\bs)$ specify a realisation of a bivariate binary process driven by a latent bivariate Gaussian process, $\bY_{t\ell}(\bs) = (\bar{Y}_{t\ell}(\bs), \munderbar{Y}_{t\ell}(\bs))^{\top}$. Now, $\bar{I}_{t\ell}(\bs) = 1$ if $\bar{Y}_{t\ell}(\bs) > 0$ ($\munderbar{I}_{t\ell}(\bs) = 1$ if $\munderbar{Y}_{t\ell}(\bs) > 0$) and $\bar{I}_{t\ell}(\bs) = 0$ ($\munderbar{I}_{t\ell}(\bs) = 0$) otherwise. Then, we would model
\begin{equation}
    \bY_{t\ell}(\bs) = \left(
    \begin{array}{c}
    \bar{\mu}_{t\ell}(\bs) + \bar{v}_{t\ell}(\bs) + \bar{\epsilon}_{t\ell}(\bs)\\
    \munderbar{\mu}_{t\ell}(\bs) + \munderbar{v}_{t\ell}(\bs) + \munderbar{\epsilon}_{t\ell}(\bs) \\
    \end{array} \right),
    \quad \text{with } \bar{\epsilon}_{t\ell}(\bs), \munderbar{\epsilon}_{t\ell}(\bs) \sim \text{i.i.d. } N(0, 1).
\end{equation}

Anticipating association between both binary processes, at the first hierarchical level of the model we consider a first-order VAR structure with shared linear predictors. In particular, $\bm{\mu}_{t\ell}(\bs) = \bB \bx_{t\ell}^{\top}(\bs)$ where $\bB = (\bar{\bm{\beta}}, \munderbar{\bm{\beta}})^{\top}$ is a $2 \times k$-matrix of autoregression and regression coefficients with $k = 8$ and 
\begin{equation}
\begin{aligned}
    \bx_{t\ell}&(\bs) = (1, \bar{I}_{t,\ell-1}(\bs), \munderbar{I}_{t,\ell-1}(\bs), \log(1+\dist(\bs)), \Phi^{-1}(1/t), \\&\Phi^{-1}(1/t) \times \bar{I}_{t,\ell-1}(\bs), \Phi^{-1}(1/t) \times \munderbar{I}_{t,\ell-1}(\bs), \Phi^{-1}(1/t) \times \log(1+\dist(\bs))).
\end{aligned}
\end{equation}
The persistence terms, $\bar{I}_{t,\ell-1}(\bs)$ and $\munderbar{I}_{t,\ell-1}$, capture first-order autoregressive dependence; the spatial term $\log(1+\dist(\bs))$, with $\dist(\bs)$ denoting the minimum distance in km between location $\bs$ and the coast, accounts for the influence of large bodies of water on the climate of an area; the long-term trend, $\Phi^{-1}(1/t)$, represents the stationary evolution of the occurrence of records; and the interactions between $\Phi^{-1}(1/t)$ and the previous effects allow them to change across years with the natural change of the occurrence of records.

At the second hierarchical level of the model, we set the two daily spatial random effects, $\bv_{t\ell}(\bs) = (\bar{v}_{t\ell}(\bs), \munderbar{v}_{t\ell}(\bs))^{\top}$, to be a bivariate correlated mean-zero Gaussian process. Using the method of coregionalisation \citep{gelfand2004}, we assume that there exist two latent mean-zero, unit-variance independent Gaussian processes with an exponential covariance function, $w_{1,t\ell}(\bs)$ and $w_{2,t\ell}(\bs)$,\footnote{More explicitly, we consider $\cov(w_{j,t\ell}(\bs), w_{j,t\ell}(\bs^{\prime})) = \sigma_{j}^{2} \exp\{-\phi_{j} \lvert\lvert \bs - \bs^{\prime} \rvert\rvert\}$ with variance $\sigma_{j}^{2} = 1$, decay $\phi_{j}$, and Euclidean distance $\lvert\lvert \bs - \bs^{\prime} \rvert\rvert$ in km between $\bs$ and $\bs^{\prime}$ using the projected coordinate reference system for peninsular Spain called Madrid 1870 (Madrid) / Spain LCC, EPSG projection 2062.} such that
\begin{equation} \label{eq:v}
    \bv_{t\ell}(\bs)
    = \bA \bw_{t\ell}(\bs), \quad
    \text{with } 
    \bA = \left(
    \begin{array}{cc}
    a_{11} & 0 \\
    a_{21} & a_{22} \\
    \end{array}
    \right)
    \text{ and }
    \bw_{t\ell}(\bs) =
    \left(
    \begin{array}{c}
    w_{1,t\ell}(\bs) \\
    w_{2,t\ell}(\bs) \\
    \end{array}
    \right).
\end{equation}
The coregionalisation matrix $\bA$ determines the correlation between the two daily spatial effects in $\bv_{t\ell}(\bs)$ and is customarily assumed to be lower-triangular. Note that $\bT = \bA \bA^{\top}$ is the constant local covariance matrix of $\bv_{t\ell}(\bs)$ that captures dependence within $\bI_{t\ell}(\bs)$. The induced correlation between $\bar{v}_{t\ell}(\bs)$ and $\munderbar{v}_{t\ell}(\bs)$ is $a = a_{11} a_{21} / (a_{11} \sqrt{a_{21}^{2} + a_{22}^{2}})$, free of $a_{11}$. The relative proportion of spatial and pure error for $\bar{I}_{t\ell}(\bs)$ and $\munderbar{I}_{t\ell}(\bs)$ can be extracted with $\bar{a} = a_{11}^{2} / (a_{11}^2 + 1)$ and $\munderbar{a} = (a_{21}^{2} + a_{22}^{2}) / (a_{21}^{2} + a_{22}^{2} + 1)$, respectively.

\subsection{Extensions of the record models}

We propose two extensions of the bivariate model that relax the assumptions of stationarity and isotropy in the Gaussian processes. Particularisations to the univariate case are immediate.

\subsubsection{Spatially varying coregionalisation} We extend the coregionalisation replacing $\bA$ by $\bA(\bs)$ and
thus substituting \eqref{eq:v} with
\begin{equation}
    \bv_{t\ell}(\bs) = \bA(\bs) \bw_{t\ell}(\bs).
\end{equation}
Again, $\bA(\bs)$ is taken to be lower triangular. Note that now the covariance matrix of $\bv_{t\ell}(\bs)$ is $\bT(\bs) = \bA(\bs) \bA(\bs)^{\top}$, a bivariate version of the case of a spatial process with a spatially varying variance, i.e., $\bv_{t\ell}(\bs)$ is no longer a stationary process. In particular, the diagonal elements, $a_{jj}(\bs)$, are modelled as a log-Gaussian process with mean $\mu_{a_{jj}}(\bs) = \bz(\bs) \bm{\beta}_{a_{jj}}$ where $\bm{\beta}_{a_{jj}}$ is a vector of regression coefficients with length $q = 2$, $\bz(\bs) = (1, \log(1+\dist(\bs)))$, and exponential covariance function with variance $\sigma_{a_{jj}}^{2}$ and decay $\phi_{a_{jj}}$. Combining distance to the coast into the Gaussian process allows to capture the spatial variations in residual variance and correlation observed in Section~\ref{sec:EDA}. Equivalently, $a_{21}(\bs)$ is a Gaussian process with mean $\mu_{a_{21}}(\bs) = \bz(\bs) \bm{\beta}_{a_{21}}$ and exponential covariance function with variance $\sigma_{a_{21}}^{2}$ and decay $\phi_{a_{21}}$.  The induced correlation, relative proportion of spatial dependence and pure error, or other quantities of interest can be averaged in space though block averages \citep[][Chapter~7]{BCG}; see Section~2.1 of the Supplementary material.

\subsubsection{Covariate information in the spatial covariance structure} 

We consider models which relax the assumption of isotropic Gaussian processes for the $w(\bs)$'s by accounting for covariate information in the covariance structure of the processes \citep{schmidt2011}.  We find that in such a geographically diverse region, the recovery of the spatial signal could be improved if in addition to using the usual Euclidean distance, one also works in an alternative climate space.

Consider $\bs,\bs^{\prime} \in D$, and let $x(\bs)$ map the locations from $D$ into a 1-dimensional covariate space. We introduce anisotropic covariance functions in $D$ using the product of two exponential correlation functions. Explicitly, we consider $\cov(w_{j,t\ell}(\bs), w_{j,t\ell}(\bs^{\prime})) = \sigma_{j}^{2} \exp\{-\phi_{j} \lvert\lvert \bs - \bs^{\prime} \rvert\rvert\} \exp\{-\phi_{x,j} \lvert x(\bs) - x(\bs^{\prime}) \rvert\}$ with variance $\sigma_{j}^{2} = 1$, and decay parameters $\phi_{j}$ and $\phi_{x,j}$ for space and covariate, respectively.  This form is valid when the support of $x(\bs)$ is $R^{1}$.

As explained in Section~\ref{sec:EDA}, we explored several options for the spatial covariate. Once the Euclidean distance between locations was considered, the residual correlation showed the most notable relationship with the differences in the climatic variable $x(\bs) = \log(s_{x}(\bs))$ where $s_{x}(\bs)$ is the standard deviation of the daily maximum temperature in JJA in the 30-year reference period of 1991--2020 at location $\bs$. While this variable helps classify the type of climate of each subregion, we want to emphasise the following concern. Since $s_{x}(\bs)$ is a product of the temperature itself, it may appear that data are being used on both the left side and right side of the modelling. Instead of using $s_{x}(\bs)$, it could be reasonable to use other covariates such as elevation or distance to the coast directly. However, this type of climatic variable is far from the response variable and allows the spatial latent processes to be related to both the orography and climatology of the region \citep[see, e.g.,][who used mean precipitation to model extreme precipitation]{cooley2007}.

\subsection{Prior specification and model fitting}

Model inference is implemented in a Bayesian framework. Adopting the data augmentation strategy, we can use the same conjugate priors as in a standard framework with normal errors. We assign proper but weakly informative independent priors as follows. Let $\bone_n$ ($\bzero_n$) be the $n$-dimensional vector of ones (zeros) and $\bI_n$ the $n \times n$-dimensional identity matrix. The regression coefficients are assigned $\bar{\bm{\beta}}, \munderbar{\bm{\beta}} \sim N_{k}(\bm{\mu}_{\bm{\beta}}, \bm{\Sigma}_{\bm{\beta}})$, where $N_{k}(\bm{\mu}_{\bm{\beta}}, \bm{\Sigma}_{\bm{\beta}})$ stands for the multivariate normal distribution of dimension $k$ with mean vector $\bm{\mu}_{\bm{\beta}} = \bzero_{k}$ and covariance matrix $\bm{\Sigma}_{\bm{\beta}} = 10^{2} \bI_{k}$. In the base model, the diagonal elements of the coregionalisation matrix $\bA$ are each assigned a half normal prior distribution or equivalently a zero-mean normal distribution truncated from below at zero and scale parameter $\sigma_{a_{jj}}^{2} = 5^{2}$. The dependence element is assigned $a_{21} \sim N(\mu_{a_{21}}, \sigma_{a_{21}}^{2})$, with mean $\mu_{a_{21}} = 0$ and variance $\sigma_{a_{21}}^{2} = 10^{2}$. Finally, after model fitting we observed that the posterior distribution of both decay parameters, the $\phi$'s, was similar. Subsequently, an inverse gamma prior was assigned to a common effective range, $3 / \phi_{j} \equiv 3 / \phi \sim IG(a_{\phi}, 3 b_{\phi})$, which indicate the distance beyond which spatial association becomes negligible. In particular, $a_{\phi} = 2$ and $b_{\phi} = 100$ impose a mean of $300$ km, roughly one-third of the maximum pairwise site distances, with infinite variance.

In the case of a spatially varying coregionalisation, the mean parameters are assigned $\bm{\beta}_{a_{jj}}, \bm{\beta}_{a_{21}} \sim N_{q}(\bm{\mu}_{a}, \bm{\Sigma}_{a})$ and the variance parameters are assigned $\sigma_{a_{jj}}^{2}, \sigma_{a_{21}}^{2} \sim IG(a_{a}, b_{a})$; with mean $\bm{\mu}_{a} = \bzero_{q}$, covariance matrix $\bm{\Sigma}_{a} = 10^{2} \bI_{q}$, shape and scale $a_{a} = b_{a} = 0.1$, respectively. Due to the usual problems in identifying both variance and decay parameters in spatial models \citep[][Chapter~6]{BCG}, the latter was assumed common for all three processes and fixed to $\phi_{a} = 1 / 300$ (a value tuned through cross validation), i.e., an effective range of $900$ km.

When we considered covariate information in the correlation structure, we also defined a common $\phi_{x,j}$ for both processes, $3 / \phi_{x,j} \equiv 3 / \phi_{x} \sim IG(a_{\phi_{x}}, 3 b_{\phi_{x}})$ with $a_{\phi_{x}} = 2$ and $b_{\phi_{x}}$ chosen by imposing a mean of one-third of the maximum pairwise covariate differences. This is roughly $b_{\phi_{x}} = 0.1$ for the covariate $x(\bs) = \log(s_{x}(\bs))$.

Computational details of the MCMC algorithm used to fit the models are given in Section~2.2 of the Supplementary material. With the exception of the variance and decay in the $\bA(\bs)$ processes, we found little inference sensitivity to the hyperparameters of the prior distribution for the regression, coregionalisation, or decay parameters.

\subsection{Prediction and model-based summary tools} \label{sec:pred_tools}

\subsubsection{Spatial prediction of record occurrence}

The inference objective is the joint posterior predictive distribution for the record indicators and their probabilities for any site in the study region within the observed time period, i.e., for $\bI_{t\ell}(\bs_0)$ and $\bp_{t\ell}(\bs_0)$ for any $\bs_0 \in D$. After fitting the bivariate model, a posterior sample from $\bI_{t\ell}(\bs_0)$ can be obtained by sampling from two independent Bernoulli distributions with probabilities $\bp_{t\ell}(\bs_0)$. A sample of $\bp_{t\ell}(\bs_0)$ is obtained through the inverse link function taking a sample of the linear predictors $\bm{\eta}_{t\ell}(\bs_0)$, which are functions of the model parameters and processes realisations. Posterior samples for the parameters are available from each MCMC iteration, and posterior samples for the spatial processes are available using posterior samples of the parameters through usual Bayesian kriging \citep[][Chapter~6]{BCG}. When we consider covariate information in the correlation structure we do not have the observed $x(\bs_{0}) = \log(s_{x}(\bs_{0}))$, so we employ simple kriging described in Section~1.4 of the Supplementary material to interpolate these values. Sampling of indicators and probabilities must be done dynamically in $\ell$ because the response variable depends on its previous day's values. 

To avoid additional modelling, starting values on the 31st of May for the autoregression, i.e., $\bI_{t,151}(\bs_{0})$, are sampled with a slightly disturbed version of the probabilities $\hat{\bp}_{t,151}$ (disturbed to avoid values of 0 or 1), where the empirical probabilities $\hat{\bp}_{t,151}$ are estimated as the respective proportion of observed records on day $151$ within year $t$ in all weather stations. Although we could consider more sophisticated strategies, the starting value has little influence on the joint inference for the series. 

Employing $G_{D}$, a fine spatial grid for $D$, posterior samples from $\bI_{t\ell}(\bs)$ and $\bp_{t\ell}(\bs)$ can be realised at each location in $G_{D}$ for every day $\ell$, $\ell \in JJA$, within year $t$, $t=2,\ldots,T$.  So, we can make inference about any feature of interest related to the occurrence of both partial and joint bivariate records.

\subsubsection{Model-based tools for assessing marginal incidence}
\label{Sec342}

First, we consider four univariate summaries proposed in \cite{castillo2024}. They can be applied equally well to maximum or minimum temperatures, so we use generic variables and add an over or under bar when appropriate. The first summary, 
\begin{equation}
    N_{t_1:t_2,\ell_1:\ell_2}(\bs) = \frac{1}{\ell_2 - \ell_1 + 1} \sum_{t=t_1}^{t_2} \sum_{\ell=\ell_1}^{\ell_2} I_{t\ell}(\bs),
\end{equation}
is the average across days from $\ell_1$ to $\ell_2$ of the cumulative number of records from year $t_1$ up to year $t_2$ at location $\bs$. The second summary, $R_{t_1:t_2,\ell_1:\ell_2}(\bs)$, is the ratio between the $N_{t_1:t_2,\ell_1:\ell_2}(\bs)$ predicted by the model and $\sum_{t=t_1}^{t_2} 1 / t$, the expected number of records under the stationary case. This summary is closely related to risk ratios using record events in \emph{extreme event attribution} \citep{naveau2020}. Computing these quantities for all locations in $G_{D}$ we can draw surfaces of their posterior mean and borders of the $90\%$ credible interval (CI). The third summary, the posterior mean of $p_{t\ell}(\bs)$ for all locations in $G_{D}$, is a surface of probabilities of a record that can be drawn to observe spatial persistence of a particular heat episode. The fourth summary is the extent of record surface (ERS) over a block $B \subseteq D$ for a given day $\ell$ within a given year $t$. Suppose we compute 
\begin{equation} \label{eq:ERS}
    \widehat{\text{ERS}}_{t\ell}(B) = \frac{1}{\lvert G_{B} \rvert} \sum_{\bs_j \in G_{B}} I_{t\ell}(\bs_j),
\end{equation}
where $G_{B}$ is a fine spatial grid of $B$ and $\lvert G_{B} \rvert$ is the number of grid cells in $G_{B}$. $\text{ERS}_{t\ell}(B)$ is defined as the limit of the average of $I_{t\ell}(\bs)$ for all locations in $G_{B}$ as the size of the grid cells goes to $0$, which in practice is approximated by \eqref{eq:ERS}. This limit can be interpreted as the proportion of $B$ which produced a record on day $\ell$ in year $t$. In this regard, we propose to use  \emph{calendar heatmaps}  that enables us to quickly identify daily and monthly patterns, and to recognise anomalies; see Section \ref{Sec345} for details.  We can average $\text{ERS}_{t\ell}(B)$ for all or some $\ell$'s, e.g., days in summer months, leading to a yearly evolution of the average number of records  all over $B$ and  plot the posterior mean of $t \times \text{ERS}_{t,JJA}(B)$ against $t=1,\ldots,T$ to assess  deviations from $1$.

Finally, while the second summary  is not useful for a joint occurrence indicator, the other summaries can be calculated by grouping joint indicators in space and time. In addition, it is also possible to group marginal record indicators into maximums and minimums in space and time as well.

\subsubsection{Model-based tools for assessing joint incidence}
\label{Sec343}

Here, interest is on joint occurrence, which cannot be directly compared with the stationary reference case $1/t^{2}$ due to the strong dependence between maximum and minimum temperatures. Instead, the objective is to measure the strength in the relationship between record-breaking events, evaluate the magnitude of persistence in space and time, compare the occurrence of records between both temperatures, and compare the joint occurrence over time between different groupings of days and years.

To measure the amount of dependence between maximum and minimum temperatures, the first summary we define is a spatial Jaccard index for joint maximum and minimum temperature records relative to a record in at least one of the temperatures, defined by
\begin{equation} \label{eq:modelbased_Jaccard}
    \frac{\Prob(\bar{I}_{t\ell}(\bs) = 1, \munderbar{I}_{t\ell}(\bs) = 1)}{\Prob(\bar{I}_{t\ell}(\bs) = 1, \munderbar{I}_{t\ell}(\bs) = 1) + \Prob(\bar{I}_{t\ell}(\bs) = 1, \munderbar{I}_{t\ell}(\bs) = 0) + \Prob(\bar{I}_{t\ell}(\bs) = 0, \munderbar{I}_{t\ell}(\bs) = 1)}.
\end{equation}
We can calculate its block average as the spatial average and average it over time periods of interest.

The second summary we propose, $\bar{N}_{t_1:t_2,\ell_1:\ell_2}(\bs) / \munderbar{N}_{t_1:t_2,\ell_1:\ell_2}(\bs)$ or $\bar{N}_{t_1:t_2,\ell_1:\ell_2}(\bs) - \munderbar{N}_{t_1:t_2,\ell_1:\ell_2}(\bs)$, considers the ratio or the difference between the number of records in the maximum and minimum temperatures in a particular period of time. It can be plotted in terms of a surface over $G_{D}$, a block average across years, or instead we can consider probability surfaces and block averages over years of quantities that cannot be estimated empirically,
\begin{equation} \label{eq:modelbased_N_max_N_min}
    \Prob(\bar{N}_{t_1:t_2,\ell_1:\ell_2}(\bs) > \munderbar{N}_{t_1:t_2,\ell_1:\ell_2}(\bs)) - \Prob(\bar{N}_{t_1:t_2,\ell_1:\ell_2}(\bs) < \munderbar{N}_{t_1:t_2,\ell_1:\ell_2}(\bs)).
\end{equation}

The third summary we propose measures the amount of change in joint occurrences across years. This is, $\bar{\munderbar{N}}_{t_1:t_2,\ell_1:\ell_2}(\bs)$ is the average across days from $\ell_1$ to $\ell_2$ of the cumulative number of joint records from year $t_1$ up to year $t_2$ at location $\bs$. We can compare these values across decades of interest  or yearly through block averages. Following the preceding ideas we can consider probability surfaces and block averages for the probabilities defined in terms of
\begin{equation} \label{eq:modelbased_N_joint}
    \Prob(\bar{\munderbar{N}}_{t_1:t_2,\ell_1:\ell_2}(\bs) > \bar{\munderbar{N}}_{t_1:t_2 - (t_2 - t_1 + 1),\ell_1:\ell_2}(\bs)) - \Prob(\bar{\munderbar{N}}_{t_1:t_2,\ell_1:\ell_2}(\bs) < \bar{\munderbar{N}}_{t_1:t_2 - (t_2 - t_1 + 1),\ell_1:\ell_2}(\bs)).
\end{equation}
The above expression measures the probability of having more joint records from $t_1$ to $t_2$ than in the immediately preceding disjoint temporal period of equal length of years.

\subsection{Model comparison and adequacy} \label{sec:metrics}

The models are compared using 10-fold cross-validation. This means that the dataset is split into 10 groups with data from four different sites for each group. Then, each group of four sites is taken as hold-out and the model is fitted to the remaining 36 sites. Finally, posterior samples from the conditional probabilities of a record for the hold-out sites are obtained using one-step ahead prediction. In the anisotropic models we also use cross-validation to krige $\log(s_{x}(\bs))$ in the hold-out sites. Indicators associated with tied records are not included in the metrics because we do not know their true value. 

The metrics used in \cite{castillo2024} for model comparison included non Bayesian versions of the area under the ROC curve (AUC) and the Brier score, and the DIC. Here, these metrics did not allow us to discriminate between the proposed models, since there is not a difference between them as large as between the models in the original paper. Consequently, we propose an alternative metric for comparing and assessing model performance. Due to the unbalanced nature of the data, it is particularly relevant to consider metrics which are primarily concerned with the occurrence of a record and where the occurrence of a record is rare relative to the absence of a record. We first consider $\bp_{t\ell}^{(b)}(\bs) = (\bar{p}_{t\ell}^{(b)}(\bs), \munderbar{p}_{t\ell}^{(b)}(\bs))^{\top}$ for $b = 1,\ldots,B$, where $b$ denotes an MCMC posterior realisation of the probabilities of record for maximum and minimum temperatures on day $\ell$, year $t$, site $\bs$. In what follows, we consider generic probabilities $p_{t\ell}(\bs)$ and indicators $I_{t\ell}(\bs)$ that, with obvious notation, can refer to three events of interest: records in the maxima $\bar{p}_{t\ell}(\bs) \equiv \Prob(\bar{I}_{t\ell}(\bs) = 1)$, records in the minima $\munderbar{p}_{t\ell}(\bs) \equiv \Prob(\munderbar{I}_{t\ell}(\bs) = 1)$, or records in both $\bar{\munderbar{p}}_{t\ell}(\bs) \equiv \Prob(\bar{\munderbar{I}}_{t\ell}(\bs) = 1) = \Prob(\bar{I}_{t\ell}(\bs) = 1, \munderbar{I}_{t\ell}(\bs) = 1)$ where $\bar{\munderbar{I}}_{t\ell}(\bs) = \bar{I}_{t\ell}(\bs) \munderbar{I}_{t\ell}(\bs)$. Note that these are all conditional probabilities given the observed previous day's indicators.

The proposed metric is a Bayesian version of the Jaccard index,
\begin{equation}
    J^{(b)} = \frac{TP^{(b)}}{TP^{(b)}+FP^{(b)}+FN^{(b)}},
\end{equation}
with $TP^{(b)} = \sum_{(t,\ell,\bs_{i}) \in A_{1}} p_{t\ell}^{(b)}(\bs_{i})$ the true positive probabilities, $FP^{(b)} = \sum_{(t,\ell,\bs_{i}) \in A_{0}} p_{t\ell}^{(b)}(\bs_{i})$ the false positive probabilities, and $FN^{(b)} = \sum_{(t,\ell,\bs_{i}) \in A_{1}} (1 - p_{t\ell}^{(b)}(\bs_{i}))$ the false negative probabilities.  Here $A_{j} = \{(t,\ell,\bs_{i}) \in \mathcal{T} \times \mathcal{L} \times \mathcal{S} \,:\, I_{t\ell}(\bs_{i}) = j\}$ is the subset of space-time indices in the period of time $\mathcal{T} \times \mathcal{L}$ and subset of weather stations $\mathcal{S}$ whose associated record indicator is $j=0$ or $j=1$. Note that $0 \leq J^{(b)} \leq 1$ measure similarity between observed and predicted values, with $1$ indicating perfect agreement. These values can be summarised by their posterior mean, which can be estimated using the empirical mean across MCMC realisations. In cross-validation we obtain the posterior mean of these metrics where $\mathcal{S}$ is each held-out group one by one.  Then we average the metric across all groups. The time period $\mathcal{T} \times \mathcal{L}$ is specified in each case when the results are shown.

Model adequacy is criticised in terms of both calibration and sharpness of the probabilistic predictions. Scatterplots showing the actual observations for specific features against summaries of their corresponding predictive distributions, based on the posterior mean and $90\%$ CI, are helpful for assessing the consistency between predictions and observations as well as the concentration of the predictive distributions.

\section{Results}\label{sec:results}

The models were fitted by scaling the covariates to have zero mean and unit variance to enhance MCMC mixing, but regression coefficients are shown on the original scale of the covariates. Two chains of the MCMC algorithm were run for each model, each starting with distinct initial values and running for $600{,}000$ iterations to collect samples from the joint posterior distribution. The first third of the samples were discarded as burn-in, and the remaining samples were thinned to $500$ from each chain. Convergence diagnostics for the MCMC samples are provided in Section~3.1 of the Supplementary material.

To present the results, we use a fine grid $G_{D}$ of points within peninsular Spain, $G_{D}$, with resolution of $25 \text{ km} \times 25 \text{ km}$ yielding $\lvert G_{D} \rvert = 790$ grid cells. Altogether, the entire posterior predictive time series result in a large dataset from which surfaces and time series are computed. For instance, we have $63$ years $\times$ $92$ days $\times$ $790$ grid centroids $\times$ $1{,}000$ replicates yielding approximately $4.6$ billion points.

\subsection{Model validation and comparison}

Table~\ref{tab:metricsKFCV} shows the metric proposed in Section~\ref{sec:metrics} for the three events of interest. Additionally, Section~3.2 of the Supplementary material presents standard classification model comparison metrics, namely AUC and DIC. The six models compared include: $M_0$, the stationary model, viewed as a baseline for improvement; $M_{1}$, univariate models maintaining the same fixed effects except for the autoregressive term of the record indicator of the other series; the intent is to demonstrate the benefits of the joint modelling; $M_{2}$, the base bivariate model; $M_{3}$, the bivariate model with spatially varying coregionalisation; $M_{4}$, the bivariate model introducing the information of the covariate $\log(s_{x}(\bs))$ into the spatial covariance structure of the spatial daily random effects; and $M_{5}$, the bivariate model bringing together the innovations of $M_{3}$ and $M_{4}$. Other models considered but not included because they presented equal or worse performance included additional interactions or other covariates in the spatial correlation function, among the variables introduced in Section~\ref{sec:EDA}. In all cases, the indicators are considered conditionally independent given the probabilities, i.e., $\Prob(\bar{I}_{t\ell}(\bs) = 1, \munderbar{I}_{t\ell}(\bs) = 1) = \bar{p}_{t\ell}(\bs) \munderbar{p}_{t\ell}(\bs)$. With regard to the cross-validation metric, two periods for the breaking of records are considered. The first employs the first 30 (non-trivial) years of the series when records are more frequent; the second employs the last 33 years of the series when records are more rare.

\begin{table}[tb]
\caption{Model comparison for records in maximum, minimum, and both temperatures. Metrics: J 1 in 1961--1990 and J 2 in 1991--2023.}
\label{tab:metricsKFCV}
\begin{tabular*}{\textwidth}{@{\extracolsep\fill}ccccccc}
  \toprule 
  & \multicolumn{6}{c}{Event / Metric} \\ 
  \cmidrule{2-7} 
  Model & \multicolumn{2}{c}{$\T_{x}$} & \multicolumn{2}{c}{$\T_{n}$} & \multicolumn{2}{c}{$\T_{x} \times \T_{n}$} \\
  \cmidrule{2-3} \cmidrule{4-5} \cmidrule{6-7} 
  & {J 1} & {J 2} & {J 1} & {J 2} & {J 1} & {J 2} \\ 
  \midrule
  $M_{0}$ & $0.115$ & $0.015$ & $0.108$ & $0.015$ & $0.068$ & $0.000$ \\
  $M_{1}$ & $0.375$ & $0.234$ & $0.273$ & $0.143$ & $0.286$ & $0.133$ \\
  $M_{2}$ & $0.375$ & $0.235$ & $0.278$ & $0.147$ & $0.289$ & $0.146$ \\
  $M_{3}$ & $0.375$ & $0.234$ & $0.281$ & $0.151$ & $0.296$ & $0.151$ \\
  $M_{4}$ & $0.382$ & $0.241$ & $0.278$ & $0.149$ & $0.292$ & $0.147$ \\
  $M_{5}$ & $0.381$ & $0.239$ & $0.281$ & $0.152$ & $0.298$ & $0.152$ \\
  \bottomrule 
\end{tabular*} 
\end{table}

The cross-validation performance in the bivariate models ($M_2$ to $M_5$) is slightly better than in the univariate models ($M_1$), especially at minimum temperatures or for joint records, while the performance of the four bivariate models is similar. In agreement with the exploratory analysis, the inclusion of anisotropy ($M_4$ and $M_5$) improves performance at maximum temperatures but barely at minimum temperatures. The opposite happens when including spatially-varying coregionalisation ($M_3$ and $M_5$), which improves performance in the minima but not in the maxima.

The posterior predictive checks assess the adequacy of the full model $M_5$ in explaining both marginal and joint record-breaking occurrences. A graphical evaluation, including plots of the number of records $N_{64,month}(\bs)$, the ratio over stationarity of the number of records in the last decade $R_{55:64,month}(\bs)$, the number of joint records in the last decade $\bar{\munderbar{N}}_{55:64,month}(\bs)$, and ERS is presented in Section~3.2 of the Supplementary material. The results demonstrate good model adequacy, although some features for minimum temperatures have a coverage slightly below the nominal level. We want to emphasise that reduced models, in particular models that did not include a spatially varying coregionalisation, suffered from a serious problem of under-dispersion for specific features in the minima or led to predictions of these quantities not so close to the observed value, despite the fact that on a daily level they offer a performance similar to $M_{5}$.

The results suggest that $M_{5}$ with spatially varying coregionalisation coefficients and anisotropy has the best overall performance and checking. The results for the selected model $M_{5}$ are shown below.

\subsection{Model parameters}

A summary of the posterior distribution of the regression coefficients of $M_{5}$ is shown in Table~\ref{tab:coefs}. It should be recalled that the interpretation of the regression coefficients is associated with the conditional distribution, both with respect to the random effects and the VAR structure.  The following results are in good agreement with the  exploratory analysis.

The model captures the persistence of the process, which yields significant autoregressive terms for both $\bar{I}_{t,\ell-1}(\bs)$ and $\munderbar{I}_{t,\ell-1}(\bs)$, except $\munderbar{I}_{t,\ell-1}(\bs)$ in $\bar{I}_{t\ell}(\bs)$, probably because its effect is already captured by $\bar{I}_{t,\ell-1}(\bs)$. All significant autoregressive terms have positive coefficients, indicating that the probability of a record is higher if in the previous day there was some type of record. The positive term $\Phi^{-1}(1/t)$ reduces the probability of record over the years, but the three interactions with the autoregressive terms that are significant have a negative coefficient indicating that this reduction is lower when in the previous day there was some type of record. Note that there is a significant effect of the autoregressive term of the other signal in both $\bar{I}_{t\ell}(\bs)$ and $\munderbar{I}_{t\ell}(\bs)$, although in the former it is in interaction with the long-term trend. The effect of the distance to the coast on $\munderbar{I}_{t\ell}(\bs)$ is not significant; however, its interaction with the long-term trend is significant in both $\bar{I}_{t\ell}(\bs)$ and $\munderbar{I}_{t\ell}(\bs)$.

\begin{table}[t]
\caption{Posterior mean and $90\%$ CI of the regression coefficients $\bar{\bm{\beta}}$ and $\protect\munderbar{\bm{\beta}}$ in $M_{5}$. The mean of the coefficients whose CI does not contain zero is marked in bold.} \label{tab:coefs}
\begin{tabular*}{\textwidth}{@{\extracolsep\fill}lrcrc}
  \toprule
  & \multicolumn{2}{c}{$\bar{\bm{\beta}}$} & \multicolumn{2}{c}{$\munderbar{\bm{\beta}}$} \\
  \cmidrule{2-3} \cmidrule{4-5}
  \multirow{-2.5}{*}{Coefficient of} & Mean & $90\%$ CI & Mean & $90\%$ CI \\ 
  \midrule
  Intercept                                           
  & $\mathbf{-0.62}$ & $(-0.73,-0.52)$ & $\mathbf{-1.08}$ & $(-1.18,-0.98)$ \\
  $\bar{I}_{t,\ell-1}(\bs)$                             
  & $\mathbf{ 0.92}$ & $( 0.83, 1.01)$ & $\mathbf{ 0.68}$ & $( 0.60, 0.76)$ \\
  $\munderbar{I}_{t,\ell-1}(\bs)$                       
  & $-0.03$ & $(-0.13, 0.08)$ & $\mathbf{ 0.79}$ & $( 0.71, 0.87)$ \\
  $\log(1 + \dist(\bs))$                                    
  & $\mathbf{-0.05}$ & $(-0.09,-0.02)$ & $ 0.00$ & $(-0.02, 0.02)$ \\
  $\Phi^{-1}(1/t)$                                      
  & $\mathbf{ 0.92}$ & $( 0.84, 0.99)$ & $\mathbf{ 0.74}$ & $( 0.68, 0.81)$ \\
  $\Phi^{-1}(1/t) \times \bar{I}_{t,\ell-1}(\bs)$       
  & $\mathbf{-0.11}$ & $(-0.17,-0.05)$ & $ 0.02$ & $(-0.03, 0.07)$ \\
  $\Phi^{-1}(1/t) \times \munderbar{I}_{t,\ell-1}(\bs)$ 
  & $\mathbf{-0.15}$ & $(-0.22,-0.09)$ & $\mathbf{-0.10}$ & $(-0.15,-0.04)$ \\
  $\Phi^{-1}(1/t) \times \log(1 + \dist(\bs))$              
  & $\mathbf{ 0.24}$ & $( 0.22, 0.27)$ & $\mathbf{ 0.09}$ & $( 0.07, 0.10)$ \\
  \bottomrule
\end{tabular*}
\end{table}

The posterior distribution of the block average of the coregionalisation coefficients from $M_{5}$, the induced correlation, and the ratio between spatial dependence and pure error in each temperature signal are shown in Table~\ref{tab:a}. The posterior mean of the correlation between the random effects for maximum and minimum signals is $0.59$, revealing the need to consider this dependence in the model. The ratio of spatial dependence to pure noise is beyond 80\% for the maxima, while for the minima it is close to 70\%. This could reflect the fact that the minimum temperature is usually more influenced by local factors, such as evening storms, sea breezes, and local wind circulation in the valley \citep{pena2015}. Additional results including maps and tables for the coregionalisation parameters are shown in Section~3.3 of the Supplementary material; we note the strong relationship between $a_{11}(\bs)$ and the distance to the coast.

\begin{table}[tb]
\caption{Posterior mean and $90\%$ CI of the block average of the coregionalisation parameters in $\bA(\bs)$, the induced correlation $a(\bs)$, and the proportion of spatial dependence and pure error in $\bar{I}_{t\ell}(\bs)$ and $\protect\munderbar{I}_{t\ell}(\bs)$, $\bar{a}(\bs)$ and $\protect\munderbar{a}(\bs)$, respectively.} \label{tab:a}
\begin{tabular*}{\textwidth}{@{\extracolsep\fill}lcc}
  \toprule
  Block average of & Mean & $90\%$ CI \\ 
  \midrule
  $a_{11}(\bs)$ & $2.35$ & $(2.28,2.44)$ \\
  $a_{22}(\bs)$ & $1.16$ & $(1.08,1.24)$ \\
  $a_{21}(\bs)$ & $0.85$ & $(0.80,0.91)$ \\
  $a(\bs)$      & $0.59$ & $(0.56,0.62)$ \\
  $\bar{a}(\bs)$ & $0.84$ & $(0.83,0.85)$ \\
  $\munderbar{a}(\bs)$ & $0.67$ & $(0.65,0.69)$ \\
  \bottomrule
\end{tabular*}
\end{table}

The decay parameters of the anisotropic spatial processes in the model impose an effective range $3 / \phi$ of $1964$ $(1886,2039)$ km and $3 / \phi_{x}$ of $5.5$ $(5.1,6.0)$ units. Note that these effective ranges are very wide due to the consideration of both types of distance, in the physical space $D$ and the covariate space, simultaneously. The anisotropic correlation pattern in $M_{5}$ is compared to the isotropic pattern in $M_{3}$ in Section~3.4 of the Supplementary material. The anisotropy allows to capture the effect of the relief in the correlation pattern.

\subsection{Inferential results over peninsular Spain}

This section shows how the fitted model can be used to offer answers to the questions raised in the Introduction which have not yet been answered, using some of the model-based tools described in Section~\ref{Sec342}.

\subsubsection{Could the actual record rates be observed without the influence of climate change?}

\begin{figure}[tb]
\centering
\includegraphics[width=0.49\textwidth]{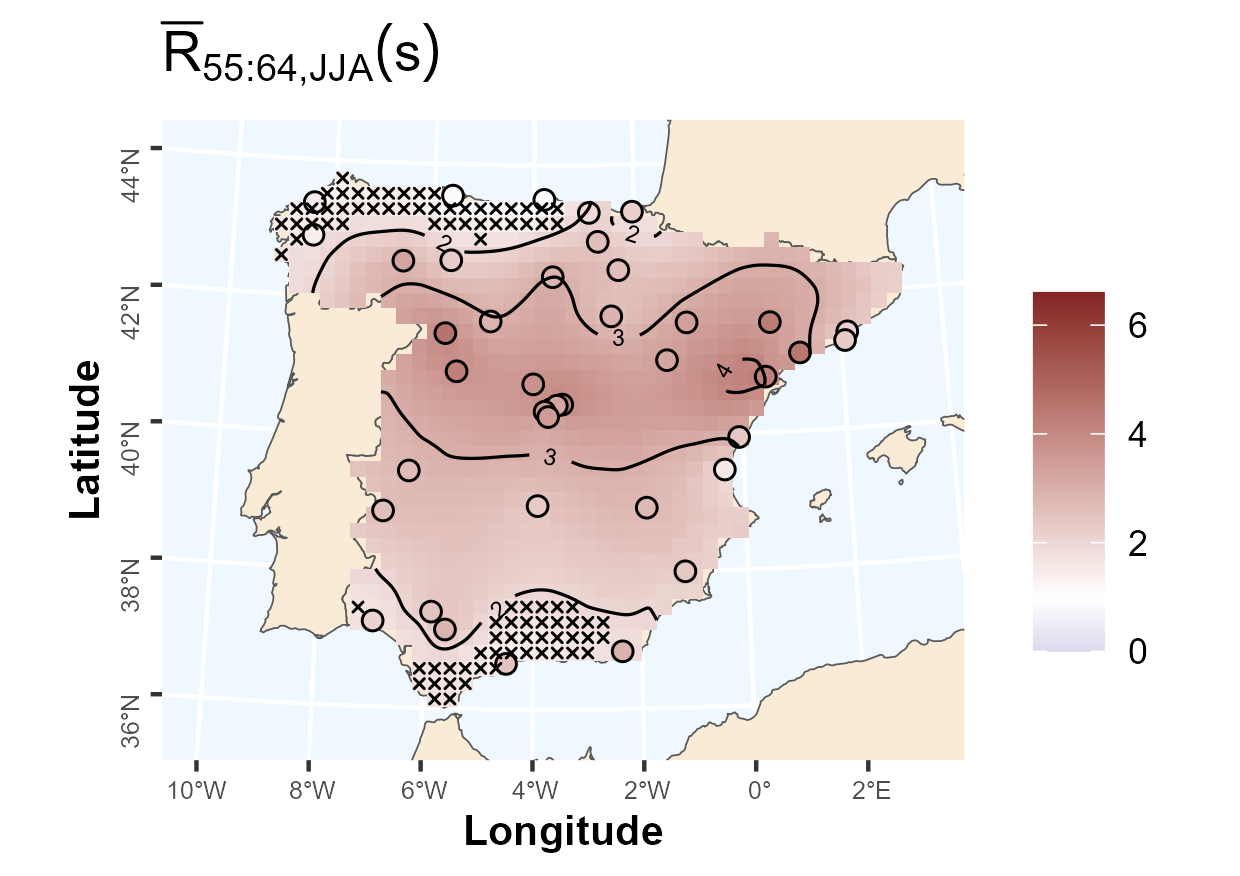}
\includegraphics[width=0.49\textwidth]{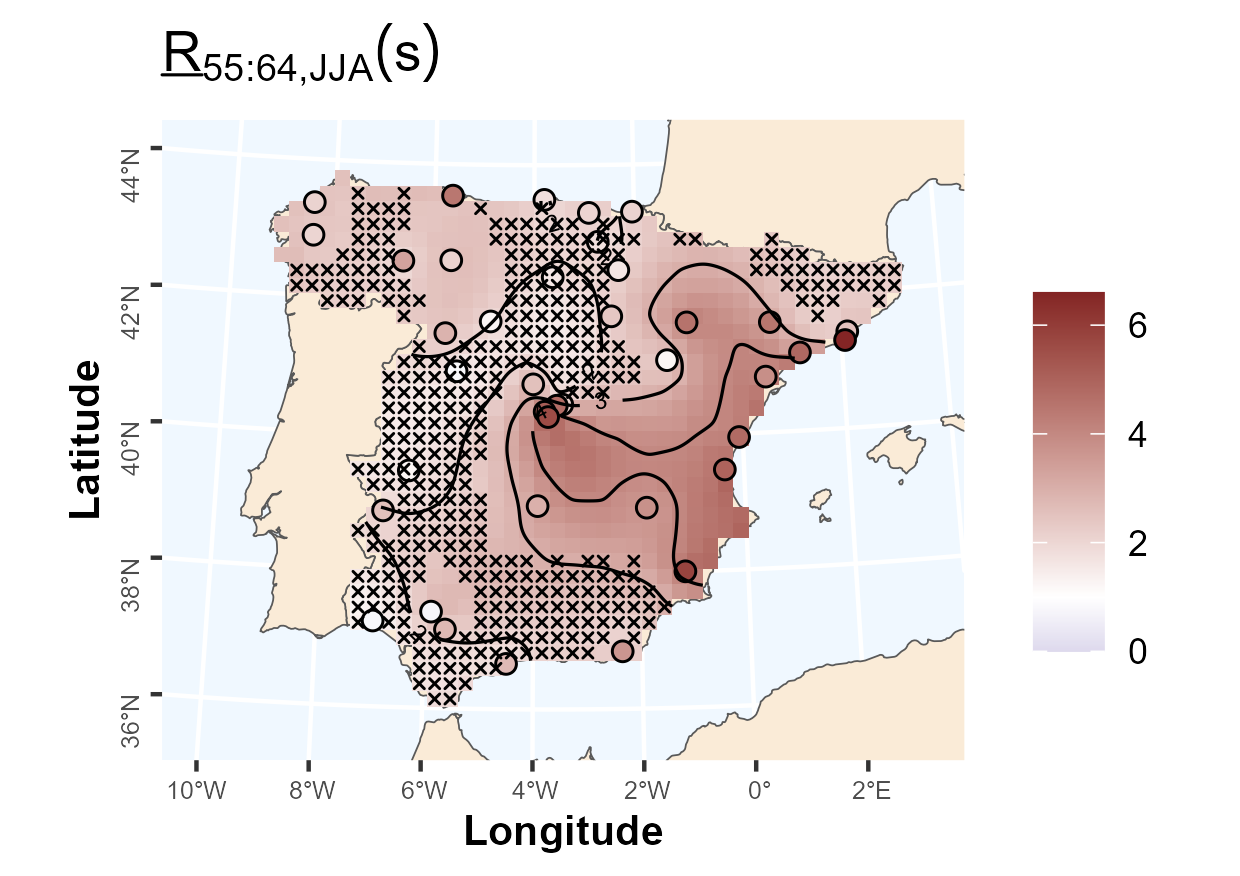}
\caption{Maps of the posterior mean of the ratios in the last decade (2014--2023), $\bar{R}_{55:64,JJA}(\bs)$ and $\protect\munderbar{R}_{55:64,JJA}(\bs)$. Grid points where the $90\%$ CI covers the stationary case are marked with a cross. Contour lines are plotted at levels 1 (CRM), 2, 3, 4. Empirical values at each location are displayed with circles using the same colour scale.}
\label{fig:R}
\end{figure}

To investigate the effects of global warming, extreme event attribution in climate science typically compares observed changes with a stationary scenario; see, e.g., \cite{naveau2020}.
An advantage of using records for this purpose is that the probability of their occurrence is known for any series of i.i.d. continuous variables, which makes it easier to evaluate deviations from stationarity.
To study these deviations at a particular location $\bs$, we can use the ratios of the number of records predicted by the model to those expected under stationarity. As an example, Figure~\ref{fig:R} shows the maps of the posterior mean of the ratios for the last decade (2014--2023) of maximum and minimum temperatures, $\bar{R}_{55:64,JJA}(\bs)$ and $\munderbar{R}_{55:64,JJA}(\bs)$, respectively. They provide evidence that the current record rates for both temperature series are higher than in a stationary climate for a large part of the region, although the spatial pattern is very different between the two signals.  While the minimum temperatures exhibit lower significance, they show the greatest warming, reaching values up to four times higher than expected in some areas, mainly in the Mediterranean region. In contrast, maximum temperatures are significant across most of the region, except  along the Cantabrian  and the southern Mediterranean coasts, where they appear consistent with a stationary climate.

\begin{figure}[tb]
\centering
\includegraphics[width=0.49\textwidth]{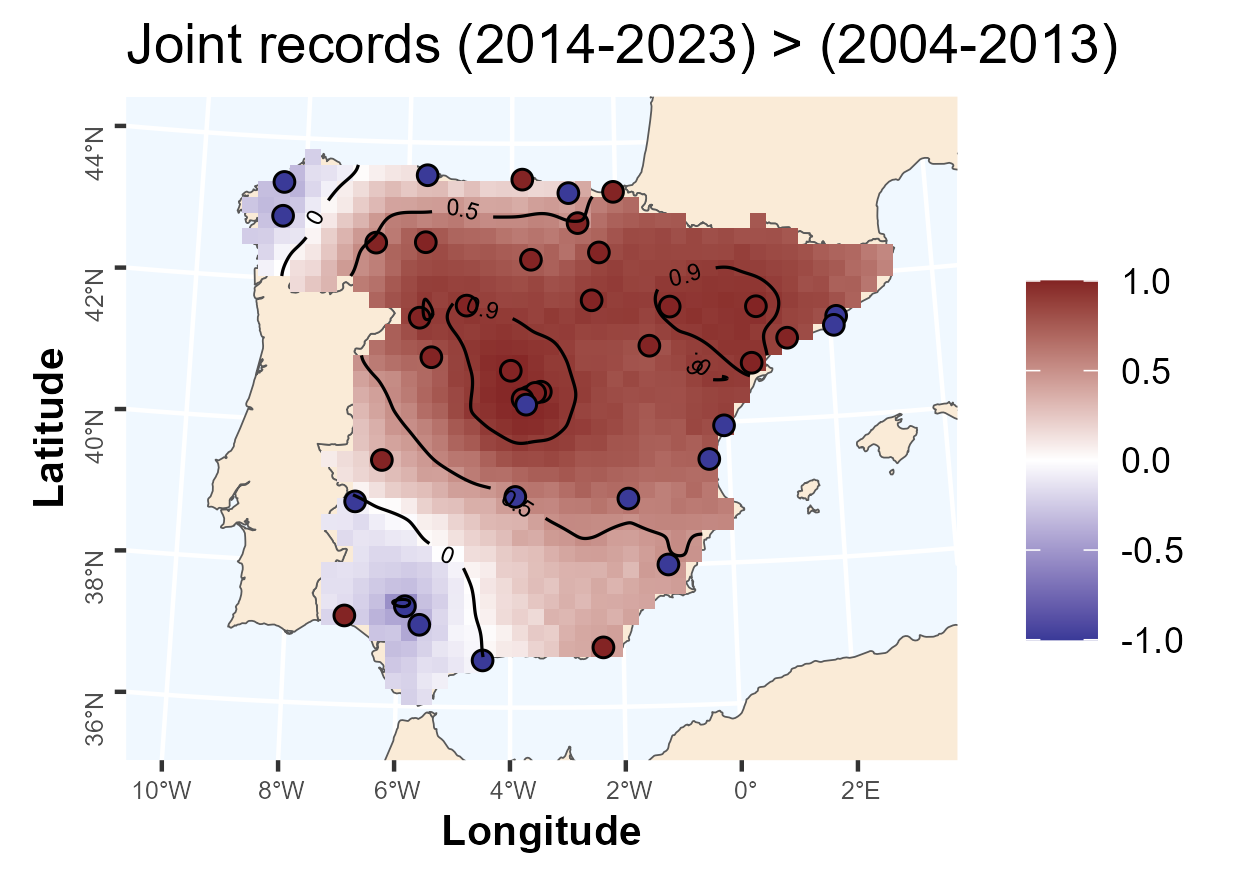}
\includegraphics[width=0.49\textwidth]{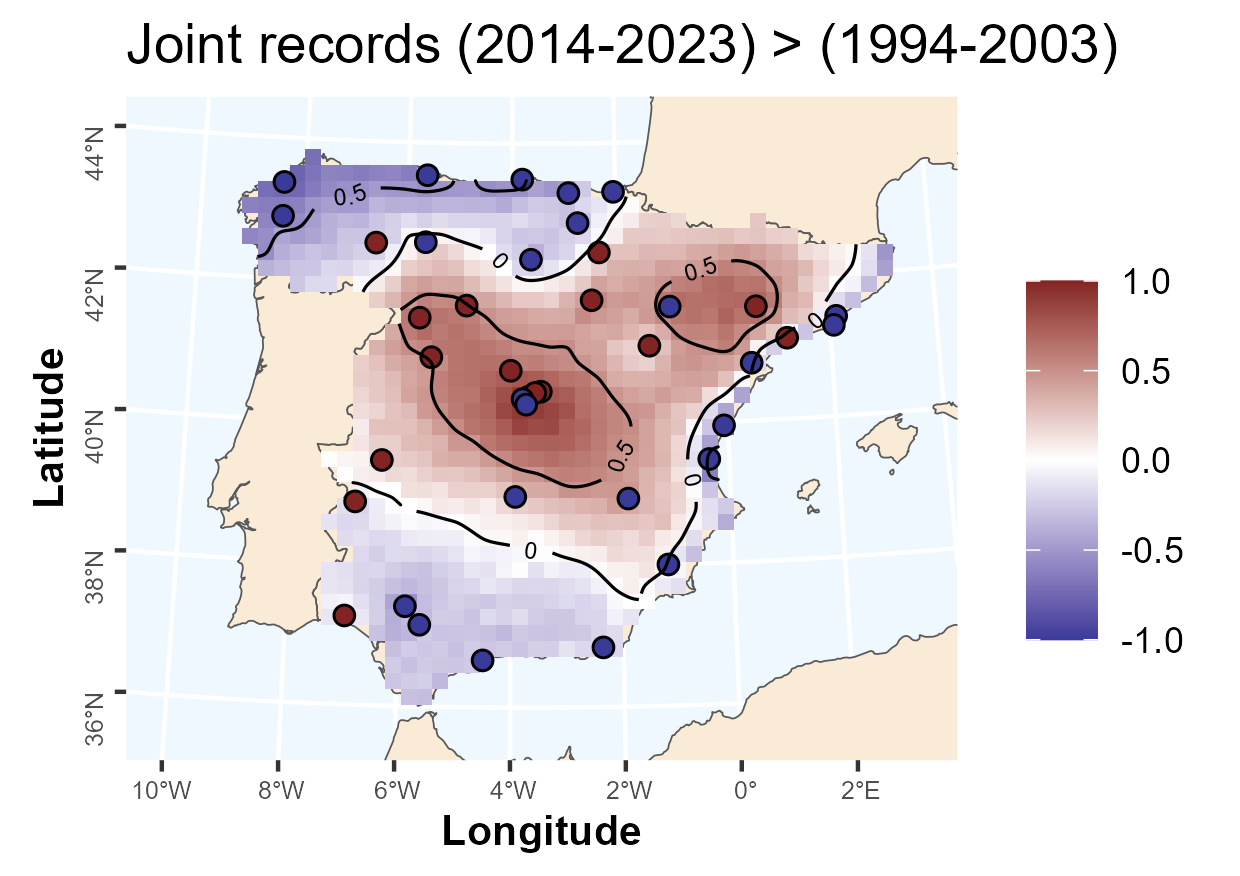}
\caption{Maps of the difference in probabilities of having more joint records in the last decade 2014--2023 than in previous decades 2004--2013 (left) and 1994--2003 (right). Contour lines are plotted at levels $-0.9, -0.5, 0, 0.5, 0.9$.  Empirical values at each location are displayed with circles using the same colour scale.}
\label{fig:diff_joint}
\end{figure}

Assessing deviation from stationarity in the occurrence of joint records is more challenging since, unlike marginal records, the probability of joint records under stationarity lacks a general expression due to the dependence between maximum and minimum temperatures.  However, model-based tools for evaluating joint incidence, such as the difference in probabilities comparing the number of joint records in two disjoint periods in \eqref{eq:modelbased_N_joint}, are useful for studying their evolution over time. Figure~\ref{fig:diff_joint}  shows the difference in probabilities of having more joint records in the last decade (2014--2023) compared to the previous decade (2004--2013) and two decades before (1994--2003). In a stationary climate, those probabilities decrease over time; however, the differences in probabilities for the previous decade remain above zero in most areas, exceeding $0.5$ in the northern half of Spain, with the exception of Galicia and some areas in the Cantabrian coast. Even the difference comparing decades 2014--2023 and 1994--2003 are above zero in the inland regions of the centre and northern half of Spain.

Another useful tool to study deviations from stationarity aggregated across space is the average of the ERS over periods of interest; this is illustrated in Section~\ref{Sec345}.

\subsubsection{How strong is the relationship between the occurrence of record-breaking maximum and minimum temperature events on the same day?}

\begin{figure}[tb]
\centering
\includegraphics[width=0.49\textwidth]{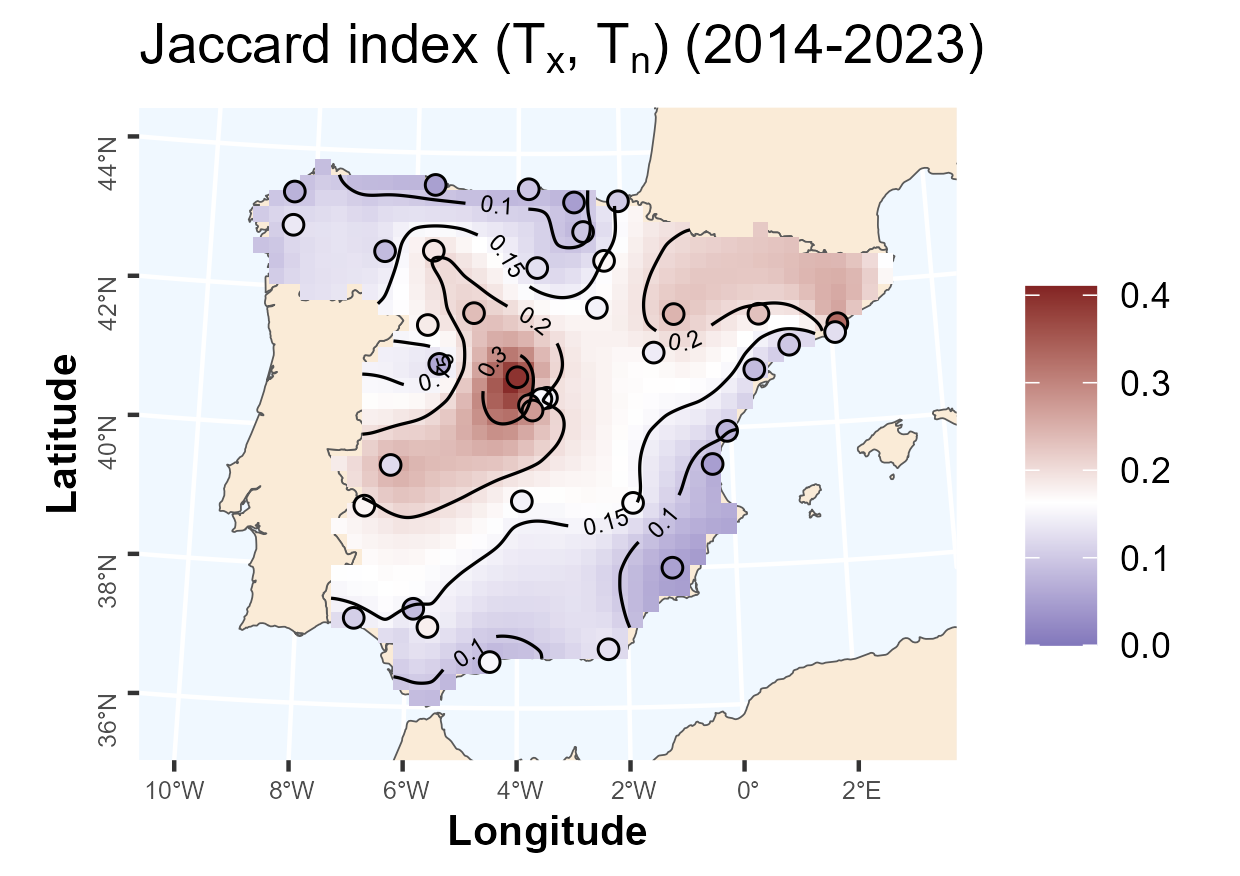}
\includegraphics[width=0.49\textwidth]{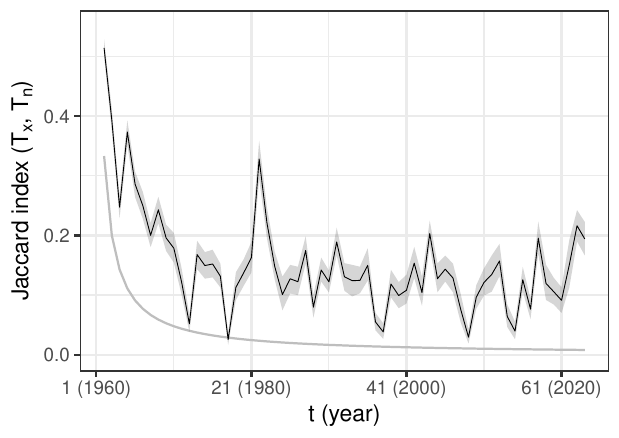}
\caption{Left: posterior mean of the Jaccard index map for the probabilities of record in 2014--2023, $\ell \in JJA$. Contour lines are plotted at levels $0.1, 0.15, 0.2, 0.3$ and empirical values at each location are displayed with circles. Right: posterior mean (black line) and $90\%$ CI (grey ribbon) of the Jaccard index time series of block averages against $t$; grey line is the Jaccard index  for two independent  stationary time series.}\label{fig:jaccard}
\end{figure}

Given that the occurrence of records in maximum and minimum temperatures is represented by binary variables, the dependence between them is evaluated using the Jaccard index in \eqref{eq:modelbased_Jaccard}. This index for two independent i.i.d. series at time $t$ is $1/(2t-1)$, but there is no general reference value for independent series that are not stationary.
The left plot in Figure~\ref{fig:jaccard} shows the spatial pattern of  the posterior mean of the  Jaccard index computed across $ \ell \in JJA$ in the last decade (2014--2023). The highest values, above  $0.15$, are observed in the Inner Plateau and the Ebro Basin, decreasing to values around $0.10$ in the Cantabrian and southern Mediterranean coast. The right plot studies the temporal evolution using the posterior mean and $90\%$ CI of the block average of the Jaccard index averaged in summer across years. The Jaccard index, after the initial period where the probabilities of records are high, remains stable around $0.12$, far from the value for two independent stationary time series.

\subsubsection{Can we learn about persistence in record-breaking from the previous day's events?}

The dependence between maximum and minimum temperatures may manifest not only in the occurrence of record-breaking events on the same day, but also on consecutive days, leading to persistence in the occurrence of records in both temperatures. It is particularly interesting to study the persistence of records in $\T_{x}$ on day $\ell-1$, followed by a record in $\T_{n}$ on day $\ell$, since consecutive hot days and nights increase the risk of health problems \citep{roye2017,he2022}.
This persistence may be measured in terms of conditional probabilities by considering the ratio between $\Prob(\munderbar{I}_{t\ell}(\bs) = 1 \mid \bar{I}_{t,\ell-1}(\bs) = 1)$ and $\Prob(\munderbar{I}_{t\ell}(\bs) = 1 \mid \bar{I}_{t,\ell-1}(\bs) = 0)$. These probabilities can be estimated for one day or for a period of interest by averaging appropriately the indicators obtained from the model. Values close to 1 show independence while values higher than 1 suggest persistence in the occurrence of records.

\begin{figure}[tb]
\centering
\includegraphics[width=0.32\textwidth]{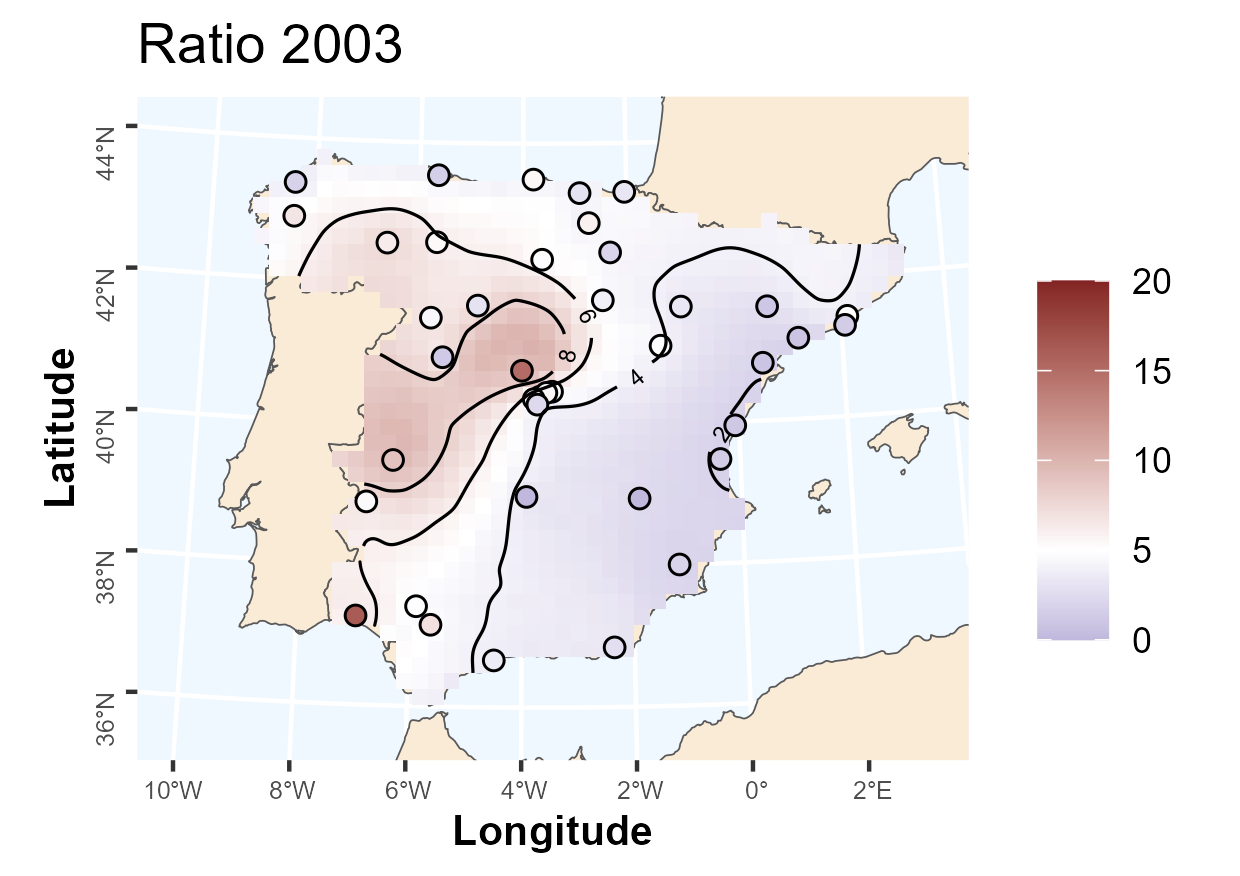}
\includegraphics[width=0.32\textwidth]{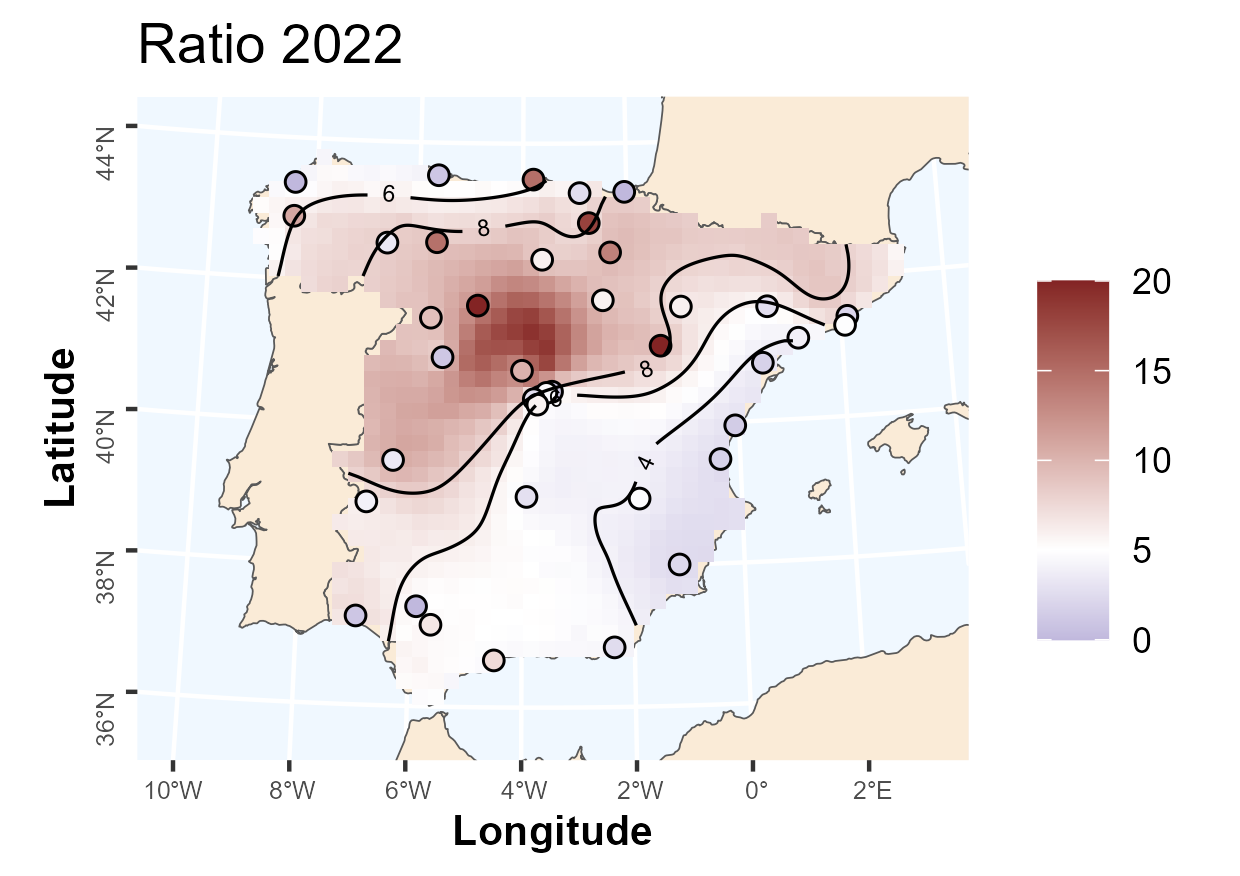}
\includegraphics[width=0.32\textwidth]{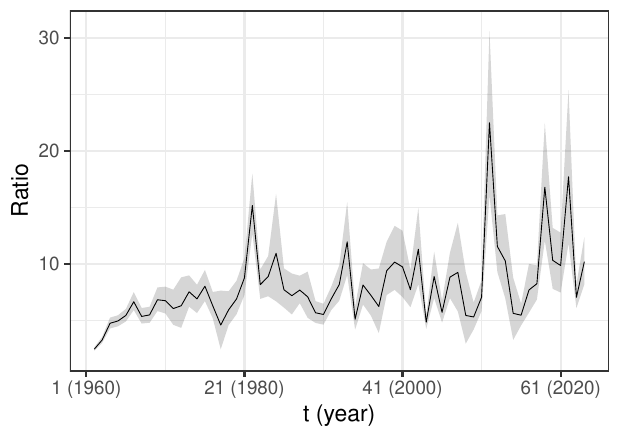}
\caption{Left and centre: maps of the ratio between $\Prob(\protect\munderbar{I}_{t\ell}(\bs) = 1 \mid \bar{I}_{t,\ell-1}(\bs) = 1)$ and $\Prob(\protect\munderbar{I}_{t\ell}(\bs) = 1 \mid \bar{I}_{t,\ell-1}(\bs) = 0)$ for $t=44$ (year 2003) and $t=63$ (2022). Contour lines are plotted at levels $2, 4, 6, 8$ and empirical values at each location are displayed with circles. Right: ratio of the yearly probabilities computed across space  against $t$.} \label{fig:persistence}
\end{figure}

Figure~\ref{fig:persistence} shows the ratio of the probabilities computed for two hot years over peninsular Spain, 2003 and 2022. The ratios in 2003 are higher than 2 in almost the entire area, providing clear evidence of persistence; however, there are significant spatial differences, with ratios lower than 4 along the Mediterranean coast and approaching 6 in the northern Inner Plateau.  By 2022, these ratios have increased further, reaching nearly 9 in the northwest region.  The increase in the persistence of record occurrences over time is confirmed by the plot of the ratio of the yearly probabilities averaged across space, also shown in Figure~\ref{fig:persistence}; the plot also indicates an increasing inter-annual variability.

\subsubsection{Is the occurrence of maximum and minimum temperature records similar over time and across space?}

\begin{figure}[tb]
\centering
\includegraphics[width=0.32\textwidth]{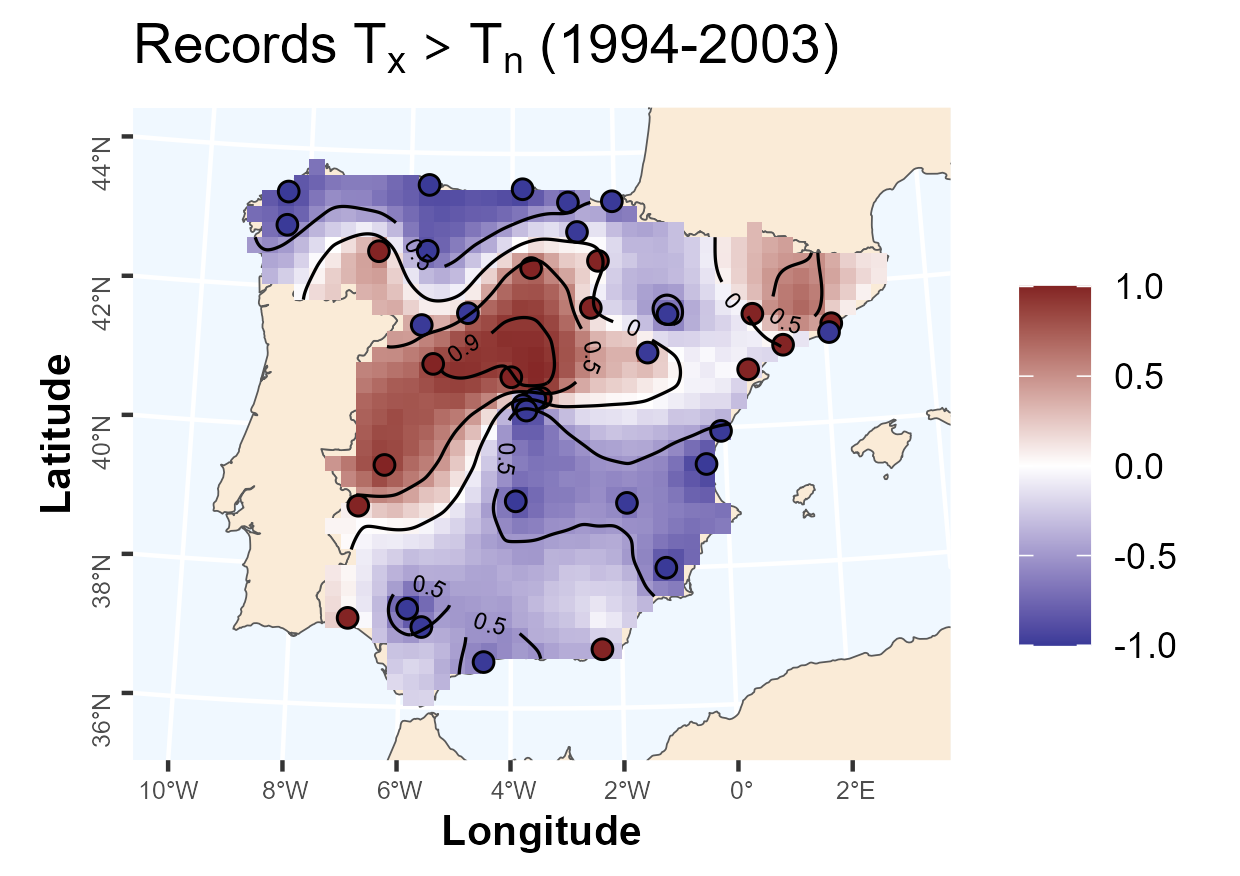}
\includegraphics[width=0.32\textwidth]{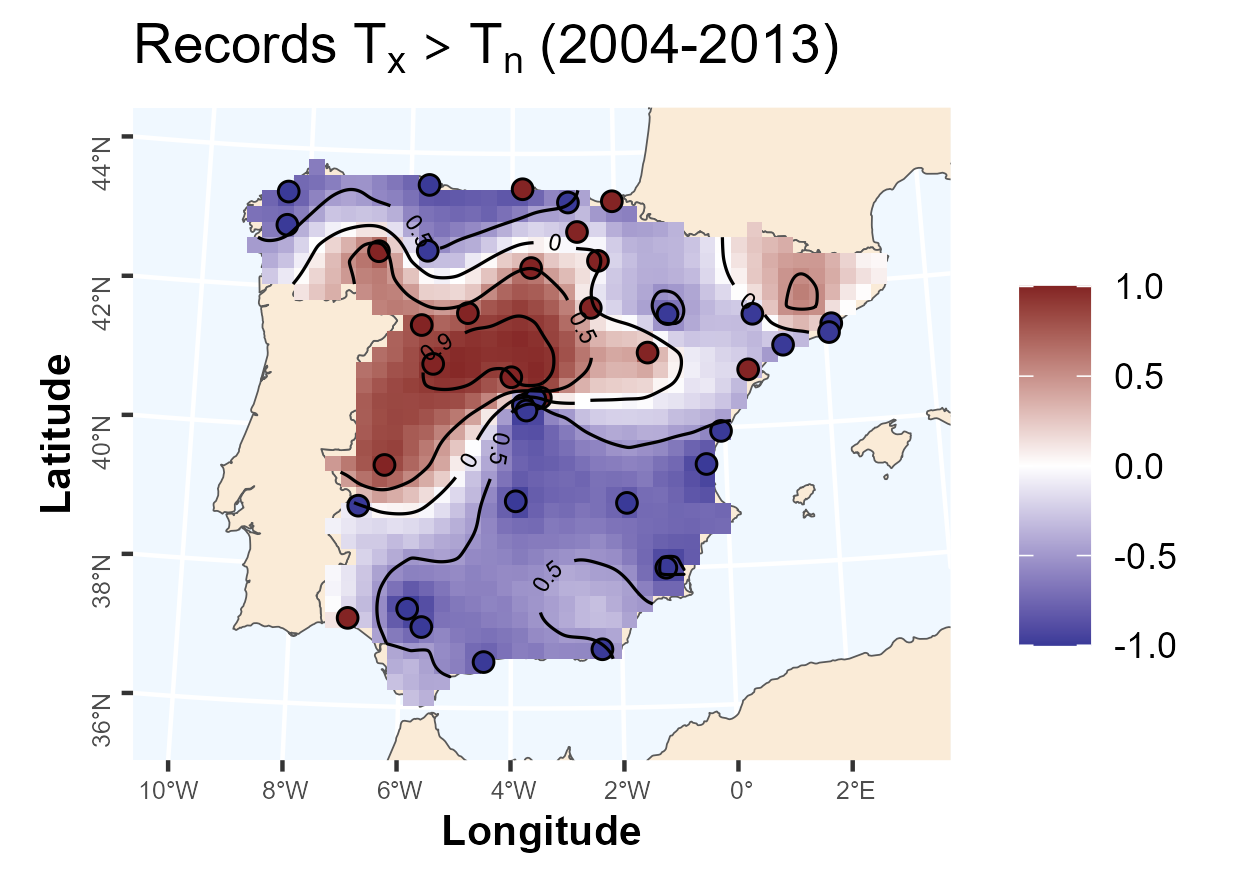}
\includegraphics[width=0.32\textwidth]{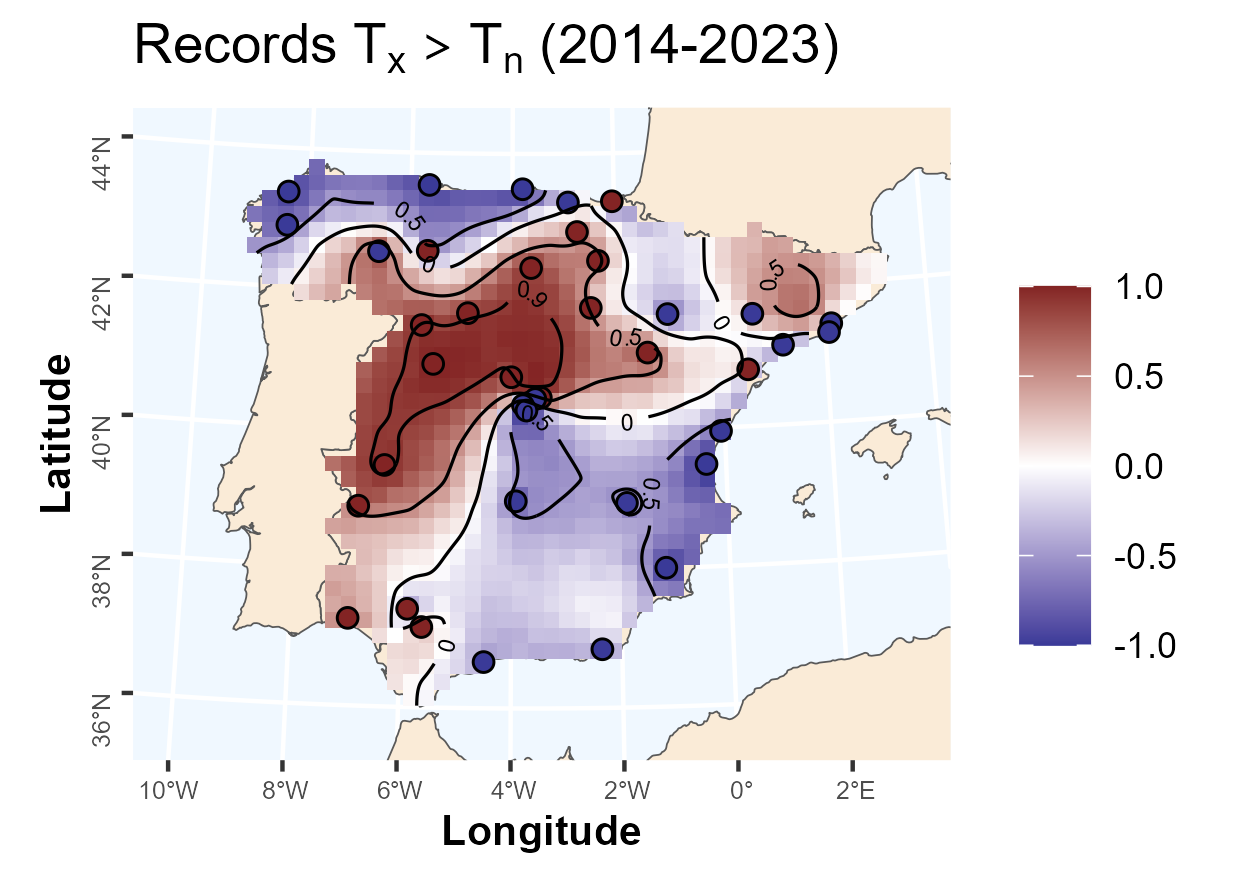}
\caption{Maps of the difference in probabilities that the number of records in maximum temperature is higher than in minimum temperature for the decades 1994--2003 (left), 2004--2013 (centre), and 2014--2023 (right). Contour lines are plotted at levels $-0.9, -0.5, 0, 0.5, 0.9$ and empirical values at each location are displayed with circles.  }\label{fig:diff_Tx_Tn}
\end{figure}

To compare the patterns over time and across space of the occurrence of records in maximum and minimum temperatures, Figure~\ref{fig:diff_Tx_Tn} shows the difference in probabilities that the number of records in maximum temperature is higher than in minimum temperature as defined in  \eqref{eq:modelbased_N_max_N_min}, for the decades 1994--2003, 2004--2013, and 2014--2023. The three decades show very similar spatial patterns, the differences are below zero across most of the area, except in some regions of the western Inner Plateau and Cataluña. This suggests that the number of records in minimum temperature tends to be higher than in maximum temperature. The differences are notably negative in areas along the Mediterranean coast, especially in the most recent decade, suggesting that in that region the number of records in minimum temperature is increasing relative to the number in maximum temperature.

\subsubsection{Can we predict joint record-breaking behaviour at unobserved locations  and develop spatial record-breaking surfaces?}

\begin{figure}[tb]
\centering
\includegraphics[width=0.24\textwidth]{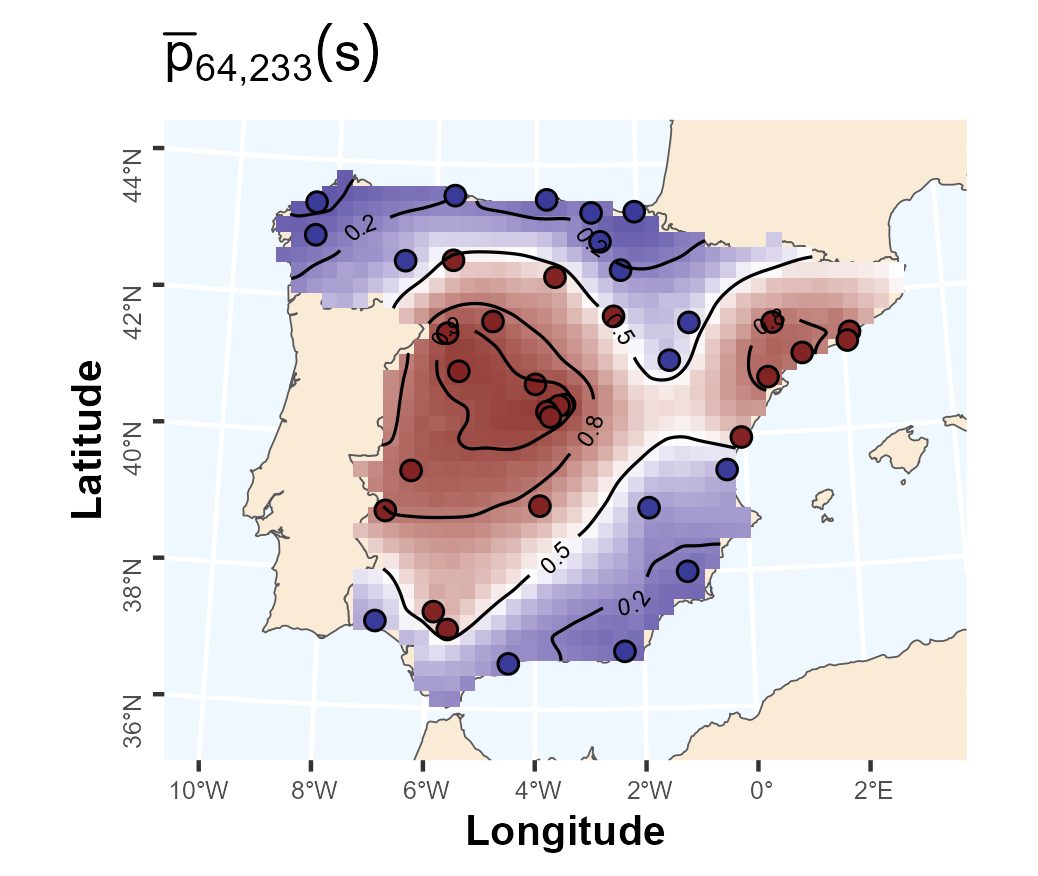}
\includegraphics[width=0.24\textwidth]{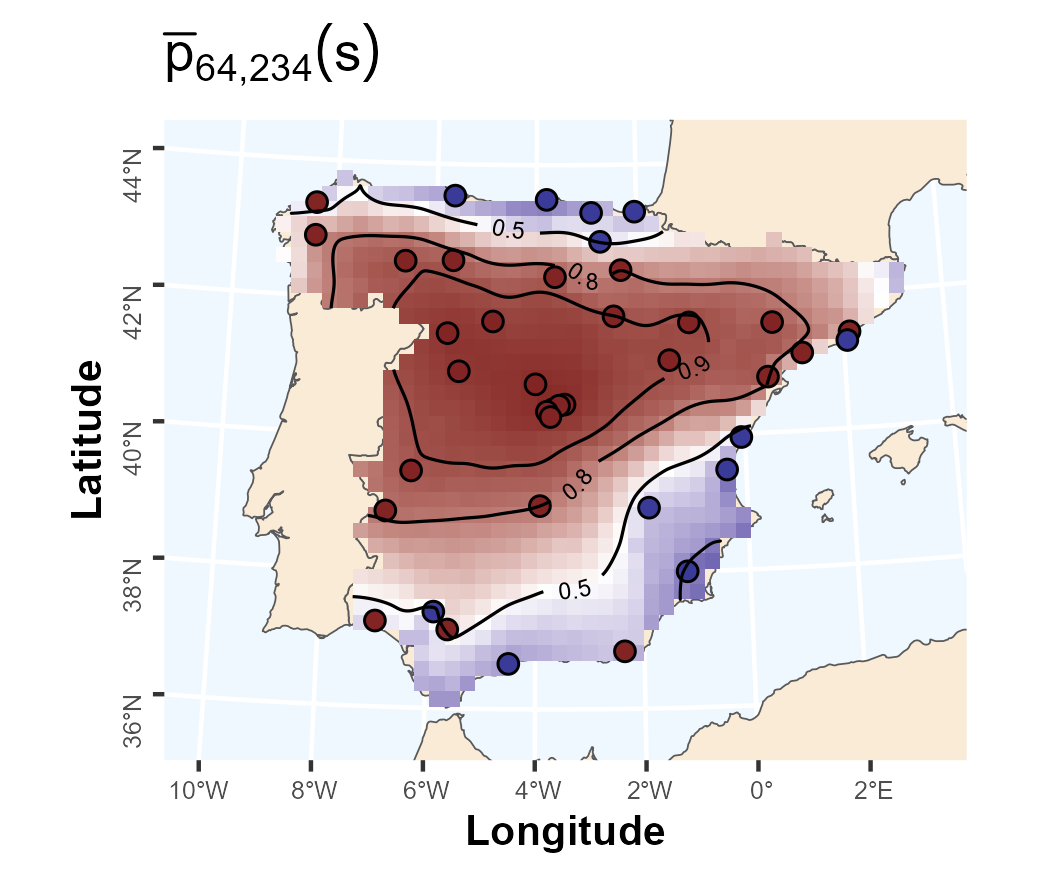}
\includegraphics[width=0.24\textwidth]{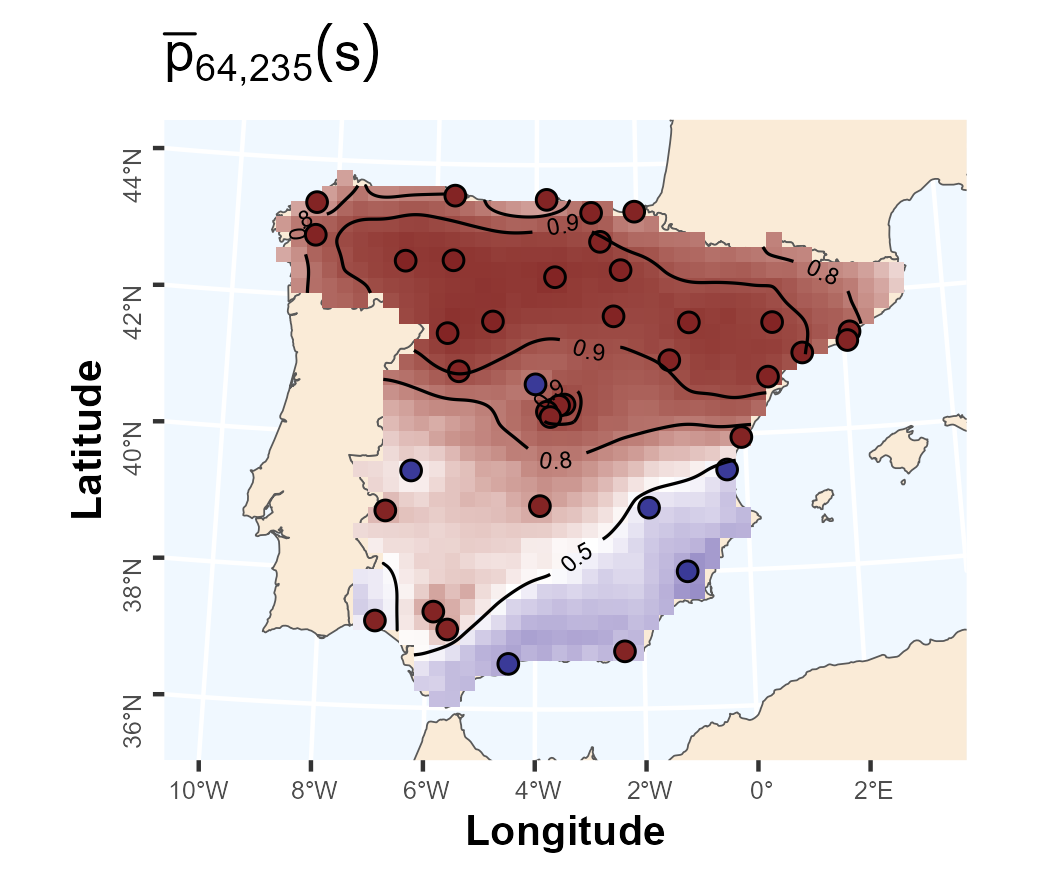}
\includegraphics[width=0.24\textwidth]{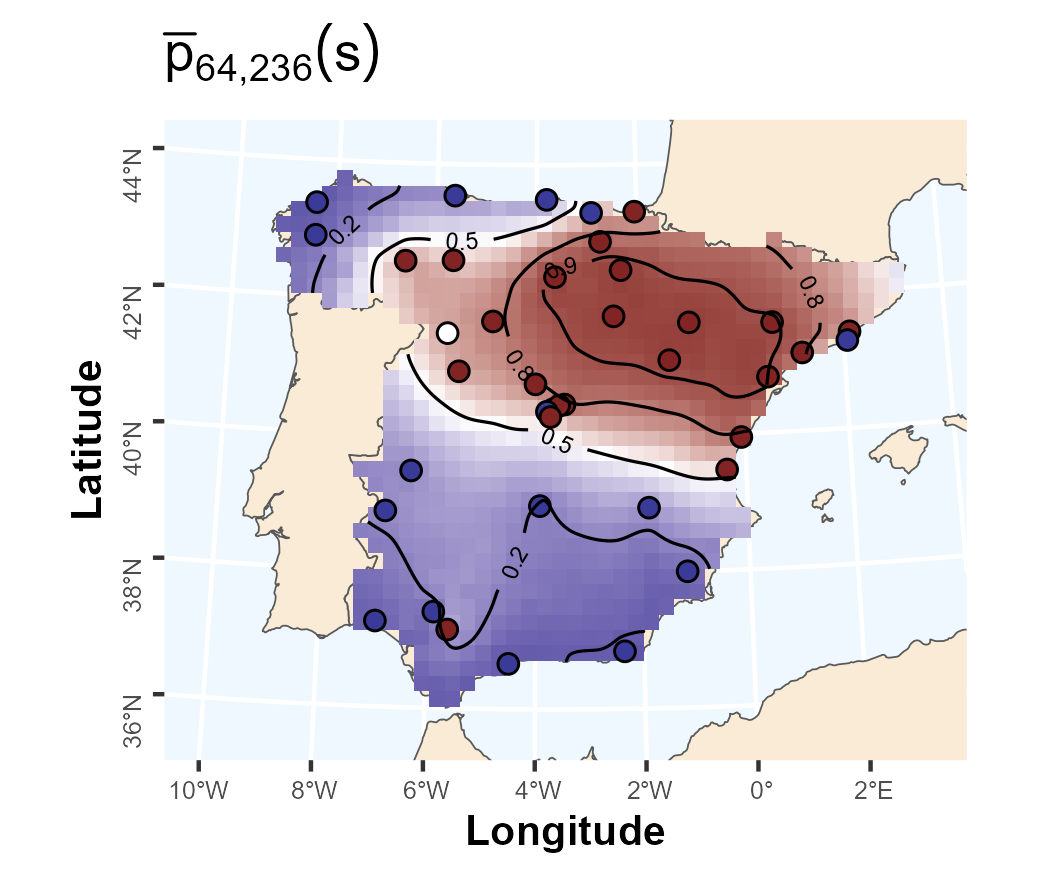}
\includegraphics[width=0.24\textwidth]{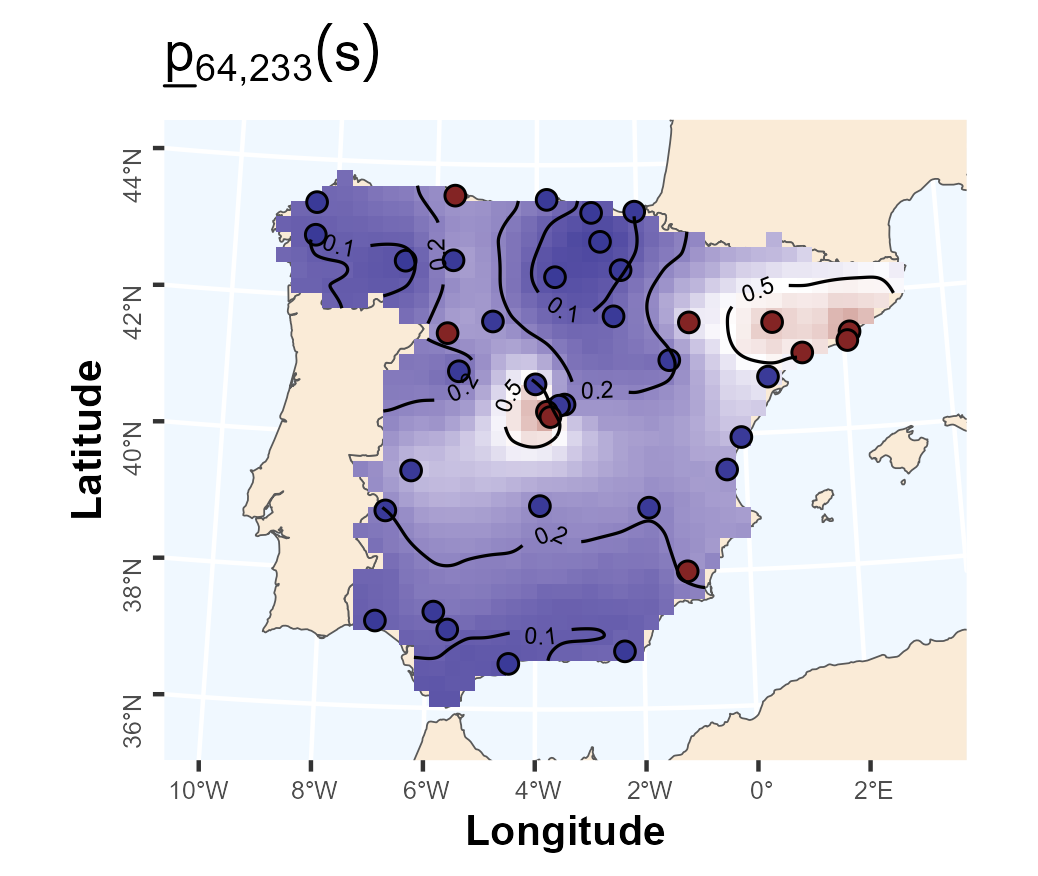}
\includegraphics[width=0.24\textwidth]{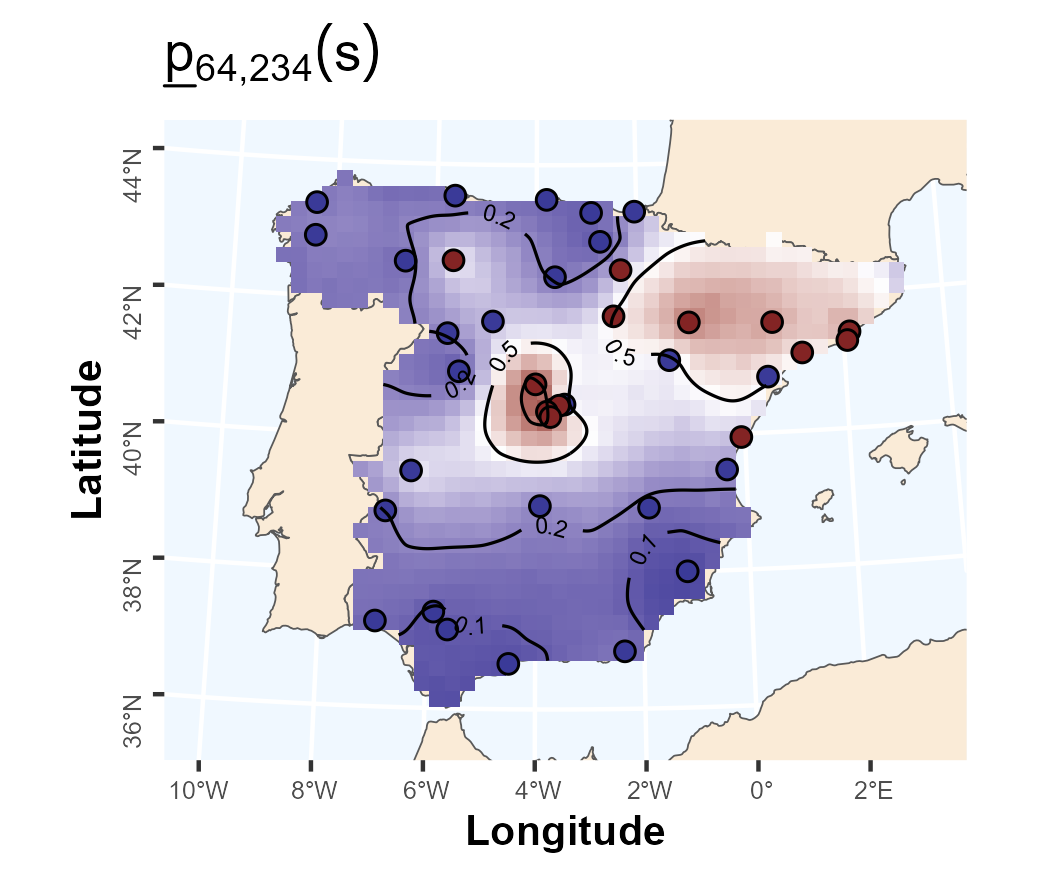}
\includegraphics[width=0.24\textwidth]{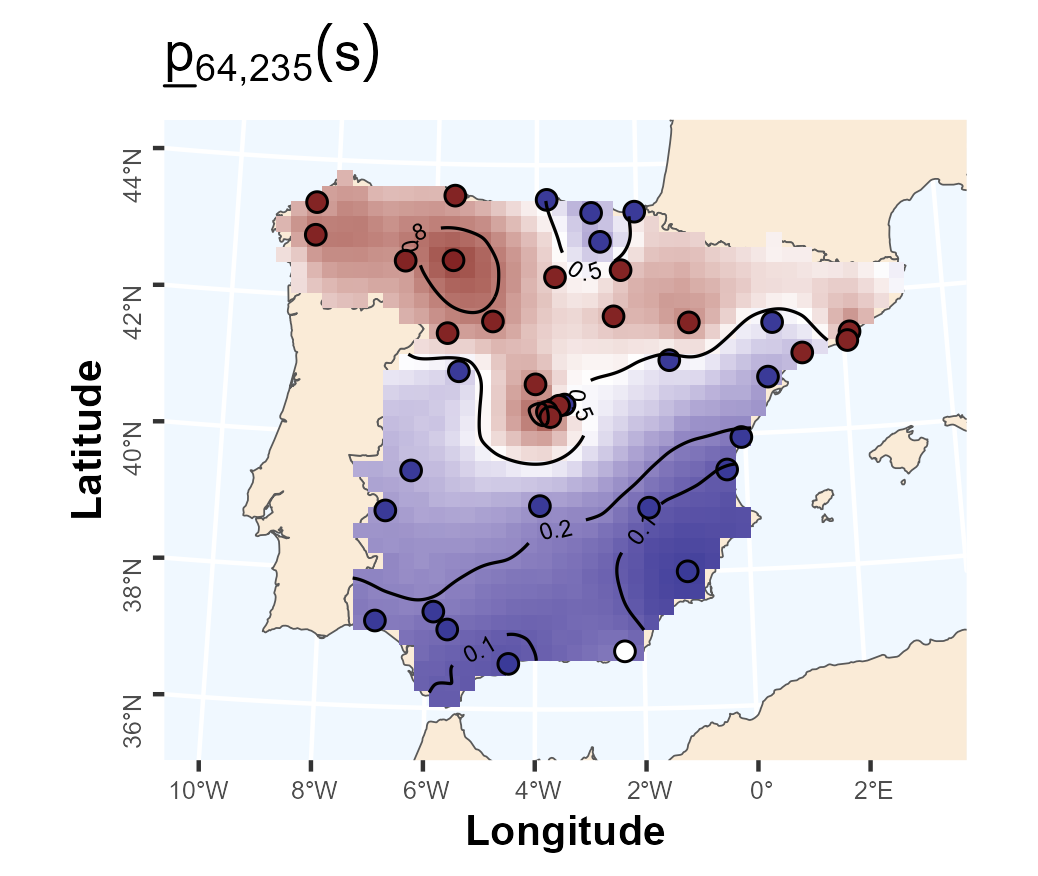}
\includegraphics[width=0.24\textwidth]{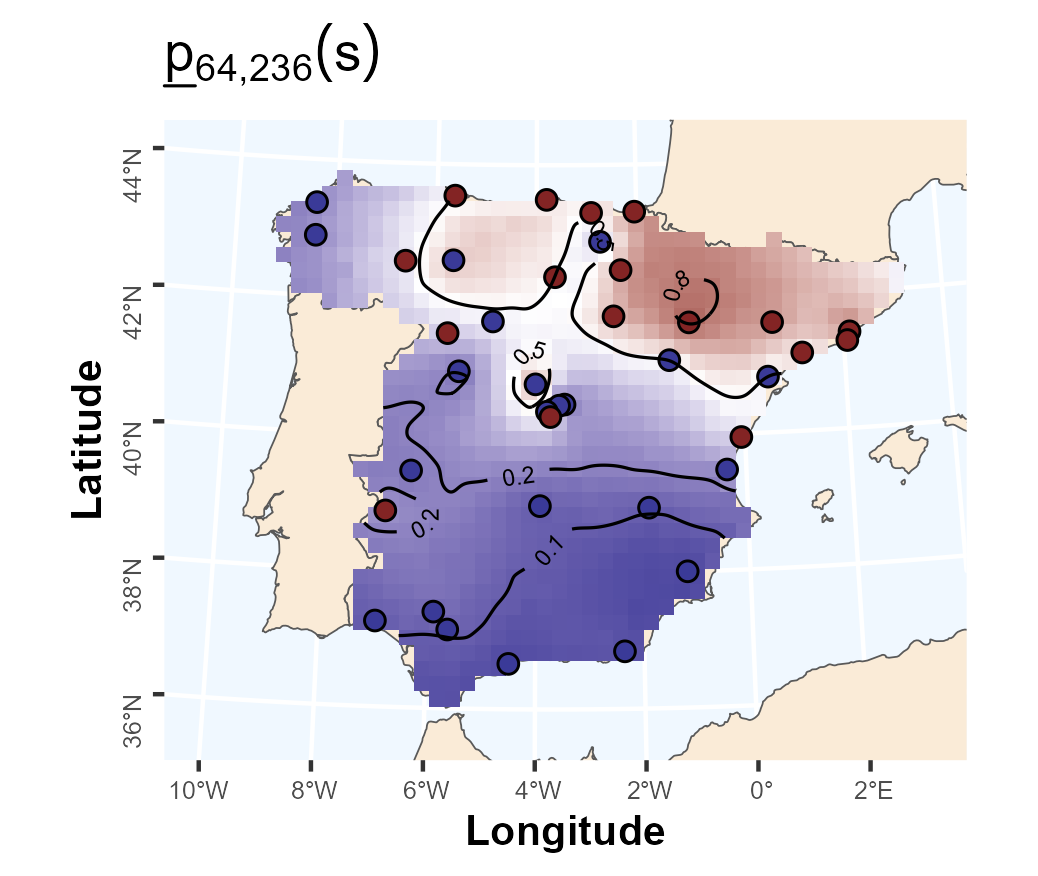}
\includegraphics[width=0.24\textwidth]{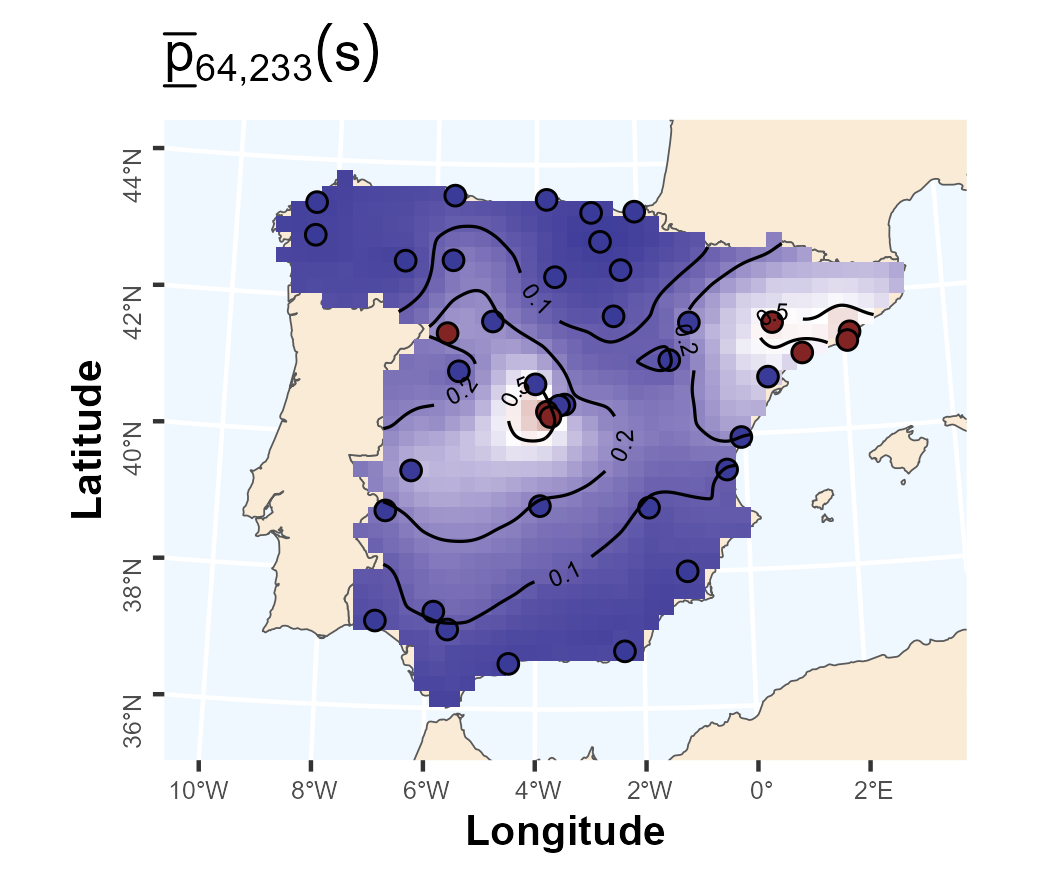}
\includegraphics[width=0.24\textwidth]{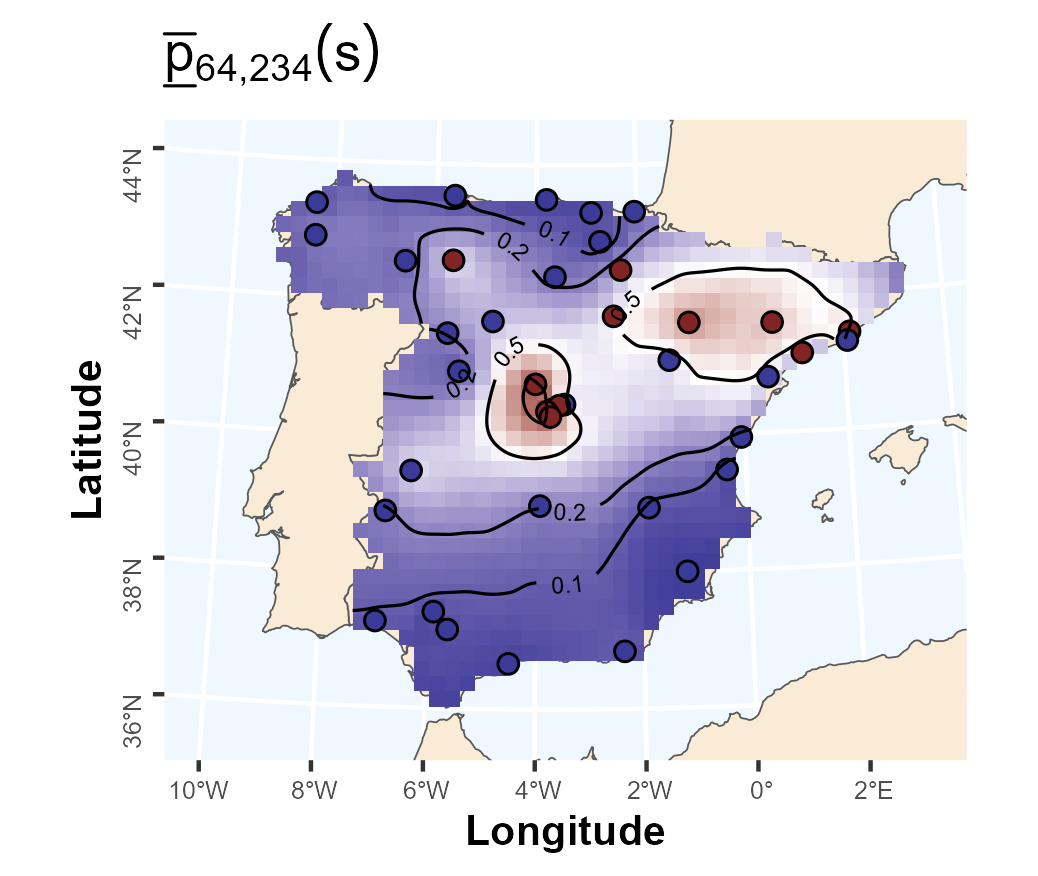}
\includegraphics[width=0.24\textwidth]{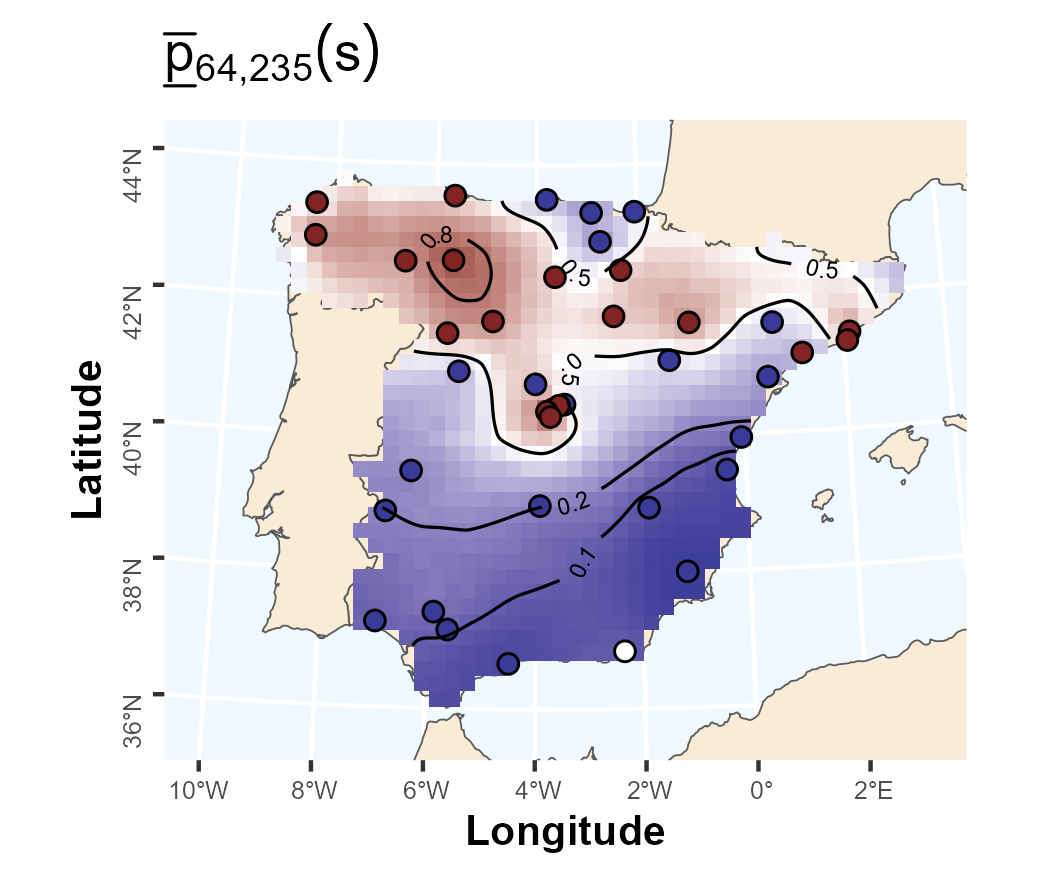}
\includegraphics[width=0.24\textwidth]{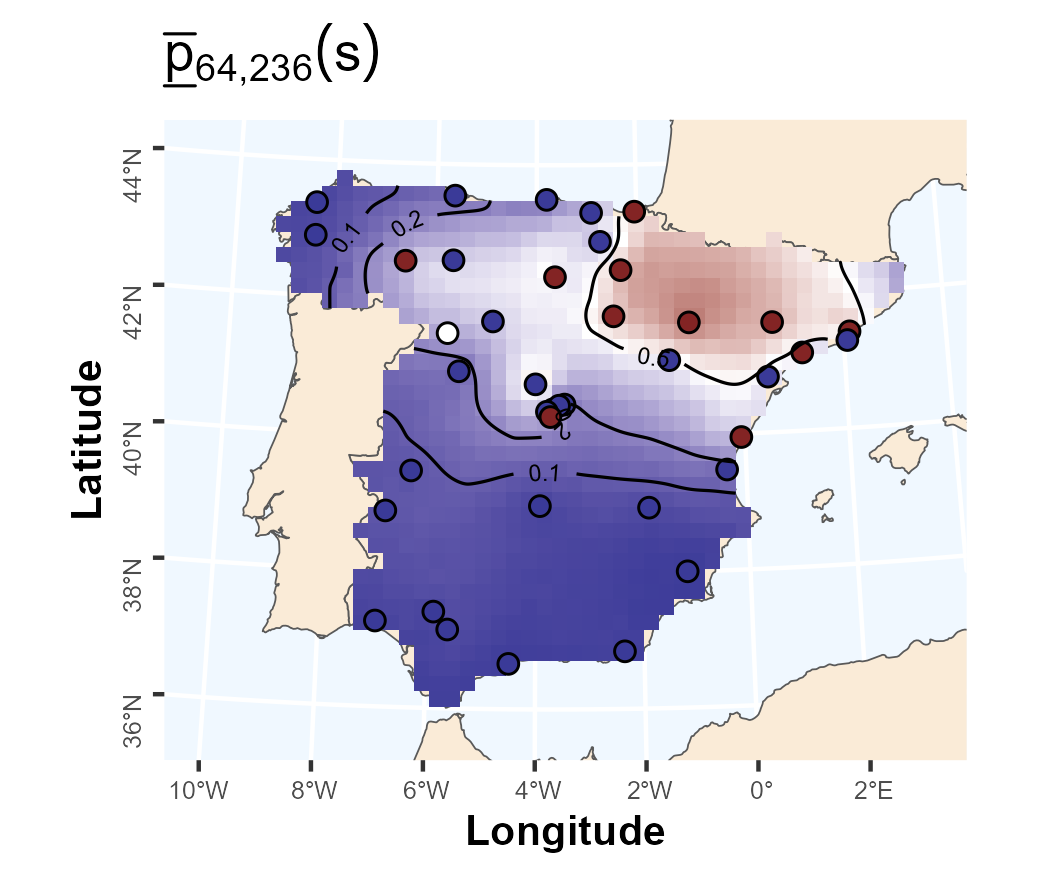}
\includegraphics[width=0.2\textwidth]{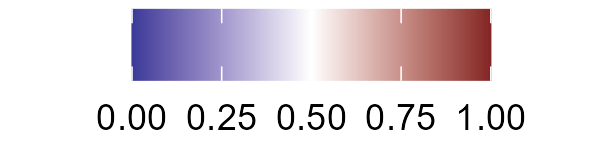}
\caption{Maps of the posterior mean of $\bar{p}_{64,\ell}(\bs)$ (first row), $\protect\munderbar{p}_{64,\ell}(\bs)$ (second row) and $\protect\munderbar{\bar p}_{64,\ell}(\bs)$ (third row) on days $\ell = 233,\ldots,236$ within year 2023; contour lines  are plotted at levels $0.1,0.2,0.5,0.8,0.9$. Points mark the locations with observed records.}
\label{fig:p}
\end{figure}

The proposed model, which includes fixed and random components of spatial and temporal structures, can be reliably used to obtain predictions at unobserved locations. More precisely, the model provides the posterior distribution not only of the marginal record-breaking probabilities but also of the probability of joint record-breaking at any point $\bs$ in the study area. Given that we can obtain predictions of these probabilities at any point, we can also develop record-breaking probability surfaces. These surfaces are useful for studying and characterizing the dynamics of record probabilities across space and time during a particular period. For illustration, we consider the 4-day period from the 21st to the 24th of August ($\ell=233,\ldots,236$) within the year 2023, when peninsular Spain suffered a heat wave. Figure~\ref{fig:p}  shows the spatial evolution of the posterior mean of the marginal and joint probabilities of record occurrences across each day of the heat wave, $\bar{p}_{64,\ell}(\bs)$, $\protect\munderbar{p}_{64,\ell}(\bs)$ and $\protect\bar{\munderbar{p}}_{64,\ell}(\bs)$.
The probability of a record in maximum temperatures on the first day was very high (above $0.8$) in the western Inner Plateau and Cataluña. On the second and third days, high probabilities extended across the entire peninsula, except for some coastal areas. On the last day, high probabilities were observed only in the northern half of the peninsula, excluding Galicia. The results emphasise the importance of distinguishing between the behaviours of maximum and minimum temperatures, as their spatial patterns and temporal evolutions differ significantly. The probabilities of record in minimum temperatures were much lower, always below $0.8$, and below $0.3$ in the southern half of the peninsula throughout the event. The probabilities for joint records were similar to those of minimum temperatures. Locations where a record was observed are marked  on the plots and  align well with the high probabilities predicted by the model.

\subsubsection{Can we analyse the extent of joint or marginal record-breaking over the study region across time?} \label{Sec345}

\begin{figure}[tb]
\centering
\includegraphics[width=0.32\textwidth]{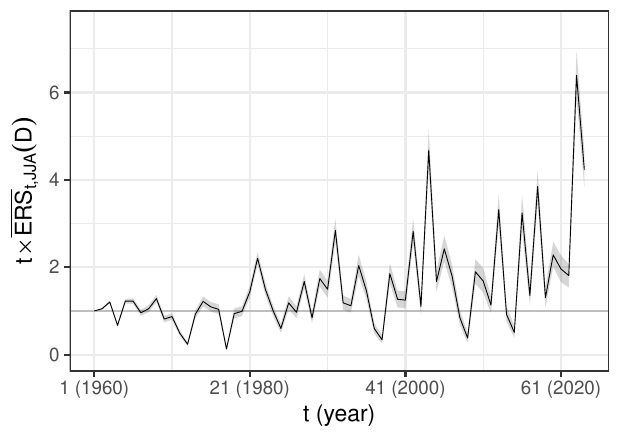}
\includegraphics[width=0.32\textwidth]{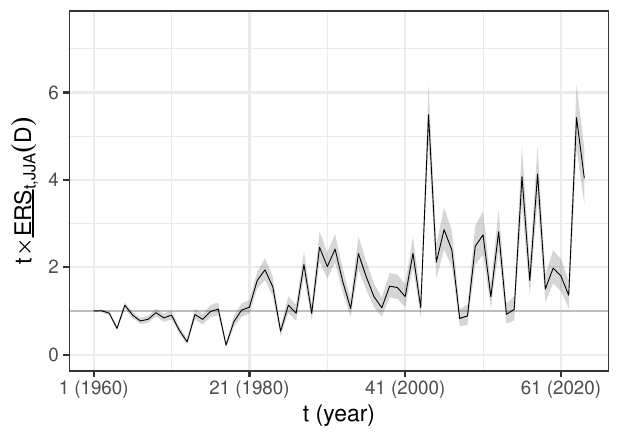}
\includegraphics[width=0.32\textwidth]{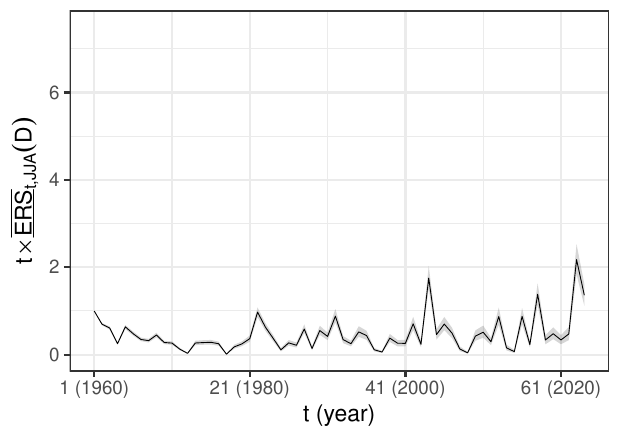}
\caption{Posterior mean (black line) and $90\%$ CI (grey ribbon) of $t \times \text{ERS}_{t,JJA}(D)$ for   maximum, minimum  and joint records against $t$. The reference value under stationarity in the plots for maximum and minimum records is $1$ (grey line).}\label{fig:ers}
\end{figure}

The $\text{ERS}_{t\ell}(D)$  described in Section~\ref{Sec342}  allows us to make inference on the proportion  of the area of peninsular Spain where a record  is observed on a particular day.  To analyse the evolution across years, Figure~\ref{fig:ers} shows the posterior mean  of the average extent for days in summer months, $t \times \text{ERS}_{t,JJA}(D)$ for maximum, minimum and joint temperatures over the years. The values are clearly above 1 from around the year 2000 for maximum  temperatures, and even earlier for minimum temperatures, showing clear deviations from stationarity.  The extent for joint records tends to take higher values during the last period, but there is no exact reference value.

\begin{figure}[tb]
\centering
\includegraphics[width=\textwidth]{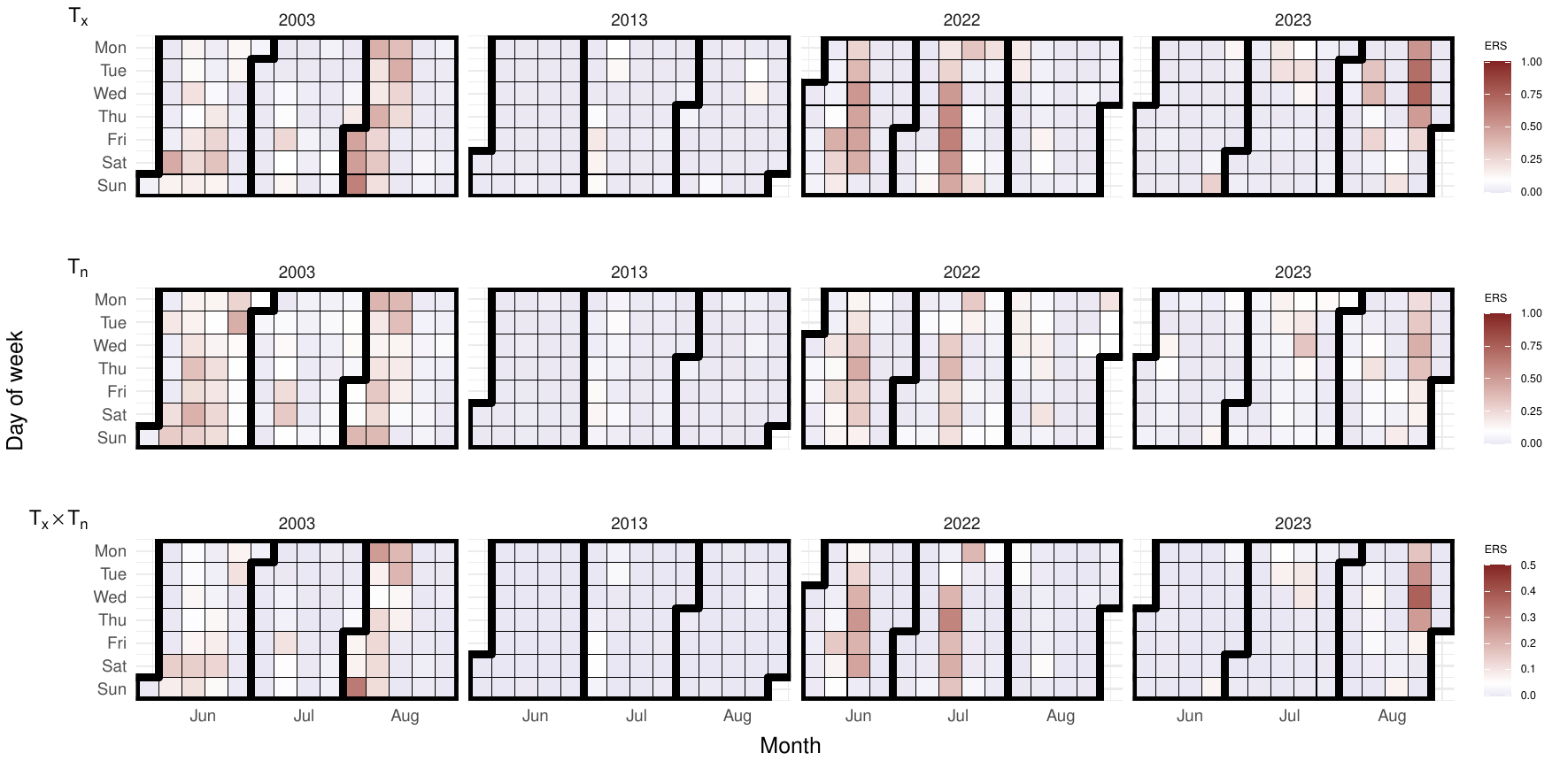}
\caption{Calendar heatmaps for the posterior mean of $\text{ERS}_{t\ell}(D)$ for the maximum (top), minimum (center), and joint (bottom) records for years 2003, 2013, 2022, and 2023 by columns across days within the summer season (from top to bottom); note that the scale for joint records ends at $0.5$.}\label{fig:calendar}
\end{figure}

To analyse potential seasonal pattern differences over time, we compute the posterior mean of the extent for each day in JJA, $\text{ERS}_{t\ell}(D)$, during three hot years (2003, 2022, and 2023) and one cooler year (2013). These posterior means for the maximum, minimum, and joint records are summarised in Figure~\ref{fig:calendar} using calendar heatmaps; these graphs use coloured cells to show the relative number of events for each day in a calendar view, arranged into columns by week and grouped by month and years.  The seasonal pattern of the extent is not consistent across years, but persistence is very strong, as records and their extents appear to cluster in space and time. Note that the highest values for both maximum and joint records occur in 2023. Section~3.5 of the Supplementary material shows calendar heatmaps for a longer period of years.

\section{Conclusions and future work}\label{sec:end}

Interest in studying the tails in both maximum and minimum temperatures within a joint framework is clear due to their impact on human health, agriculture, and biodiversity. The analysis of record-breaking temperatures offers a valuable approach in this context.  Although they do not directly correspond to the analysis of excessive heat events, they can be viewed as a manifestation of climate change.   Further, their use  enables fair comparisons across time series with differing distributions  and provides a robust theoretical framework to detect and quantify deviations from climate stationarity. Thus, we have proposed the first substantial investigation of joint record-breaking for daily maximum and daily minimum temperatures in the literature, using bivariate time series modelling. Our modelling introduced, in the mean, an autoregressive specification given the previous day's binary indicators as well as a long-term trend and environmental predictors, along with interaction terms. Most novel, with regard to uncertainty, our modelling employed a flexible spatially varying coregionalisation to capture non stationary spatial dependence and introduced a climatic variable in the correlation function to capture anisotropy. These novelties relax the assumptions of stationarity and isotropy, which may be too restrictive for daily temperature data, and provide a more flexible spatial dependence specification.

This approach has been applied to the spatial modelling of bivariate daily record-breaking indicators over the summer season for peninsular Spain, covering a period of 64 years. The proposed bivariate specification allows us to capture short and long-term trends and concurrence or persistence between records in both maximum and minimum temperatures, enabling us to address the questions proposed at the beginning of the paper.  

Record-breaking at different time scales, such as weekly, monthly, or seasonally, could be of interest.  Although the model presented here specifically analyses daily temperatures across peninsular Spain, the proposed modelling approach provides a roadmap for modelling maximum and minimum temperature records at other scales or in other regions. In fact, this approach could be applied to model records of another pair of dependent signals. For example, other problem of interest in climate is the analysis of records of an atmospheric variable measured at different geopotential levels, such as 300 hPa and 500 hPa. Another line of future work could generalise the bivariate model to the multivariate modelling of compound extreme events defined by daily rainfall and temperature, or compound events defined in higher dimensions. A different potential future research direction related to daily maximum and minimum temperatures is the analysis of DTR record events.

\section*{Acknowledgments}
This work was supported by MCIN/AEI/10.13039/501100011033 and Uni\'on Europea NextGenerationEU under Grants PID2020-116873GB-I00, TED2021-130702B-I00, and PID2023-150234NB-I00; and Gobierno de Arag\'on under Research Group E46\_23R: Modelos Estoc\'asticos. J. C.-M. was supported by Gobierno de Arag\'on under Doctoral Scholarship ORDEN CUS/581/2020. Z. G.-T. was supported by MIU and NGEU under Margarita Salas Scholarship RD 289/2021 UNI/551/2021.

\section*{Conflict of interest}

None declared.

\section*{Data and code availability}

The data supporting the findings of this study are openly available from the ECA\&D at \url{https://www.ecad.eu/}. The code to replicate this work, along with the functions used to fit all models, is provided as Supplementary material and is also available in the \texttt{R} package \texttt{spbrom} on GitHub at \url{https://github.com/JorgeCastilloMateo/spbrom}.

\bibliographystyle{abbrvnat}

\begin{thebibliography}{46}
	\providecommand{\natexlab}[1]{#1}
	\providecommand{\url}[1]{\texttt{#1}}
	\expandafter\ifx\csname urlstyle\endcsname\relax
	\providecommand{\doi}[1]{doi: #1}\else
	\providecommand{\doi}{doi: \begingroup \urlstyle{rm}\Url}\fi
	
	\bibitem[Abaurrea et~al.(2018)Abaurrea, Asín, and Cebrián]{abaurrea2018}
	J.~Abaurrea, J.~Asín, and A.~C. Cebrián.
	\newblock Modelling the occurrence of heat waves in maximum and minimum
	temperatures over {Spain} and projections for the period 2031--60.
	\newblock \emph{Global and Planetary Change}, 161:\penalty0 244--260, 2018.
	\newblock \doi{10.1016/j.gloplacha.2017.11.015}.
	
	\bibitem[Albert and Chib(1993)]{albert1993}
	J.~H. Albert and S.~Chib.
	\newblock Bayesian analysis of binary and polychotomous response data.
	\newblock \emph{Journal of the American Statistical Association}, 88\penalty0
	(422):\penalty0 669--679, 1993.
	\newblock \doi{10.1080/01621459.1993.10476321}.
	
	\bibitem[Arnold et~al.(1998)Arnold, Balakrishnan, and Nagaraja]{arnold1998}
	B.~C. Arnold, N.~Balakrishnan, and H.~N. Nagaraja.
	\newblock \emph{Records}.
	\newblock Wiley Series in Probability and Statistics. John Wiley \& Sons, New
	York, 1998.
	\newblock \doi{10.1002/9781118150412}.
	
	\bibitem[Balakrishnan et~al.(2020)Balakrishnan, Stepanov, and
	Nevzorov]{balakrishnan2020}
	N.~Balakrishnan, A.~Stepanov, and V.~B. Nevzorov.
	\newblock North-east bivariate records.
	\newblock \emph{Metrika}, 83:\penalty0 961--976, 2020.
	\newblock \doi{10.1007/s00184-020-00766-2}.
	
	\bibitem[Banerjee et~al.(2014)Banerjee, Carlin, and Gelfand]{BCG}
	S.~Banerjee, B.~P. Carlin, and A.~E. Gelfand.
	\newblock \emph{Hierarchical Modeling and Analysis for Spatial Data}.
	\newblock Chapman and Hall/CRC, New York, NY, 2 edition, 2014.
	\newblock \doi{10.1201/b17115}.
	
	\bibitem[Battisti and Naylor(2009)]{Battisti2009}
	D.~S. Battisti and R.~L. Naylor.
	\newblock Historical warnings of future food insecurity with unprecedented
	seasonal heat.
	\newblock \emph{Science}, 323\penalty0 (5911):\penalty0 240--244, 2009.
	\newblock \doi{10.1126/science.1164363}.
	
	\bibitem[Benestad(2003)]{benestad2003}
	R.~E. Benestad.
	\newblock How often can we expect a record event?
	\newblock \emph{Climate Research}, 25\penalty0 (1):\penalty0 3--13, 2003.
	\newblock \doi{10.3354/cr025003}.
	
	\bibitem[Castillo-Mateo et~al.(2022)Castillo-Mateo, Lafuente, As{\'\i}n,
	Cebri{\'a}n, Gelfand, and Abaurrea]{castillo2022b}
	J.~Castillo-Mateo, M.~Lafuente, J.~As{\'\i}n, A.~C. Cebri{\'a}n, A.~E. Gelfand,
	and J.~Abaurrea.
	\newblock Spatial modeling of day-within-year temperature time series: An
	examination of daily maximum temperatures in {Arag{\'o}n}, {Spain}.
	\newblock \emph{Journal of Agricultural, Biological and Environmental
		Statistics}, 27\penalty0 (3):\penalty0 487--505, 2022.
	\newblock \doi{10.1007/s13253-022-00493-3}.
	
	\bibitem[Castillo-Mateo et~al.(2023)Castillo-Mateo, Cebri\'an, and
	As\'in]{castillo2023}
	J.~Castillo-Mateo, A.~C. Cebri\'an, and J.~As\'in.
	\newblock {RecordTest}: An {R} package to analyse non-stationarity in the
	extremes based on record-breaking events.
	\newblock \emph{Journal of Statistical Software}, 106\penalty0 (5):\penalty0
	1--28, 2023.
	\newblock \doi{10.18637/jss.v106.i05}.
	
	\bibitem[Castillo-Mateo et~al.(2024)Castillo-Mateo, Gelfand, Gracia-Tabuenca,
	As\'in, and Cebri\'an]{castillo2024}
	J.~Castillo-Mateo, A.~E. Gelfand, Z.~Gracia-Tabuenca, J.~As\'in, and A.~C.
	Cebri\'an.
	\newblock Spatio-temporal modeling for record-breaking temperature events in
	{Spain}.
	\newblock \emph{Journal of the American Statistical Association}, 2024.
	\newblock \doi{10.1080/01621459.2024.2427430}.
	
	\bibitem[Cebri\'an et~al.(2022)Cebri\'an, Castillo-Mateo, and
	As\'in]{cebrian2022a}
	A.~C. Cebri\'an, J.~Castillo-Mateo, and J.~As\'in.
	\newblock Record tests to detect non-stationarity in the tails with an
	application to climate change.
	\newblock \emph{Stochastic Environmental Research and Risk Assessment},
	36\penalty0 (2):\penalty0 313--330, 2022.
	\newblock \doi{10.1007/s00477-021-02122-w}.
	
	\bibitem[Cooley et~al.(2007)Cooley, Nychka, and Naveau]{cooley2007}
	D.~Cooley, D.~Nychka, and P.~Naveau.
	\newblock Bayesian spatial modeling of extreme precipitation return levels.
	\newblock \emph{Journal of the American Statistical Association}, 102\penalty0
	(479):\penalty0 824--840, 2007.
	\newblock \doi{10.1198/016214506000000780}.
	
	\bibitem[Correa et~al.(2024)Correa, Dorta, López-Díez, and
	Díaz-Pacheco]{correa2024}
	J.~Correa, P.~Dorta, A.~López-Díez, and J.~Díaz-Pacheco.
	\newblock Analysis of tropical nights in {Spain} (1970--2023): Minimum
	temperatures as an indicator of climate change.
	\newblock \emph{International Journal of Climatology}, 2024.
	\newblock \doi{10.1002/joc.8510}.
	
	\bibitem[Daramola et~al.(2024)Daramola, Li, and Xu]{daramola2024}
	M.~T. Daramola, R.~Li, and M.~Xu.
	\newblock Increased diurnal temperature range in global drylands in more recent
	decades.
	\newblock \emph{International Journal of Climatology}, 44\penalty0
	(2):\penalty0 521--533, 2024.
	\newblock \doi{10.1002/joc.8341}.
	
	\bibitem[Davison and Huser(2015)]{davison2015}
	A.~C. Davison and R.~Huser.
	\newblock Statistics of extremes.
	\newblock \emph{Annual Review of Statistics and Its Application}, 2:\penalty0
	203--235, 2015.
	\newblock \doi{10.1146/annurev-statistics-010814-020133}.
	
	\bibitem[De~Oliveira(2020)]{deoliveira2020}
	V.~De~Oliveira.
	\newblock Models for geostatistical binary data: Properties and connections.
	\newblock \emph{The American Statistician}, 74\penalty0 (1):\penalty0 72--79,
	2020.
	\newblock \doi{10.1080/00031305.2018.1444674}.
	
	\bibitem[Diggle et~al.(1998)Diggle, Tawn, and Moyeed]{diggle1998}
	P.~J. Diggle, J.~A. Tawn, and R.~A. Moyeed.
	\newblock Model-based geostatistics.
	\newblock \emph{Journal of the Royal Statistical Society: Series C (Applied
		Statistics)}, 47\penalty0 (3):\penalty0 299--350, 1998.
	\newblock \doi{10.1111/1467-9876.00113}.
	
	\bibitem[Fill(2023)]{fill2023}
	J.~A. Fill.
	\newblock Breaking multivariate records.
	\newblock \emph{Electronic Journal of Probability}, 28:\penalty0 1--27, 2023.
	\newblock \doi{10.1214/23-EJP968}.
	
	\bibitem[Gelfand et~al.(2004)Gelfand, Schmidt, Banerjee, and
	Sirmans]{gelfand2004}
	A.~E. Gelfand, A.~M. Schmidt, S.~Banerjee, and C.~F. Sirmans.
	\newblock Nonstationary multivariate process modeling through spatially varying
	coregionalization.
	\newblock \emph{Test}, 13\penalty0 (2):\penalty0 263--312, 2004.
	\newblock \doi{10.1007/BF02595775}.
	
	\bibitem[Hajat and Kosatky(2010)]{hajat2006}
	S.~Hajat and T.~Kosatky.
	\newblock Heat-related mortality: a review and exploration of heterogeneity.
	\newblock \emph{Journal of Epidemiology \& Community Health}, 64\penalty0
	(9):\penalty0 753--760, 2010.
	\newblock \doi{10.1136/jech.2009.087999}.
	
	\bibitem[He et~al.(2022)He, Kim, Hashizume, Lee, Honda, Kim, Kinney, Schneider,
	Zhang, Zhu, et~al.]{he2022}
	C.~He, H.~Kim, M.~Hashizume, W.~Lee, Y.~Honda, S.~E. Kim, P.~L. Kinney,
	A.~Schneider, Y.~Zhang, Y.~Zhu, et~al.
	\newblock The effects of night-time warming on mortality burden under future
	climate change scenarios: a modelling study.
	\newblock \emph{The Lancet Planetary Health}, 6\penalty0 (8):\penalty0
	e648--e657, 2022.
	\newblock \doi{10.1016/S2542-5196(22)00139-5}.
	
	\bibitem[Healy et~al.(2025)Healy, Tawn, Thorne, and Parnell]{healy2021}
	D.~Healy, J.~Tawn, P.~Thorne, and A.~Parnell.
	\newblock Inference for extreme spatial temperature events in a changing
	climate with application to {Ireland}.
	\newblock \emph{Journal of the Royal Statistical Society: Series C},
	74\penalty0 (2):\penalty0 275--299, 2025.
	\newblock \doi{10.1093/jrsssc/qlae047}.
	
	\bibitem[Held and Holmes(2006)]{held2006}
	L.~Held and C.~C. Holmes.
	\newblock Bayesian auxiliary variable models for binary and multinomial
	regression.
	\newblock \emph{Bayesian Analysis}, 1\penalty0 (1):\penalty0 145--168, 2006.
	\newblock \doi{10.1214/06-BA105}.
	
	\bibitem[Huang et~al.(2023)Huang, Dunn, Li, McVicar, Azorin-Molina, and
	Zeng]{huang2023}
	X.~Huang, R.~J. Dunn, L.~Z. Li, T.~R. McVicar, C.~Azorin-Molina, and Z.~Zeng.
	\newblock Increasing global terrestrial diurnal temperature range for
	1980--2021.
	\newblock \emph{Geophysical Research Letters}, 50\penalty0 (11):\penalty0
	e2023GL103503, 2023.
	\newblock \doi{10.1029/2023GL103503}.
	
	\bibitem[Kemalbay and Bayramo{\u{g}}lu(2019)]{kemalbay2019}
	G.~Kemalbay and {\.I}.~Bayramo{\u{g}}lu.
	\newblock On distribution of upper marginal records in bivariate random
	sequences.
	\newblock \emph{Turkish Journal of Mathematics}, 43\penalty0 (3):\penalty0
	1474--1491, 2019.
	\newblock \doi{10.3906/mat-1901-54}.
	
	\bibitem[Kleiber et~al.(2013)Kleiber, Katz, and Rajagopalan]{Kleiber2013}
	W.~Kleiber, R.~W. Katz, and B.~Rajagopalan.
	\newblock Daily minimum and maximum temperature simulation over complex
	terrain.
	\newblock \emph{Annals of Applied Statistics}, 7\penalty0 (1):\penalty0
	588--612, 2013.
	\newblock \doi{10.1214/12-AOAS602}.
	
	\bibitem[Klein~Tank et~al.(2002)Klein~Tank, Wijngaard, Können, Böhm,
	Demarée, Gocheva, Mileta, Pashiardis, Hejkrlik, Kern-Hansen, Heino,
	Bessemoulin, Müller-Westermeier, Tzanakou, Szalai, Pálsdóttir, Fitzgerald,
	Rubin, Capaldo, Maugeri, Leitass, Bukantis, Aberfeld, van Engelen, Forland,
	Mietus, Coelho, Mares, Razuvaev, Nieplova, Cegnar, Antonio~López,
	Dahlström, Moberg, Kirchhofer, Ceylan, Pachaliuk, Alexander, and
	Petrovic]{tank2002}
	A.~M.~G. Klein~Tank, J.~B. Wijngaard, G.~P. Können, R.~Böhm, G.~Demarée,
	A.~Gocheva, M.~Mileta, S.~Pashiardis, L.~Hejkrlik, C.~Kern-Hansen, R.~Heino,
	P.~Bessemoulin, G.~Müller-Westermeier, M.~Tzanakou, S.~Szalai,
	T.~Pálsdóttir, D.~Fitzgerald, S.~Rubin, M.~Capaldo, M.~Maugeri, A.~Leitass,
	A.~Bukantis, R.~Aberfeld, A.~F.~V. van Engelen, E.~Forland, M.~Mietus,
	F.~Coelho, C.~Mares, V.~Razuvaev, E.~Nieplova, T.~Cegnar, J.~Antonio~López,
	B.~Dahlström, A.~Moberg, W.~Kirchhofer, A.~Ceylan, O.~Pachaliuk, L.~V.
	Alexander, and P.~Petrovic.
	\newblock Daily dataset of 20th-century surface air temperature and
	precipitation series for the {European Climate Assessment}.
	\newblock \emph{International Journal of Climatology}, 22\penalty0
	(12):\penalty0 1441--1453, 2002.
	\newblock \doi{10.1002/joc.773}.
	
	\bibitem[Krock et~al.(2022)Krock, Bessac, Stein, and Monahan]{krock2022}
	M.~Krock, J.~Bessac, M.~L. Stein, and A.~H. Monahan.
	\newblock Nonstationary seasonal model for daily mean temperature distribution
	bridging bulk and tails.
	\newblock \emph{Weather and Climate Extremes}, 36:\penalty0 100438, 2022.
	\newblock \doi{10.1016/j.wace.2022.100438}.
	
	\bibitem[Lewis and King(2015)]{lewis2015}
	S.~C. Lewis and A.~D. King.
	\newblock Dramatically increased rate of observed hot record breaking in recent
	{Australian} temperatures.
	\newblock \emph{Geophysical Research Letters}, 42\penalty0 (18):\penalty0
	7776--7784, 2015.
	\newblock \doi{10.1002/2015GL065793}.
	
	\bibitem[Liu et~al.(2024)Liu, Guo, Xia, Liu, Song, Yang, and Zhang]{liu2024}
	G.~Liu, Y.~Guo, H.~Xia, X.~Liu, H.~Song, J.~Yang, and Y.~Zhang.
	\newblock Increase asymmetric warming rates between daytime and nighttime
	temperatures over global land during recent decades.
	\newblock \emph{Geophysical Research Letters}, 51\penalty0 (24):\penalty0
	e2024GL112832, 2024.
	\newblock \doi{10.1029/2024GL112832}.
	
	\bibitem[Naveau et~al.(2018)Naveau, Ribes, Zwiers, Hannart, Tuel, and
	Yiou]{naveau2018}
	P.~Naveau, A.~Ribes, F.~Zwiers, A.~Hannart, A.~Tuel, and P.~Yiou.
	\newblock Revising return periods for record events in a climate event
	attribution context.
	\newblock \emph{Journal of Climate}, 31\penalty0 (9):\penalty0 3411--3422,
	2018.
	\newblock \doi{10.1175/JCLI-D-16-0752.1}.
	
	\bibitem[Naveau et~al.(2020)Naveau, Hannart, and Ribes]{naveau2020}
	P.~Naveau, A.~Hannart, and A.~Ribes.
	\newblock Statistical methods for extreme event attribution in climate science.
	\newblock \emph{Annual Review of Statistics and Its Application}, 7:\penalty0
	89--110, 2020.
	\newblock \doi{10.1146/annurev-statistics-031219-041314}.
	
	\bibitem[North et~al.(2021)North, Schliep, and Wikle]{North2021}
	J.~S. North, E.~M. Schliep, and C.~K. Wikle.
	\newblock On the spatial and temporal shift in the archetypal seasonal
	temperature cycle as driven by annual and semi-annual harmonics.
	\newblock \emph{Environmetrics}, 32\penalty0 (6):\penalty0 e2665, 2021.
	\newblock \doi{10.1002/env.2665}.
	
	\bibitem[Noubary and Noubary(2004)]{noubary2004}
	F.~Noubary and R.~Noubary.
	\newblock On survival times of sport records.
	\newblock \emph{Journal of Computational and Applied Mathematics}, 169\penalty0
	(1):\penalty0 227--234, 2004.
	\newblock \doi{10.1016/j.cam.2003.12.022}.
	
	\bibitem[Pena-Angulo et~al.(2015)Pena-Angulo, Cortesi, Brunetti, and
	González-Hidalgo]{pena2015}
	D.~Pena-Angulo, N.~Cortesi, M.~Brunetti, and J.~C. González-Hidalgo.
	\newblock Spatial variability of maximum and minimum monthly temperature in
	{Spain} during 1981--2010 evaluated by correlation decay distance {(CDD)}.
	\newblock \emph{Theoretical and Applied Climatology}, 122\penalty0
	(1--2):\penalty0 35--45, 2015.
	\newblock \doi{10.1007/s00704-014-1277-x}.
	
	\bibitem[Plummer et~al.(1999)Plummer, Salinger, Nicholls, Suppiah, Hennessy,
	Leighton, Trewin, Page, and Lough]{plummer1999}
	N.~Plummer, M.~Salinger, N.~Nicholls, R.~Suppiah, K.~Hennessy, R.~Leighton,
	B.~Trewin, C.~Page, and J.~Lough.
	\newblock Changes in climate extremes over the {Australian} region and {New
		Zealand} during the twentieth century.
	\newblock \emph{Climatic Change}, 42:\penalty0 183--202, 1999.
	\newblock \doi{10.1023/A:1005472418209}.
	
	\bibitem[Polson et~al.(2013)Polson, Scott, and Windle]{polson2013}
	N.~G. Polson, J.~G. Scott, and J.~Windle.
	\newblock Bayesian inference for logistic models using {P}\'olya--gamma latent
	variables.
	\newblock \emph{Journal of the American Statistical Association}, 108\penalty0
	(504):\penalty0 1339--1349, 2013.
	\newblock \doi{10.1080/01621459.2013.829001}.
	
	\bibitem[Prabhjyot-Kaur and Hundal(2010)]{kaur2010}
	Prabhjyot-Kaur and S.~S. Hundal.
	\newblock Global climate change vis-à-vis crop productivity.
	\newblock In \emph{Natural and Anthropogenic Disasters}, pages 413--431.
	Springer, Dordrecht, 2010.
	\newblock \doi{10.1007/978-90-481-2498-5_18}.
	
	\bibitem[Roy{\'e}(2017)]{roye2017}
	D.~Roy{\'e}.
	\newblock The effects of hot nights on mortality in {Barcelona}, {Spain}.
	\newblock \emph{International Journal of Biometeorology}, 61\penalty0
	(12):\penalty0 2127--2140, 2017.
	\newblock \doi{10.1007/s00484-017-1416-z}.
	
	\bibitem[S{\'a}ez et~al.(2012)S{\'a}ez, Barcel{\'o}, Tob{\'\i}as, Varga,
	Oca{\~n}a, Juan, and Mateu]{saez2012}
	M.~S{\'a}ez, M.~A. Barcel{\'o}, A.~Tob{\'\i}as, D.~Varga, R.~Oca{\~n}a,
	P.~Juan, and J.~Mateu.
	\newblock Space-time interpolation of daily air temperatures.
	\newblock \emph{Journal of Environmental Statistics}, 3\penalty0 (5):\penalty0
	1--15, 2012.
	
	\bibitem[Schmidt et~al.(2011)Schmidt, Guttorp, and O'Hagan]{schmidt2011}
	A.~M. Schmidt, P.~Guttorp, and A.~O'Hagan.
	\newblock Considering covariates in the covariance structure of spatial
	processes.
	\newblock \emph{Environmetrics}, 22\penalty0 (4):\penalty0 487--500, 2011.
	\newblock \doi{10.1002/env.1101}.
	
	\bibitem[Serinaldi and Kilsby(2018)]{serinaldi2018}
	F.~Serinaldi and C.~G. Kilsby.
	\newblock Unsurprising surprises: The frequency of record-breaking and
	overthreshold hydrological extremes under spatial and temporal dependence.
	\newblock \emph{Water Resources Research}, 54\penalty0 (9):\penalty0
	6460--6487, 2018.
	\newblock \doi{10.1029/2018WR023055}.
	
	\bibitem[Serrano-Notivoli et~al.(2023)Serrano-Notivoli, Tejedor, Sarricolea,
	Meseguer-Ruiz, {de Luis}, Ángel Saz, Longares, and Olcina]{serrrano2023}
	R.~Serrano-Notivoli, E.~Tejedor, P.~Sarricolea, O.~Meseguer-Ruiz, M.~{de Luis},
	M.~Ángel Saz, L.~A. Longares, and J.~Olcina.
	\newblock Unprecedented warmth: A look at {Spain}'s exceptional summer of 2022.
	\newblock \emph{Atmospheric Research}, 293:\penalty0 106931, 2023.
	\newblock \doi{10.1016/j.atmosres.2023.106931}.
	
	\bibitem[Tryhorn and Risbey(2006)]{tryhorn2006}
	L.~Tryhorn and J.~Risbey.
	\newblock On the distribution of heatwaves over the {Australian} region.
	\newblock \emph{Australian Meteorological Magazine}, 55\penalty0 (3):\penalty0
	169--82, 2006.
	
	\bibitem[Wergen(2014)]{wergen2014b}
	G.~Wergen.
	\newblock Modeling record-breaking stock prices.
	\newblock \emph{Physica A: Statistical Mechanics and its Applications},
	396:\penalty0 114--133, 2014.
	\newblock \doi{https://doi.org/10.1016/j.physa.2013.11.001}.
	
	\bibitem[Wergen et~al.(2014)Wergen, Hense, and Krug]{wergen2014a}
	G.~Wergen, A.~Hense, and J.~Krug.
	\newblock Record occurrence and record values in daily and monthly
	temperatures.
	\newblock \emph{Climate Dynamics}, 42\penalty0 (5):\penalty0 1275--1289, 2014.
	\newblock \doi{10.1007/s00382-013-1693-0}.
	
\end{thebibliography}



\section*{Supplementary material}

# Supplementary material of the Manuscript: "Joint space-time modelling for upper daily maximum and minimum temperature record-breaking"

Jorge Castillo-Mateo
Department of Statistical Methods, University of Zaragoza
Zeus Gracia-Tabuenca
Department of Statistical Methods, University of Zaragoza
Jesús Asín
Department of Statistical Methods, University of Zaragoza
Ana C. Cebrián
Department of Statistical Methods, University of Zaragoza
Alan E. Gelfand
Department of Statistical Science, Duke University

This Supplementary material offers additional exploratory analyses, a detailed data augmentation MCMC algorithm, convergence diagnostics, and further results complementing the Main manuscript.

## 1 More about the data and exploratory analysis

### 1.1 Provenance of the dataset

The dataset of the 40 weather stations used in this work was provided by the ECA&D repository. It was based on previous work that required a minimum of 99.5% reliable data and relative homogeneity as an inclusion criteria (Castillo-Mateo et al., 2023, 2024). We targeted the daily maximum and minimum temperature data in the JJA months from the 1960–2023 period. In the exploratory analysis we did not include four of the five stations located in the Madrid metropolitan area to limit the effect of its over-representation: Barajas, Cuatrovientos, Getafe, and Torrejon. Additionally, according to the World Meteorological Office (WMO, 2022), this dataset includes four centennial stations: Barcelona (Fabra), Daroca, Madrid (Retiro), and Tortosa. Table 1 describes the 40 stations grouped by geographical zones.

Table 1: Descriptive statistics of the stations downloaded from the ECA&D grouped by geographical zones. Statistics for the maximum ($T_x$) and minimum ($T_n$) temperatures in the JJA 1991–2020 reference period: sample mean ($\bar{x}$), standard deviation ($s$), kurtosis ($K$), and 99th percentile ($q^{99}$).

| Name | $\bar{x}_x$ | $\bar{x}_n$ | $s_x$ | $s_n$ | $K_x$ | $K_n$ | $q_x^{99}$ | $q_n^{99}$ |
|---|---|---|---|---|---|---|---|---|
| **Zone Andalusia (A)** | | | | | | | | |
| Almería | 29.89 | 21.56 | 3.65 | 2.56 | 2.74 | 3.00 | 38.60 | 26.30 |
| Huelva | 31.72 | 18.45 | 4.04 | 2.37 | 2.55 | 3.08 | 40.44 | 23.90 |
| Málaga | 30.33 | 20.60 | 3.54 | 2.47 | 3.51 | 3.06 | 40.00 | 26.24 |
| Morón | 34.20 | 18.13 | 4.36 | 2.80 | 2.82 | 3.04 | 42.64 | 24.34 |
| Sevilla | 35.11 | 20.05 | 4.05 | 2.54 | 2.98 | 3.06 | 43.20 | 25.65 |
| **Zone Cantabrian (C)** | | | | | | | | |
| Bilbao | 25.06 | 15.27 | 4.22 | 2.67 | 3.77 | 2.80 | 37.24 | 20.84 |
| Gijón | 22.16 | 16.41 | 2.25 | 2.29 | 4.08 | 3.10 | 27.04 | 20.80 |
| San Sebastián | 21.75 | 15.70 | 3.97 | 2.32 | 4.95 | 2.93 | 34.50 | 20.94 |
| Santander | 22.02 | 16.54 | 2.65 | 2.06 | 5.24 | 2.89 | 30.38 | 20.74 |
| **Zone Ebro (E)** | | | | | | | | |
| Daroca | 30.00 | 14.40 | 4.82 | 3.57 | 2.91 | 2.98 | 38.60 | 22.14 |
| Lleida | 32.13 | 17.05 | 3.81 | 3.00 | 3.41 | 2.82 | 39.40 | 22.30 |
| Logroño | 29.47 | 14.98 | 5.01 | 2.85 | 2.55 | 2.90 | 38.80 | 20.84 |
| Vitoria | 25.54 | 11.91 | 5.28 | 3.26 | 2.39 | 2.60 | 36.80 | 18.24 |
| Zaragoza | 31.78 | 17.87 | 4.46 | 2.78 | 2.76 | 2.78 | 39.94 | 23.04 |
| **Zone Galician (G)** | | | | | | | | |
| Coruña | 22.16 | 15.78 | 2.70 | 1.69 | 4.49 | 3.54 | 30.44 | 19.44 |
| Santiago | 24.06 | 12.69 | 4.56 | 2.43 | 2.84 | 2.96 | 35.44 | 17.80 |
| **Zone Levante (L)** | | | | | | | | |
| Albacete | 32.10 | 16.26 | 4.09 | 2.94 | 3.90 | 3.23 | 39.24 | 21.60 |
| Barcelona (Aeropuerto) | 27.59 | 19.98 | 2.82 | 2.83 | 3.33 | 2.86 | 33.64 | 25.10 |
| Barcelona (Fabra) | 27.73 | 18.80 | 3.51 | 2.97 | 3.27 | 3.09 | 35.10 | 25.00 |
| Castellón | 29.70 | 20.41 | 2.57 | 2.48 | 3.91 | 3.20 | 35.80 | 24.90 |
| Murcia | 33.46 | 20.44 | 3.21 | 2.51 | 3.65 | 3.21 | 41.00 | 25.34 |
| Reus | 29.12 | 18.77 | 2.87 | 2.81 | 3.49 | 2.78 | 35.30 | 24.15 |
| Tortosa | 32.03 | 19.78 | 3.16 | 2.52 | 4.14 | 2.76 | 38.20 | 24.30 |
| Valencia | 29.34 | 21.14 | 2.72 | 2.44 | 4.70 | 3.01 | 37.20 | 25.40 |
| **Zone North Plateau (NP)** | | | | | | | | |
| Burgos | 26.73 | 11.01 | 5.24 | 3.00 | 2.57 | 2.99 | 36.64 | 17.20 |
| León | 26.28 | 11.61 | 4.59 | 3.28 | 2.72 | 2.79 | 34.60 | 18.80 |
| Ponferrada | 28.66 | 13.59 | 4.88 | 2.79 | 2.71 | 2.93 | 37.60 | 19.30 |
| Salamanca | 29.41 | 11.90 | 4.51 | 2.76 | 2.89 | 2.97 | 37.60 | 18.00 |
| Soria | 27.53 | 11.76 | 4.92 | 3.00 | 2.90 | 2.82 | 35.98 | 17.60 |
| Valladolid | 27.93 | 11.73 | 4.66 | 3.24 | 2.76 | 3.07 | 36.68 | 19.00 |
| Zamora | 29.84 | 14.32 | 4.58 | 3.05 | 2.73 | 2.72 | 38.64 | 20.60 |
| **Zone South Plateau (SP)** | | | | | | | | |
| Badajoz | 33.95 | 17.02 | 4.29 | 2.74 | 2.97 | 2.87 | 41.90 | 23.48 |
| Cáceres | 32.67 | 18.00 | 4.47 | 3.33 | 2.95 | 2.68 | 40.60 | 25.24 |
| Ciudad Real | 33.41 | 18.38 | 4.31 | 3.12 | 3.47 | 3.01 | 40.60 | 24.70 |
| Madrid (Barajas) | 32.46 | 16.35 | 4.29 | 3.11 | 3.37 | 2.94 | 39.90 | 22.80 |
| Madrid (Cuatrovientos) | 32.00 | 17.76 | 4.30 | 3.39 | 3.33 | 2.81 | 39.40 | 24.20 |
| Madrid (Getafe) | 32.14 | 18.30 | 4.28 | 3.30 | 3.33 | 2.89 | 39.60 | 24.60 |
| Madrid (Retiro) | 31.26 | 18.49 | 4.31 | 3.30 | 3.25 | 2.78 | 38.60 | 24.70 |
| Madrid (Torrejón) | 32.45 | 16.18 | 4.35 | 3.14 | 3.42 | 2.97 | 39.74 | 22.34 |
| Navacerrada | 21.52 | 10.72 | 4.88 | 4.21 | 3.31 | 2.65 | 29.94 | 18.60 |

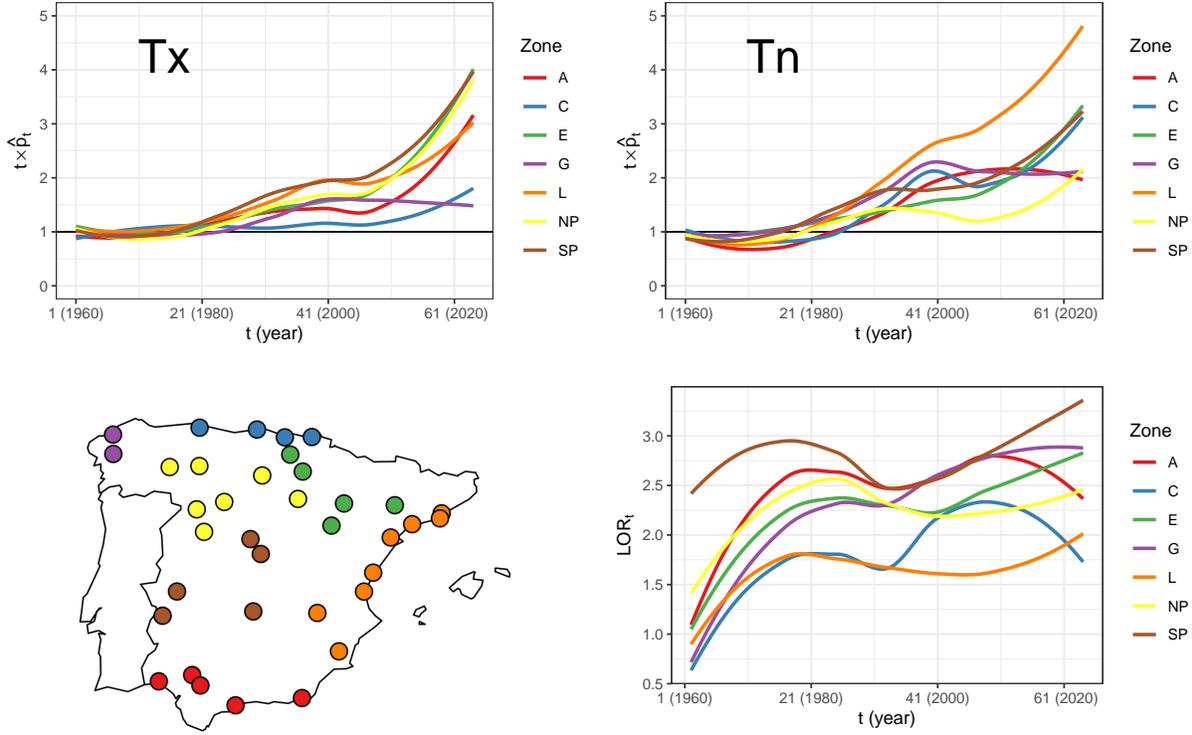

Figure 1: Top: LOESS curve of $t \times \hat{p}_t$ against $t$ with reference value 1, for $\bar{I}_{t\ell}(\mathbf{s})$ (left) and $\underline{I}_{t\ell}(\mathbf{s})$ (right) at each geographical zone. Bottom-left: map of the 36 weather stations clustered by colour for each geographical zone. Bottom-right: LOESS curve of the $LOR_t$'s against $t$ for the $\bar{I}_{t\ell}(\mathbf{s})$ and $\underline{I}_{t\ell}(\mathbf{s})$ concurrence at each geographical zone.

## 1.2 Identifying potential explanatory variables

The first objective was to distinguish potential variables that explain the temporal and spatial patterns that affect the occurrence and co-occurrence of records in $T_x$ and $T_n$, such as long-term trends, persistence, and regional variability.

The plots at the top in Figure 1 show $t \times \hat{p}_t$ against $t$ for each geographical zone, whose expected value under stationarity is 1. There is variability among the zones, but for all the cases the trend is higher than 1 at the end of the period; in agreement with the general pattern observed in the left plot in Figure 2 of the Main manuscript.

The joint distribution $[\bar{I}_{t,\ell}(\mathbf{s}), \underline{I}_{t,\ell}(\mathbf{s}), \bar{I}_{t,\ell-1}(\mathbf{s}), \underline{I}_{t,\ell-1}(\mathbf{s})]$ can be expressed in terms of $2 \times 2 \times 2 \times 2$ tables. To analyse its evolution across years, we calculate a table for each $t$, obtained by adding in space and days within the year. These tables are used to estimate LOR's for the marginal probabilities of records in $T_x$ or $T_n$ on a day given the occurrence or not of a record in $T_x$ or $T_n$ the same day or the day before. The LOR's are calculated following (2) and (3) of the Main manuscript. Concurrence between $\bar{I}_{t,\ell}(\mathbf{s})$ and $\underline{I}_{t,\ell}(\mathbf{s})$ increases over the first two decades showing a strong dependence in the right plot of Figure 2 of the Main manuscript for the entire dataset. The bottom right plot in Figure 1 shows the same but separating by geographic zone, where substantial differences between geographical zones are observed.

3

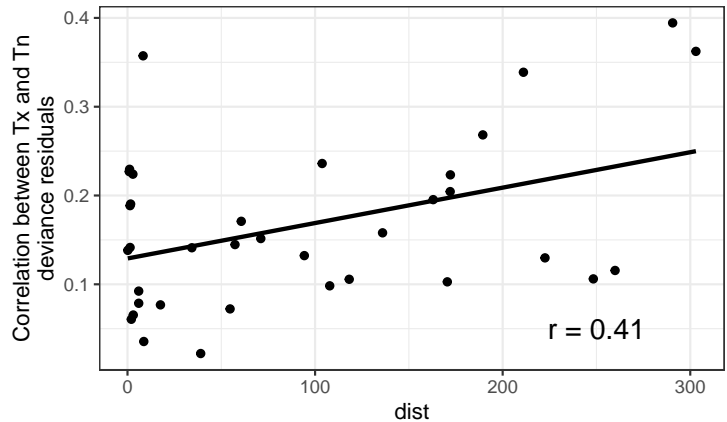

Figure 2: Scatterplot of the correlation between $\bar{I}_{t,\ell}(\mathbf{s})$ and $\underline{I}_{t,\ell}(\mathbf{s})$ univariate local probit models residuals against the distance to the coast.

## 1.3 Analysis of residual variability

We use the deviance residuals of each of the local probit models to study the heterogeneity of the correlation of the residuals across locations. Figure 2 shows the correlation between the residuals in local $\bar{I}_{t,\ell}(\mathbf{s})$ and $\underline{I}_{t,\ell}(\mathbf{s})$ models against distance to the coast, for which a positive linear correlation is observed.

Next, the aim is to study whether the spatial correlation between the residuals of local probit models could be associated with a spatial or climatic covariate. The special features of peninsular Spain as a coast separated from the plateau by high mountains, suggests that an anisotropic spatial dependence might be more suitable to capture the different meteorological conditions between the coast and the interior. We calculated the correlation between the residual series of every pair of stations, for $\bar{I}_{t,\ell}(\mathbf{s})$ and for $\underline{I}_{t,\ell}(\mathbf{s})$. Then, 630 pairwise correlations were used.

Additional exploratory linear models are fitted in order to explain the logarithm of the spatial correlation using geographical and climate distances. The logarithm of the residual correlations for each pair of stations were calculated; there were 37 and 35 negative correlations considered not available in this analysis, for maximum and minimum temperatures respectively. Several linear models were fitted using as covariates the pairwise Euclidean distance, $||\mathbf{s} - \mathbf{s}'||$, and absolute differences in the logarithm of one plus distance to the coast, $|\log(1 + \text{dist}(\mathbf{s})) - \log(1 + \text{dist}(\mathbf{s}'))|$, or in the logarithm of one plus elevation with analogous notation. Also climate information, as absolute difference in the average or in the logarithm of the standard deviation or in the annual linear trend of $T_x$ with notation $\bar{x}_x(\mathbf{s})$, $\log(s_x(\mathbf{s}))$, and $\theta_x(\mathbf{s})$, respectively, and obvious notation for their absolute differences; all of them obtained during JJA in the 30-year reference period of 1991–2020. Here, $\log(s_{x,cv}(\mathbf{s}))$ represents $\log(s_x(\mathbf{s}))$ in cross-validation explained below.

Table 2 reports a summary of the $R^2$ and $R^2_{adj}$ of these fitted linear models. The comparison of models for spatial correlation between residuals of $T_n$ shows that the Euclidean distance explains near of the 30% of variability but adding climate information with $\log(s_x)$ increases to 53% the explanation. None of the other climate o geographical distances relevantly improves the explanation. Respect to patterns in spatial correlation between residuals of $T_n$, the best $R^2_{adj}$ is reached also by the model including Euclidean dis-

4

Table 2: Model comparison based on $R^2$ and $R^2_{adj}$ for the logarithm of the correlation between locations $\mathbf{s}$ and $\mathbf{s}'$ of the deviance residuals for $\bar{I}_{t\ell}(\mathbf{s})$ and $\underline{I}_{t\ell}(\mathbf{s})$. The linear predictor includes the Euclidean distance $||\mathbf{s}-\mathbf{s}'||$ and absolute differences in the other variables.

| | Event / Metric | | | |
|---|---|---|---|---|
| | $\mathrm{T}_x$ | | $\mathrm{T}_n$ | |
| Linear predictor | $R^2$ | $R^2_{adj}$ | $R^2$ | $R^2_{adj}$ |
| $||\mathbf{s}-\mathbf{s}'||$ | 0.393 | 0.392 | 0.296 | 0.295 |
| $||\mathbf{s}-\mathbf{s}'||, \bar{x}_x(\mathbf{s})$ | 0.398 | 0.395 | 0.297 | 0.294 |
| $||\mathbf{s}-\mathbf{s}'||, \log(s_x(\mathbf{s}))$ | 0.535 | 0.534 | 0.321 | 0.319 |
| $||\mathbf{s}-\mathbf{s}'||, \log(s_{x,cv}(\mathbf{s}))$ | 0.501 | 0.500 | 0.315 | 0.312 |
| $||\mathbf{s}-\mathbf{s}'||, \theta_x(\mathbf{s})$ | 0.434 | 0.432 | 0.297 | 0.294 |
| $||\mathbf{s}-\mathbf{s}'||, \bar{x}_n(\mathbf{s})$ | 0.399 | 0.397 | 0.315 | 0.312 |
| $||\mathbf{s}-\mathbf{s}'||, \log(s_n(\mathbf{s}))$ | 0.401 | 0.399 | 0.299 | 0.297 |
| $||\mathbf{s}-\mathbf{s}'||, \theta_n(\mathbf{s})$ | 0.407 | 0.405 | 0.313 | 0.311 |
| $||\mathbf{s}-\mathbf{s}'||, \log(1+\mathrm{dist}(\mathbf{s}))$ | 0.455 | 0.453 | 0.300 | 0.298 |
| $||\mathbf{s}-\mathbf{s}'||, \log(1+\mathrm{elev}(\mathbf{s}))$ | 0.470 | 0.468 | 0.314 | 0.312 |
| $||\mathbf{s}-\mathbf{s}'||, \log(1+\mathrm{dist}(\mathbf{s})), \log(1+\mathrm{elev}(\mathbf{s}))$ | 0.482 | 0.479 | 0.315 | 0.311 |
| $||\mathbf{s}-\mathbf{s}'||, \log(s_x(\mathbf{s})), \log(1+\mathrm{elev}(\mathbf{s}))$ | 0.541 | 0.539 | 0.325 | 0.321 |
| $||\mathbf{s}-\mathbf{s}'||, \log(s_{x,cv}(\mathbf{s})), \log(1+\mathrm{elev}(\mathbf{s}))$ | 0.507 | 0.504 | 0.318 | 0.315 |

tance and $\log(s_x)$. Note that both the correlation of $\mathrm{T}_x$ and $\mathrm{T}_n$ showed a relationship with differences in this covariate but did not show an additional relationship with differences in the standard deviation associated with $\mathrm{T}_n$. Although differences in other covariates revealed a relationship with correlation, this disappeared when considering distance and difference in $\log(s_x(\mathbf{s}))$.

### 1.4 Kriging the logarithm of the standard deviation

The logarithm of the standard deviation of $\mathrm{T}_x$ during JJA in the 30-year reference period of 1991–2020, $\log(s_x(\mathbf{s}))$, is recovered for the grid of Spain using universal kriging via the R package `gstat` (Pebesma, 2004) with model

$$\log(s_x(\mathbf{s})) = \beta_0 + \beta_1 \log(1+\mathrm{elev}(\mathbf{s})) + \beta_2 \log(1+\mathrm{dist}(\mathbf{s})) + w(\mathbf{s}),$$

where $w(\mathbf{s})$ follows a mean-zero stationary spatial Gaussian process with an exponential correlation function and no nugget effect. The best error structure including the choice of the correlation function or the inclusion of a nugget effect was chosen by cross validation, with cross validated $R^2$'s beyond 0.75. Figure 3 shows a map of the covariates in this model as well as the kriged values of $\log(s_x(\mathbf{s}))$.

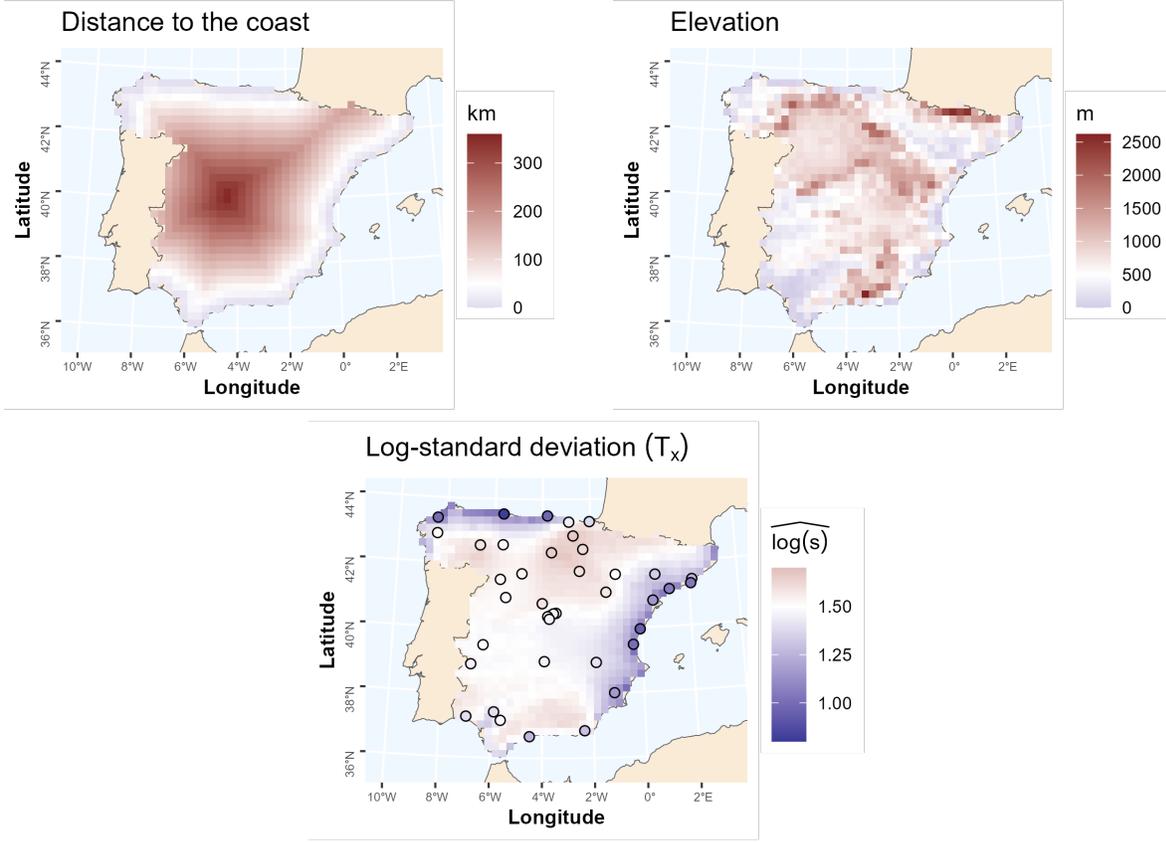

Figure 3: Maps of distance to the coast in km, elevation in m, and $\widehat{\log(s_x(\mathbf{s}))}$ for temperatures in °C (true values overlapped) over peninsular Spain.

## 2 More about the modelling framework

### 2.1 Block average

The overall behaviour of a continuous process in space like the $a(\mathbf{s})$'s can be summarised by surfaces and block averages (Banerjee et al., 2014, Chapter 7). For example, the spatial average induced correlation between $\bar{v}_{t\ell}(\mathbf{s})$ and $\underline{v}_{t\ell}(\mathbf{s})$ in $D$ is the block average

$$a(D) = \frac{1}{|D|} \int_D a(\mathbf{s})\, d\mathbf{s} = \frac{1}{|D|} \int_D \frac{a_{11}(\mathbf{s}) a_{21}(\mathbf{s})}{a_{11}(\mathbf{s}) \sqrt{a_{21}(\mathbf{s})^2 + a_{22}(\mathbf{s})^2}}\, d\mathbf{s},$$

where $|D|$ is the area of $D$. This is a stochastic integral that can be approximated through Monte Carlo integration. The block average can be obtained for other continuous processes of interest; say $\bar{a}(\mathbf{s})$, $\underline{a}(\mathbf{s})$, or the $a(\mathbf{s})$'s themselves. In a multivariate setting, other summary maps might be of interest, e.g., with $\mathbf{T}(\mathbf{s})$ as the spatially varying covariance matrix, we could obtain $|\mathbf{T}(\mathbf{s})|$ or $\mathrm{tr}(\mathbf{T}(\mathbf{s}))$ surfaces or block averages.

## 2.2 Distributions for Gibbs sampling

It is convenient to denote parameters in vector notation as follows. First, let $\bar{\mathbf{I}}^*$ and $\underline{\mathbf{I}}^*$ denote the vectors of indicators associated with any $r$-tied record. Let the vectors of latent variables be $\bar{\mathbf{Y}}_{t\ell} = (\bar{Y}_{t\ell}(\mathbf{s}_1), \ldots \bar{Y}_{t\ell}(\mathbf{s}_n))^\top$, equivalently define the vectors $\underline{\mathbf{Y}}_{t\ell}, \bar{\mathbf{V}}_{t\ell}, \underline{\mathbf{V}}_{t\ell}, \mathbf{W}_{j,t\ell}$ for $j = 1, 2$, or any other lowercase letters in bold not explicitly defined. Define the matrices $\mathbf{X}_{t\ell} = (\mathbf{x}_{t\ell}^\top(\mathbf{s}_1), \ldots, \mathbf{x}_{t\ell}^\top(\mathbf{s}_n))^\top$ and $\mathbf{Z} = (\mathbf{z}^\top(\mathbf{s}_1), \ldots, \mathbf{z}^\top(\mathbf{s}_n))^\top$. The steps that we follow in every iteration of the MCMC algorithm are:

1. We sample the values of $\bar{\mathbf{I}}^*$ and $\underline{\mathbf{I}}^*$ independently from Bernoulli$(1/r)$.

2. We sample the values of the latent variables $\bar{Y}_{t\ell}(\mathbf{s}_i)$ and $\underline{Y}_{t\ell}(\mathbf{s})$ for every $t = 2, \ldots, T$, $\ell \in JJA$, and $i = 1, \ldots, n$, from their truncated normal full conditional distribution,

$$\bar{Y}_{t\ell}(\mathbf{s}_i) \mid \cdots \sim TN\left(\mathbf{x}_{t\ell}(\mathbf{s}_i)\bar{\boldsymbol{\beta}} + \bar{v}_{t\ell}(\mathbf{s}_i), 1, (\bar{a}, \bar{b})\right),$$
$$\underline{Y}_{t\ell}(\mathbf{s}_i) \mid \cdots \sim TN\left(\mathbf{x}_{t\ell}(\mathbf{s}_i)\underline{\boldsymbol{\beta}} + \underline{v}_{t\ell}(\mathbf{s}_i), 1, (\underline{a}, \underline{b})\right),$$

where $(\bar{a}, \bar{b}) = (0, \infty)$ if $\bar{I}_{t\ell}(\mathbf{s}_i) = 1$ and $(\bar{a}, \bar{b}) = (-\infty, 0)$ otherwise, analogous for $(\underline{a}, \underline{b})$ with $\underline{I}_{t\ell}(\mathbf{s}_i)$. We sample these truncated normals with an algorithm that efficiently mixes the usual inverse transform method and the method by Robert (1995).

3. Given the normal prior for the regression coefficients, $\bar{\boldsymbol{\beta}}, \underline{\boldsymbol{\beta}} \sim N_k(\boldsymbol{\mu}_{\boldsymbol{\beta}}, \boldsymbol{\Sigma}_{\boldsymbol{\beta}})$, we sample them from their normal full conditional distribution,

$$\bar{\boldsymbol{\beta}} \mid \cdots \sim N_k(\hat{\bar{\boldsymbol{\beta}}}, \boldsymbol{\Omega}_{\boldsymbol{\beta}});$$
$$\hat{\bar{\boldsymbol{\beta}}} = \boldsymbol{\Omega}_{\boldsymbol{\beta}} \left( \sum_{t=2}^{T} \sum_{\ell \in JJA} \mathbf{X}_{t\ell}^\top \left(\bar{\mathbf{Y}}_{t\ell} - \bar{\mathbf{V}}_{t\ell}\right) + \boldsymbol{\Sigma}_{\boldsymbol{\beta}}^{-1} \boldsymbol{\mu}_{\boldsymbol{\beta}} \right),$$
$$\boldsymbol{\Omega}_{\boldsymbol{\beta}}^{-1} = \sum_{t=2}^{T} \sum_{\ell \in JJA} \mathbf{X}_{t\ell}^\top \mathbf{X}_{t\ell} + \boldsymbol{\Sigma}_{\boldsymbol{\beta}}^{-1};$$

analogous for $\underline{\boldsymbol{\beta}}$.

4. The Gaussian process prior on $w_{j,t\ell}(\mathbf{s})$ for $j = 1, 2$ defines them as stochastic processes for which the values at the $n$ observed locations come from the normal distribution, $\mathbf{W}_{j,t\ell} \sim N_n(\mathbf{0}_n, \mathbf{R}_{\phi_j})$, where $\mathbf{R}_\phi$ is the correlation matrix with the $i$'th element in the $i$th row being $\exp\{-\phi\|\mathbf{s}_i - \mathbf{s}_{i'}\|\}$. Then, for every $t = 2, \ldots, T$ and $\ell \in JJA$, we sample the processes from their normal full conditional distribution,

$$\mathbf{W}_{1,t\ell} \mid \cdots \sim N_n(\widehat{\mathbf{W}}_{1,t\ell}, \boldsymbol{\Omega}_{\mathbf{W}_1});$$
$$\widehat{\mathbf{W}}_{1,t\ell} = \boldsymbol{\Omega}_{\mathbf{W}_1} \left( a_{11} \left(\bar{\mathbf{Y}}_{t\ell} - \mathbf{X}_{t\ell}\bar{\boldsymbol{\beta}}\right) + a_{21} \left(\underline{\mathbf{Y}}_{t\ell} - \mathbf{X}_{t\ell}\underline{\boldsymbol{\beta}} - a_{22}\mathbf{W}_{2,t\ell}\right)\right),$$
$$\boldsymbol{\Omega}_{\mathbf{W}_1}^{-1} = \left(a_{11}^2 + a_{21}^2\right)\mathbf{I}_n + \mathbf{R}_{\phi_1}^{-1};$$

and

$$\mathbf{W}_{2,t\ell} \mid \cdots \sim N_n(\widehat{\mathbf{W}}_{2,t\ell}, \boldsymbol{\Omega}_{\mathbf{W}_2});$$
$$\widehat{\mathbf{W}}_{2,t\ell} = \boldsymbol{\Omega}_{\mathbf{W}_2} \left(a_{22}\left(\underline{\mathbf{Y}}_{t\ell} - \mathbf{X}_{t\ell}\underline{\boldsymbol{\beta}} - a_{21}\mathbf{W}_{1,t\ell}\right)\right),$$
$$\boldsymbol{\Omega}_{\mathbf{W}_2}^{-1} = a_{22}^2\mathbf{I}_n + \mathbf{R}_{\phi_2}^{-1}.$$

5 (a). We sample the values of the decay parameters $\phi_j$ for $j = 1, 2$ from their non-standard full conditional distribution,

$$[\phi_j \mid \cdots] \propto [\phi_j] \times |\mathbf{R}_{\phi_j}^{-1}|^{\frac{(T-1)L}{2}} \exp\left\{\frac{-1}{2} \sum_{t=2}^{T} \sum_{\ell \in JJA} \mathbf{W}_{j,t\ell}^{\top} \mathbf{R}_{\phi_j}^{-1} \mathbf{W}_{j,t\ell}\right\},$$

where $[\phi_j]$ is the prior distribution. We update this parameter incorporating a random walk Metropolis step in the logarithmic scale. This entails using a normal proposal with a variance parameter tuned to achieve an acceptance rate of approximately 33%. During the initial burn-in iterations, we use a version of the method outlined in Section 3 of Roberts and Rosenthal (2009) to automatically reach the target acceptance rate without the need of manual tuning. The particular case when both decays are the same, $\phi$, is analogous by multiplying the likelihood part of both individual decays.

5 (b). If we consider covariate information in the spatial covariance structure, we also have to sample the values of the decay parameters $\phi_{x,j}$ for $j = 1, 2$. The $\mathbf{R}_{\phi_j}$ correlation matrix in Step 4 becomes $\mathbf{R}_{\phi_j, \phi_{x,j}}$, where the $i'$th element in the $i$th row of $\mathbf{R}_{\phi, \phi_x}$ is now $\exp\{-\phi\|\mathbf{s}_i - \mathbf{s}_{i'}\|\} \exp\{-\phi_x |x(\mathbf{s}_i) - x(\mathbf{s}_{i'})|\}$. The full conditionals are equivalent to that of case (a) and the sampling procedure is the same.

6 (a). Next, we focus on the elements of the coregionalisation matrix $\mathbf{A}$. Given the half normal prior distribution of the diagonal elements or equivalently their zero-mean normal distribution truncated from below at zero and scale parameter $\sigma_{a_{jj}}^2$ for $j = 1, 2$, we sample them from their truncated normal full conditional distribution,

$$a_{11} \mid \cdots \sim TN\left(\hat{a}_{11}, \omega_{a_{11}}^2, (0, \infty)\right); \qquad a_{22} \mid \cdots \sim TN\left(\hat{a}_{22}, \omega_{a_{22}}^2, (0, \infty)\right);$$

$$\hat{a}_{11} = \omega_{a_{11}}^2 \sum_{t=2}^{T} \sum_{\ell \in JJA} \mathbf{W}_{1,t\ell}^{\top} \left(\bar{\mathbf{Y}}_{t\ell} - \mathbf{X}_{t\ell}\bar{\boldsymbol{\beta}}\right), \quad \hat{a}_{22} = \omega_{a_{22}}^2 \sum_{t=2}^{T} \sum_{\ell \in JJA} \mathbf{W}_{2,t\ell}^{\top} \left(\underline{\mathbf{Y}}_{t\ell} - \mathbf{X}_{t\ell}\underline{\boldsymbol{\beta}} - a_{21}\mathbf{W}_{1,t\ell}\right),$$

$$\omega_{a_{11}}^{-2} = \sum_{t=2}^{T} \sum_{\ell \in JJA} \mathbf{W}_{1,t\ell}^{\top} \mathbf{W}_{1,t\ell} + \frac{1}{\sigma_{a_{11}}^2}, \qquad \omega_{a_{22}}^{-2} = \sum_{t=2}^{T} \sum_{\ell \in JJA} \mathbf{W}_{2,t\ell}^{\top} \mathbf{W}_{2,t\ell} + \frac{1}{\sigma_{a_{22}}^2}.$$

Given the normal prior for the dependence element, $a_{21} \sim N(\mu_{a_{21}}, \sigma_{a_{21}}^2)$, we sample $a_{21}$ from its normal full conditional distribution,

$$a_{21} \mid \cdots \sim N\left(\hat{a}_{21}, \omega_{a_{21}}^2\right);$$

$$\hat{a}_{21} = \omega_{a_{21}}^2 \left(\sum_{t=2}^{T} \sum_{\ell \in JJA} \mathbf{W}_{1,t\ell}^{\top} \left(\underline{\mathbf{Y}}_{t\ell} - \mathbf{X}_{t\ell}\underline{\boldsymbol{\beta}} - a_{22}\mathbf{W}_{2,t\ell}\right) + \frac{\mu_{a_{21}}}{\sigma_{a_{21}}^2}\right),$$

$$\omega_{a_{21}}^{-2} = \sum_{t=2}^{T} \sum_{\ell \in JJA} \mathbf{W}_{1,t\ell}^{\top} \mathbf{W}_{1,t\ell} + \frac{1}{\sigma_{a_{21}}^2}.$$

6 (b). When the coregionalisation matrix $\mathbf{A}(\mathbf{s})$ is spatially varying, the full conditionals in Step 4 should be adapted to account for different $a$'s in space. The log-Gaussian process prior on $a_{jj}(\mathbf{s})$ imposes $\log\{\mathbf{a}_{jj}\} \sim N_n(\boldsymbol{\mu}_{\mathbf{a}_{jj}}, \boldsymbol{\Sigma}_{\mathbf{a}_{jj}})$ with mean vector $\boldsymbol{\mu}_{\mathbf{a}_{jj}} = \mathbf{Z}\boldsymbol{\beta}_{a_{jj}}$ and covariance matrix $\boldsymbol{\Sigma}_{\mathbf{a}_{jj}} = \sigma_{a_{jj}}^2 \mathbf{R}_{\phi_a}$ for $j = 1, 2$. We sample the values of the processes at every observed location $i = 1, \ldots, n$ from their non-standard

full conditional distributions using Metropolis steps analogous to those of the decay parameters outlined in Step 5 (a),

$$[a_{11}(\mathbf{s}_i) \mid \cdots] \propto N\left(\log\{a_{11}(\mathbf{s}_i)\} \mid \mu_{a_{11}}(\mathbf{s}_i) + \frac{\sum_{k \neq i} r_{ik}(\mu_{a_{11}}(\mathbf{s}_k) - \log\{a_{11}(\mathbf{s}_k)\})}{r_{ii}}, \frac{\sigma_{a_{11}}^2}{r_{ii}}\right)$$

$$\times N\left(a_{11}(\mathbf{s}_i) \mid \frac{\sum_{t=2}^{T} \sum_{\ell \in JJA} w_{1,t\ell}(\mathbf{s}_i) \left(\bar{Y}_{t\ell}(\mathbf{s}_i) - \mathbf{x}_{t\ell}^\top(\mathbf{s}_i)\bar{\boldsymbol{\beta}}\right)}{\sum_{t=2}^{T} \sum_{\ell \in JJA} w_{1,t\ell}^2(\mathbf{s}_i)}, \frac{1}{\sum_{t=2}^{T} \sum_{\ell \in JJA} w_{1,t\ell}^2(\mathbf{s}_i)}\right),$$

and

$$[a_{22}(\mathbf{s}_i) \mid \cdots] \propto N\left(\log\{a_{22}(\mathbf{s}_i)\} \mid \mu_{a_{22}}(\mathbf{s}_i) + \frac{\sum_{k \neq i} r_{ik}(\mu_{a_{22}}(\mathbf{s}_k) - \log\{a_{22}(\mathbf{s}_k)\})}{r_{ii}}, \frac{\sigma_{a_{22}}^2}{r_{ii}}\right)$$

$$\times N\left(a_{22}(\mathbf{s}_i) \mid \frac{\sum_{t=2}^{T} \sum_{\ell \in JJA} w_{2,t\ell}(\mathbf{s}_i) \left(Y_{t\ell}(\mathbf{s}_i) - \mathbf{x}_{t\ell}^\top(\mathbf{s}_i)\boldsymbol{\beta} - a_{21}(\mathbf{s}_i)w_{1,t\ell}(\mathbf{s}_i)\right)}{\sum_{t=2}^{T} \sum_{\ell \in JJA} w_{2,t\ell}^2(\mathbf{s}_i)}, \frac{1}{\sum_{t=2}^{T} \sum_{\ell \in JJA} w_{2,t\ell}^2(\mathbf{s}_i)}\right),$$

where for notational convenience $r_{ik} \equiv r^*_{\phi_a, ik}$ is the $k$th element in the $i$th row of $\mathbf{R}_{\phi_a}^{-1}$. Equivalently, the Gaussian process prior for $\mathbf{a}_{21}$ imposes $\mathbf{a}_{21} \sim N_n(\boldsymbol{\mu}_{\mathbf{a}_{21}}, \boldsymbol{\Sigma}_{\mathbf{a}_{21}})$ with mean vector $\boldsymbol{\mu}_{\mathbf{a}_{21}} = \mathbf{Z}\boldsymbol{\beta}_{a_{21}}$ and covariance matrix $\boldsymbol{\Sigma}_{\mathbf{a}_{21}} = \sigma_{a_{21}}^2 \mathbf{R}_{\phi_a}$. We sample this process from its normal full conditional distribution,

$$\mathbf{a}_{21} \mid \cdots \sim N_n(\hat{\mathbf{a}}_{21}, \boldsymbol{\Omega}_{\mathbf{a}_{21}});$$

$$\hat{\mathbf{a}}_{21} = \boldsymbol{\Omega}_{\mathbf{a}_{21}} \left(\sum_{t=2}^{T} \sum_{\ell \in JJA} \text{diag}\{\mathbf{W}_{1,t\ell}\} \left(\mathbf{Y}_{t\ell} - \mathbf{X}_{t\ell}\boldsymbol{\beta} - \text{diag}\{\mathbf{a}_{22}\}\mathbf{W}_{2,t\ell}\right) + \boldsymbol{\Sigma}_{\mathbf{a}_{21}}^{-1}\boldsymbol{\mu}_{\mathbf{a}_{21}}\right),$$

$$\boldsymbol{\Omega}_{\mathbf{a}_{21}}^{-1} = \sum_{t=2}^{T} \sum_{\ell \in JJA} \text{diag}\{\mathbf{W}_{1,t\ell}\}^2 + \boldsymbol{\Sigma}_{\mathbf{a}_{21}}^{-1}.$$

7. For the spatially varying coregionalisation coefficients we also have to sample their hyperparameters corresponding to the mean and variance, the decay $\phi_a$ is considered fixed and common for convenience. For example, given the prior specification of the hyperparameters of $\mathbf{a}_{21}$, $\boldsymbol{\beta}_{a_{21}} \sim N_q(\boldsymbol{\mu}_a, \boldsymbol{\Sigma}_a)$ and $1/\sigma_{a_{21}}^2 \sim G(a_a, b_a)$, we sample them from their normal and gamma full conditional distribution,

$$\boldsymbol{\beta}_{a_{21}} \mid \cdots \sim N_q(\hat{\boldsymbol{\beta}}_{a_{21}}, \boldsymbol{\Omega}_{\boldsymbol{\beta}_{a_{21}}});$$

$$\hat{\boldsymbol{\beta}}_{a_{21}} = \boldsymbol{\Omega}_{\boldsymbol{\beta}_{a_{21}}} \left(\frac{\mathbf{Z}^\top \mathbf{R}_{\phi_a}^{-1} \mathbf{a}_{21}}{\sigma_{a_{21}}^2} + \boldsymbol{\Sigma}_a^{-1}\boldsymbol{\mu}_a\right),$$

$$\boldsymbol{\Omega}_{\boldsymbol{\beta}_{a_{21}}}^{-1} = \frac{\mathbf{Z}^\top \mathbf{R}_{\phi_a}^{-1} \mathbf{Z}}{\sigma_{a_{21}}^2} + \boldsymbol{\Sigma}_a^{-1},$$

and

$$1/\sigma_{a_{21}}^2 \mid \cdots \sim G\left(\frac{n}{2} + a_a, \frac{1}{2}(\mathbf{a}_{21} - \boldsymbol{\mu}_{\mathbf{a}_{21}})^\top \mathbf{R}_{\phi_a}^{-1}(\mathbf{a}_{21} - \boldsymbol{\mu}_{\mathbf{a}_{21}}) + b_a\right).$$

Analogous for the hyperparameters of $\log\{\mathbf{a}_{jj}\}$, $\boldsymbol{\beta}_{a_{jj}}$ and $1/\sigma_{a_{jj}}^2$, for $j = 1, 2$.

# 3 More about the results

## 3.1 MCMC convergence diagnostics

Convergence was monitored by marginal and multivariate potential scale reduction factors, and usual traceplots (see Figures 4 to 8 for the selected model $M_5$). The plots show good agreement between the two chains in black and grey. The potential scale reduction factors were obtained via the R package `coda` (Plummer et al., 2006) and values below 1.1 are in agreement with the convergence of the chains.

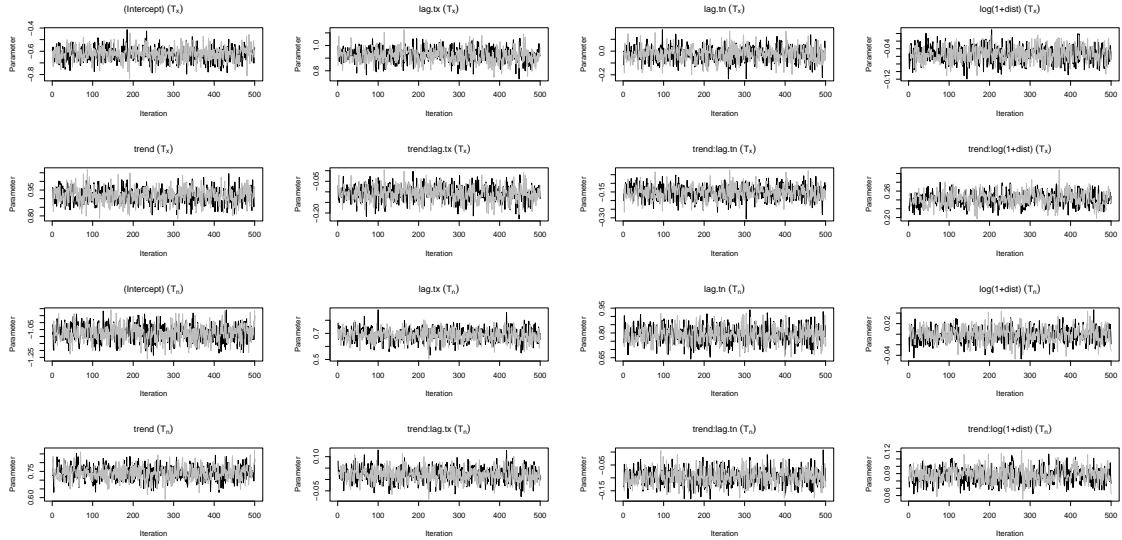

Figure 4: Traceplots for the regression coefficients $\bar{\boldsymbol{\beta}}$ ($T_x$) and $\underline{\boldsymbol{\beta}}$ ($T_n$) in $M_5$.

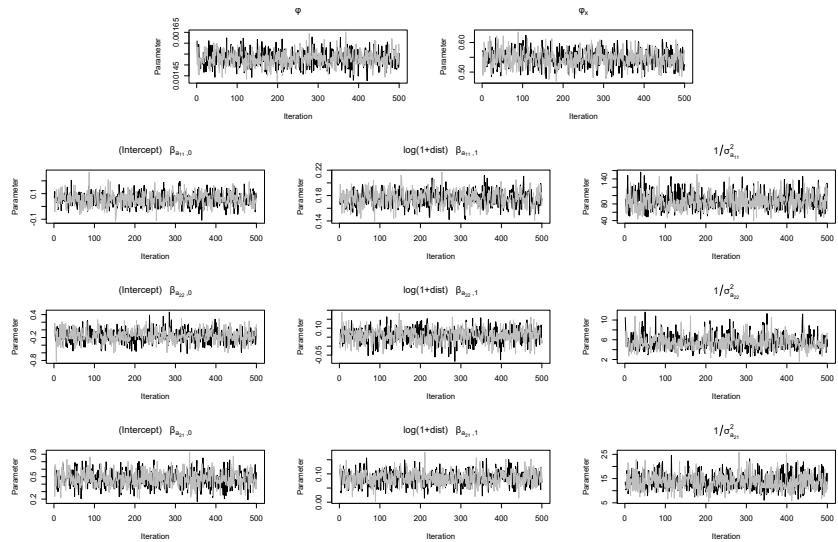

Figure 5: Traceplots for the decay parameters $\phi$ and $\phi_x$, and for the coregionalisation parameters $\boldsymbol{\beta}_{a_{11}}$, $1/\sigma^2_{a_{11}}$, $\boldsymbol{\beta}_{a_{22}}$, $1/\sigma^2_{a_{22}}$, $\boldsymbol{\beta}_{a_{21}}$, and $1/\sigma^2_{a_{21}}$ in $M_5$.

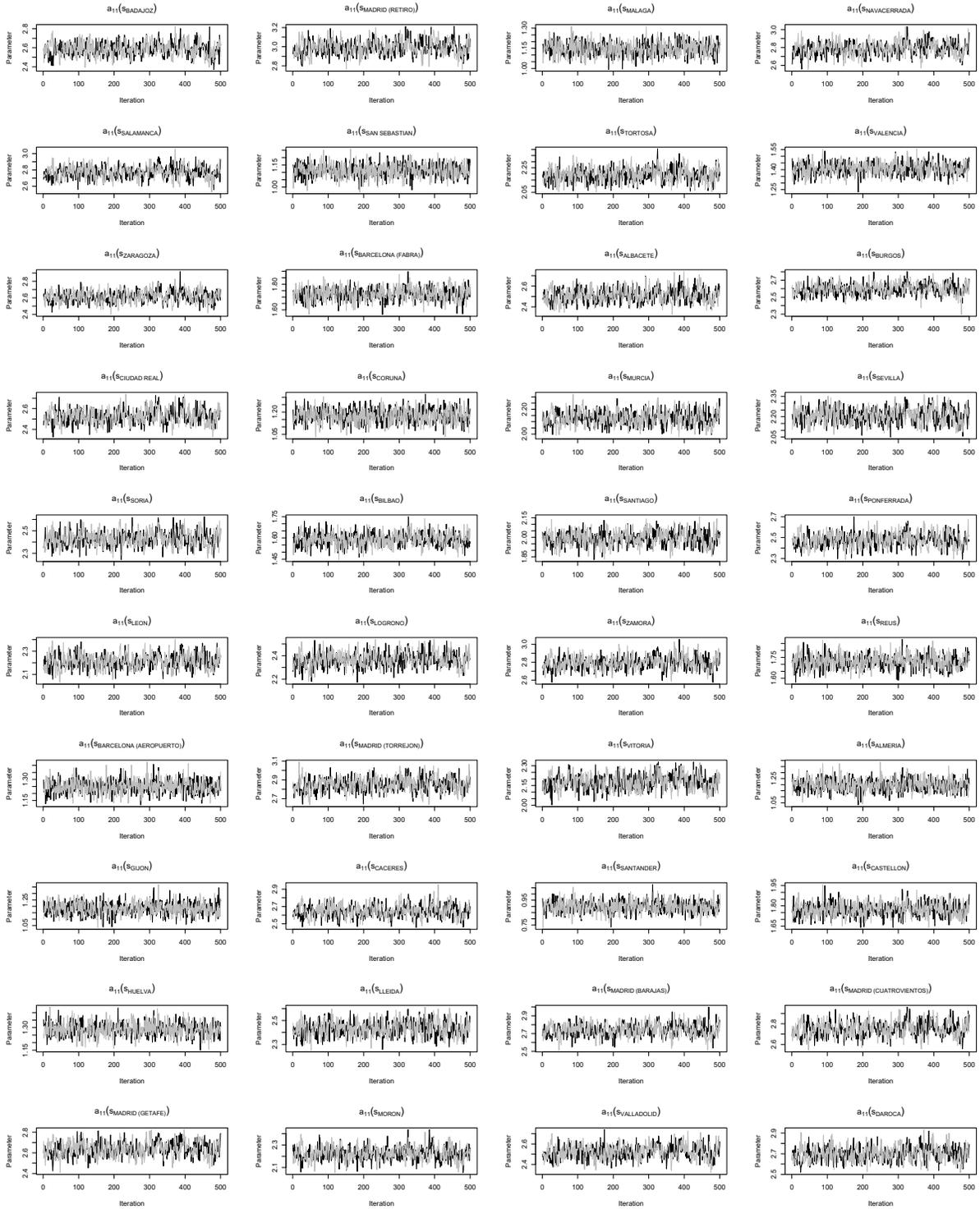

Figure 6: Traceplots for the spatially varying coregionalisation coefficients $a_{11}(\mathbf{s})$ in the weather stations in $M_5$.

11

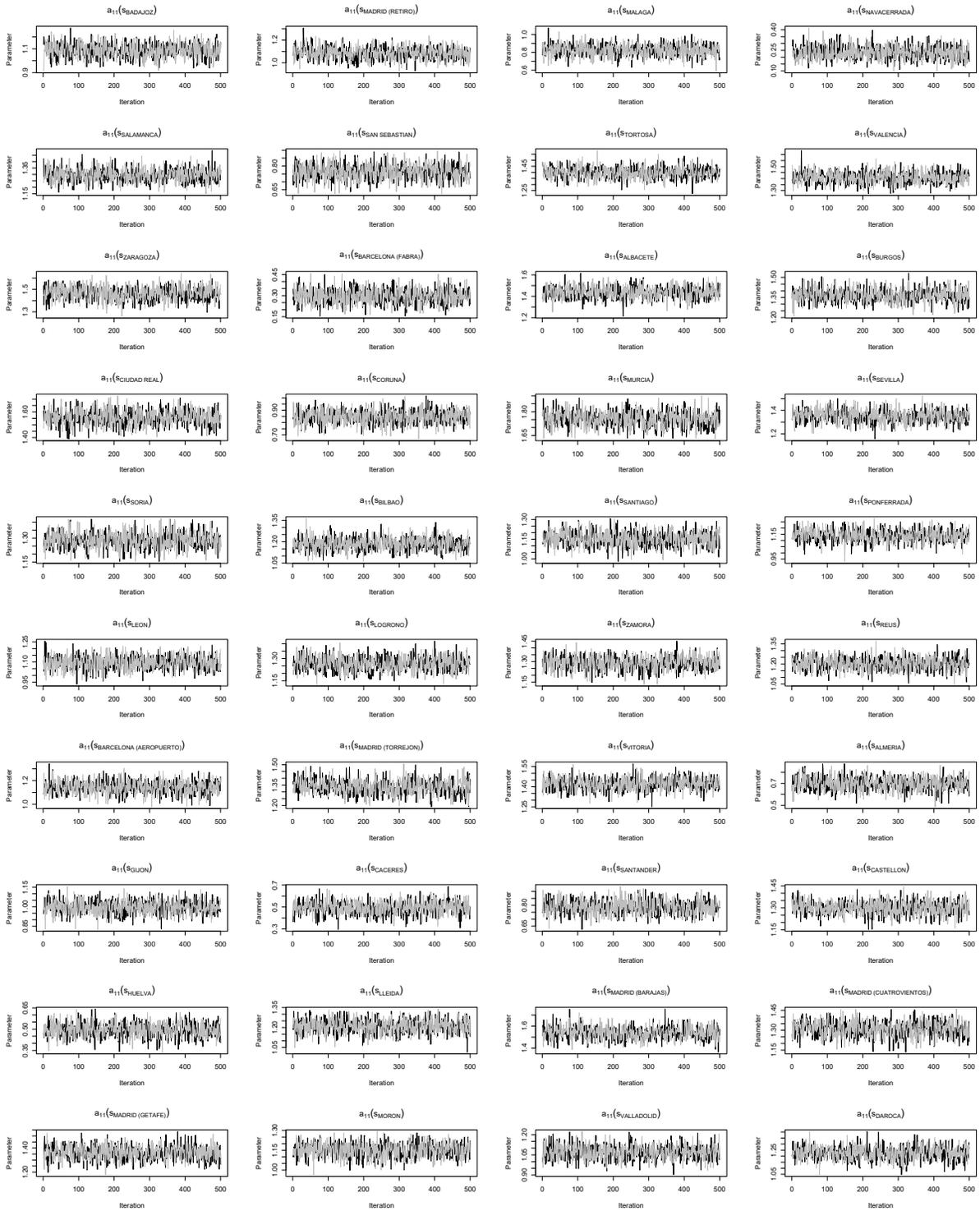

Figure 7: Traceplots for the spatially varying coregionalisation coefficients $a_{22}(\mathbf{s})$ in the weather stations in $M_5$.

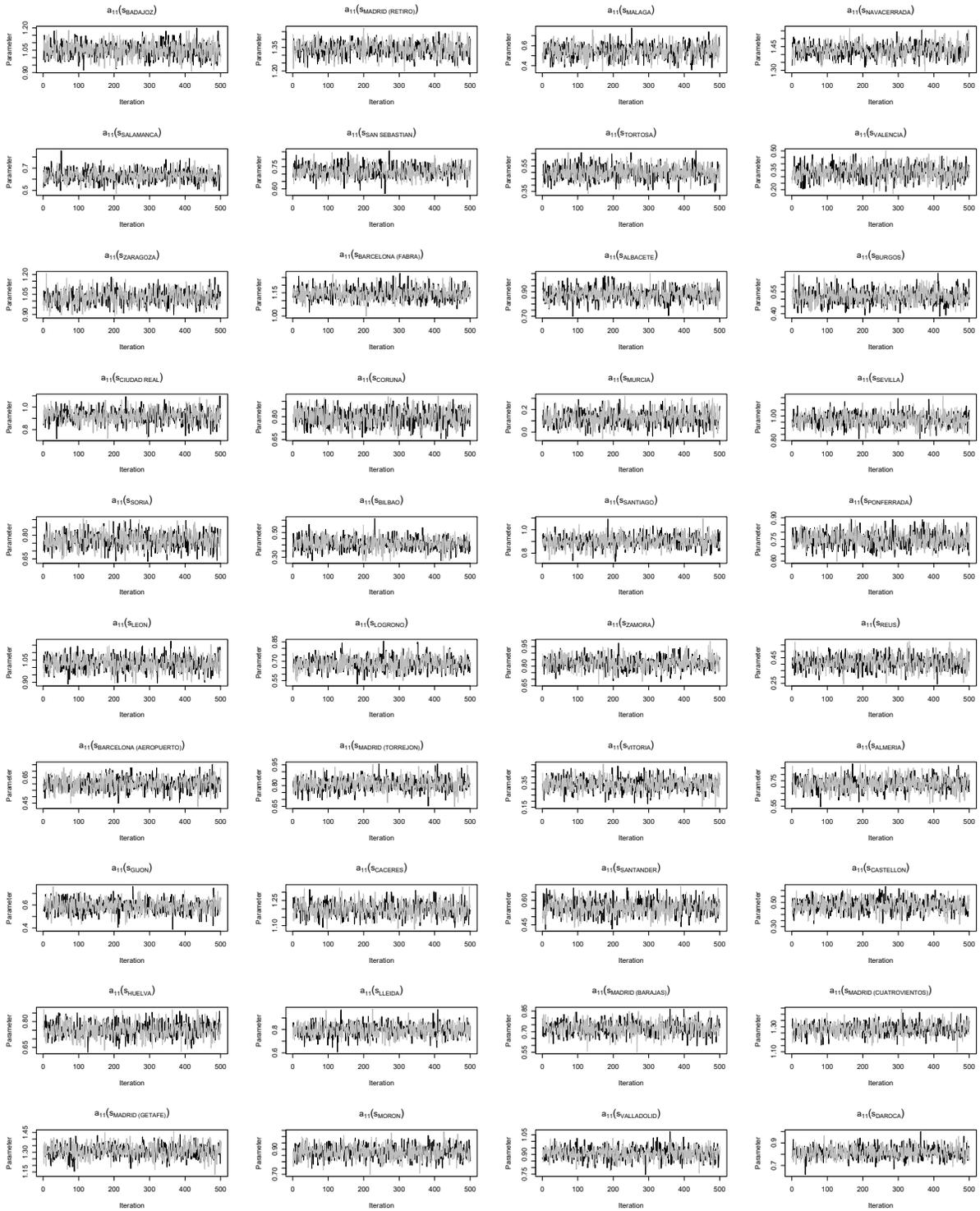

Figure 8: Traceplots for the spatially varying coregionalisation coefficients $a_{21}(\mathbf{s})$ in the weather stations in $M_5$.

## 3.2 Model comparison and adequacy

In addition to the Jaccard index, we also report the Bayesian area under the ROC curve (AUC) and the deviance information criterion (DIC; Spiegelhalter et al., 2002), as these metrics are commonly used in Bayesian model evaluation.

Following the same notation as in the definition of the Jaccard index, each posterior sample of the AUC, $AUC^{(b)}$, is obtained via the R package `pROC` (Robin et al., 2011), using an MCMC joint sample of the probabilities of record for $(t, \ell, \mathbf{s}_i) \in \mathcal{T} \times \mathcal{L} \times \mathcal{S}$. Note that $0 \leq AUC^{(b)} \leq 1$, which measures the similarity between observed values and predicted probabilities, with 1 indicating perfect agreement.

Next, the DIC is the sum of two components, one for quantifying the model fit and the other for penalising its complexity. Given the data $\mathbf{I}$ with associated probabilities $\mathbf{p}$ and an univariate Bernoulli likelihood,

$$[\mathbf{I} \mid \mathbf{p}] = \prod_{i=1}^{n} \prod_{t=2}^{T} \prod_{\ell \in JJA} p_{t\ell}(\mathbf{s}_i)^{I_{t\ell}(\mathbf{s}_i)} (1 - p_{t\ell}(\mathbf{s}_i))^{1 - I_{t\ell}(\mathbf{s}_i)},$$

the deviance $D(\mathbf{p}) = -2 \log([\mathbf{I} \mid \mathbf{p}])$ of the model,

$$D(\mathbf{p}) = -2 \sum_{i=1}^{n} \sum_{t=2}^{T} \sum_{\ell \in JJA} \left[ I_{t\ell}(\mathbf{s}_i) \log p_{t\ell}(\mathbf{s}_i) + (1 - I_{t\ell}(\mathbf{s}_i)) \log(1 - p_{t\ell}(\mathbf{s}_i)) \right],$$

measures the variability associated with the likelihood. Formally, we could calculate the true deviance of the bivariate model by adding the univariate deviance of the maximums and minimums. We can calculate the deviance for the simultaneous occurrence of records in both series using the associated indicators and probabilities in the expression of the univariate deviance. Tied records are removed from the computation of the deviance because these values do not contribute to assess how well the model fits the occurrence of records. The deviance is a random variable that we summarise by the posterior mean deviance $\hat{D} = E[D(\mathbf{p}) \mid \mathbf{I}]$.

The DIC is the sum of two components, one for quantifying the model fit and the other for penalising its complexity. The first component is measured through $\hat{D}$ while the model complexity is measured through the effective number of parameters,

$$p_D = E[D(\mathbf{p}) \mid \mathbf{I}] - D(E[\mathbf{p} \mid \mathbf{I}]) = \hat{D} - D(\hat{\mathbf{p}}),$$

and the DIC is defined as
$$\text{DIC} = \hat{D} + p_D.$$

We compute the DIC for the probabilities of record in $T_x$, in $T_n$, and a pseudo-DIC for the probabilities of joint records. A smaller DIC is associated with a better model fit.

Table 3 reports the values of the AUC's and DIC's following the same notation as in Table 1 of the Main manuscript. As can be seen, differences in AUC are minimal, and the conclusions drawn from the DIC do not necessarily align with those based on the Jaccard index. Nevertheless, the most complex model ($M_5$) appears to offer a good trade-off across all metrics and events compared to the other models.

Figure 9 shows the scatterplots of the observed values of three quantities of primary interest for the occurrence of records versus a summary of their corresponding predictive

Table 3: Model comparison for records in maximum, minimum, and both temperatures. Metrics: J and AUC 1 in 1961–1990 and 2 in 1991–2023, and DIC.

| Event | Model | Metric | | | | |
|---|---|---|---|---|---|---|
| | | J 1 | J 2 | AUC 1 | AUC 2 | DIC |
| $T_x$ | $M_0$ | 0.115 | 0.015 | 0.728 | 0.495 | 109426 |
| | $M_1$ | 0.375 | 0.234 | 0.892 | 0.867 | 44921 |
| | $M_2$ | 0.375 | 0.235 | 0.895 | 0.874 | 45236 |
| | $M_3$ | 0.375 | 0.234 | 0.905 | 0.882 | 48241 |
| | $M_4$ | 0.382 | 0.241 | 0.895 | 0.874 | 41628 |
| | $M_5$ | 0.381 | 0.239 | 0.904 | 0.882 | 45281 |
| $T_n$ | $M_0$ | 0.108 | 0.015 | 0.705 | 0.504 | 110155 |
| | $M_1$ | 0.273 | 0.143 | 0.859 | 0.809 | 62588 |
| | $M_2$ | 0.278 | 0.147 | 0.869 | 0.823 | 64010 |
| | $M_3$ | 0.281 | 0.151 | 0.869 | 0.826 | 60251 |
| | $M_4$ | 0.278 | 0.149 | 0.867 | 0.822 | 62753 |
| | $M_5$ | 0.281 | 0.152 | 0.867 | 0.824 | 59126 |
| $T_x \times T_n$ | $M_0$ | 0.068 | 0.000 | 0.786 | 0.492 | 56584 |
| | $M_1$ | 0.286 | 0.133 | 0.938 | 0.920 | 21651 |
| | $M_2$ | 0.289 | 0.146 | 0.937 | 0.922 | 21406 |
| | $M_3$ | 0.296 | 0.151 | 0.942 | 0.930 | 21089 |
| | $M_4$ | 0.292 | 0.147 | 0.936 | 0.921 | 20582 |
| | $M_5$ | 0.298 | 0.152 | 0.941 | 0.929 | 20301 |

distribution. In particular, it shows the scatterplots for each observed location and *month* going through the three summer months of the total number of records $N_{64,month}(\mathbf{s})$ for maximum, minimum and joint records; the ratio over stationarity of the number of records in the last decade $R_{55:64,month}(\mathbf{s})$ for maximum and minimum temperatures; and the number of joint records in the last decade $\bar{N}_{55:64,month}(\mathbf{s})$. The same is shown for $t \times \text{ERS}_{t,month}(D)$ averaging the ERS (defined over the grid of 40 observed locations) for $t = 2, \ldots, T$ going through the three summer months.

It is clear from the scatterplots that the predictions are very close to the observed values with sharp CI's, with some nuances outlined below. The total numbers of records show a coverage close to the 90% nominal level, 92.5%, 95.8%, and 91.7%, respectively. It seems that most of the numbers of records in the last decade are well-calibrated, with a 90% coverage of 91.7%, 79.2%, and 88.3%, respectively; although for the minima there is a reduced coverage level, which could be caused by a small bias in observations with lower values. Most of the ERS's are also well-calibrated, with a 90% coverage of 97.9%, 92.1%, and 92.6%, respectively; which suggest a slight over-dispersion for the maxima, which could be explained by the already remarkably sharp CI's.

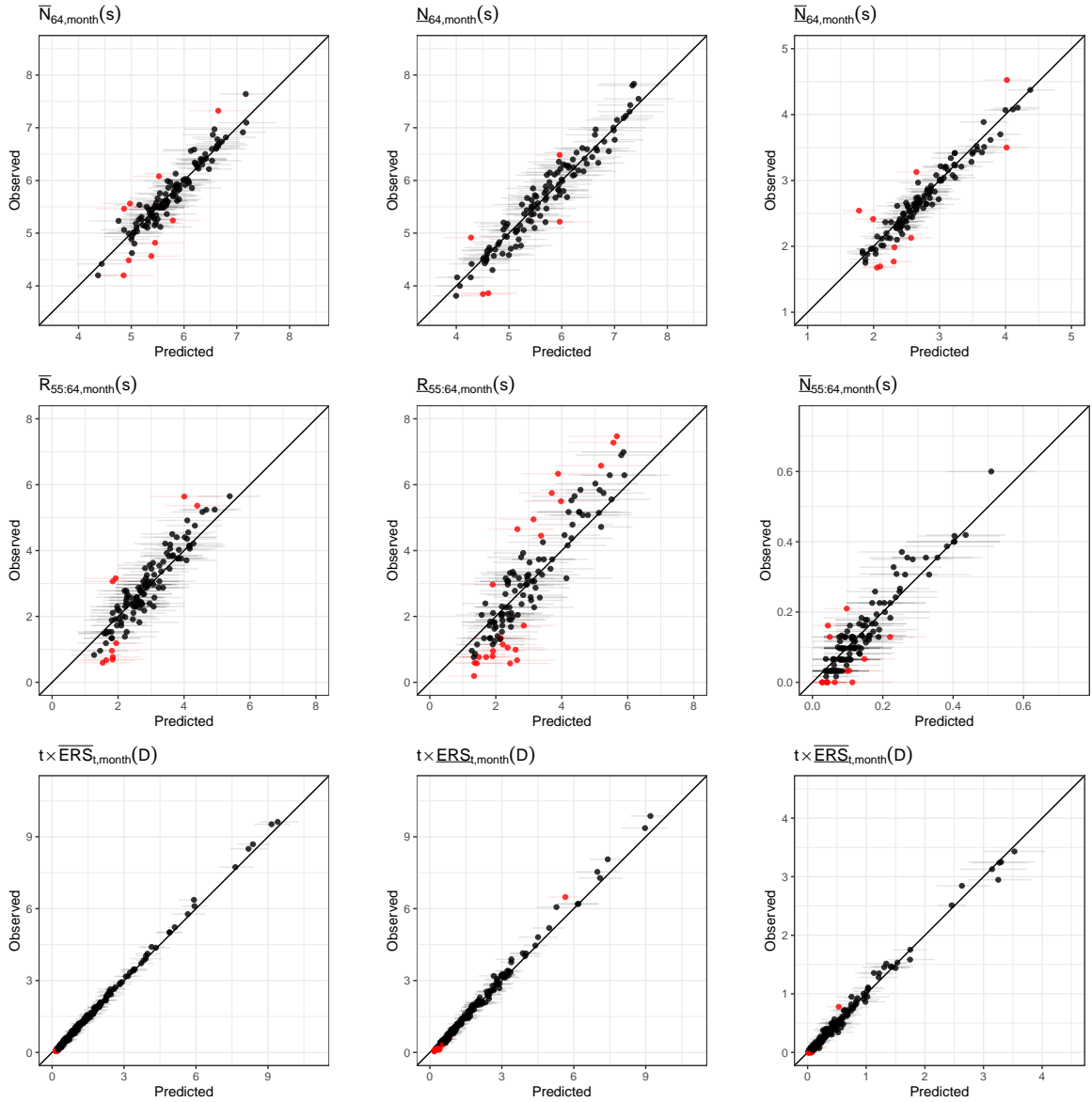

Figure 9: Scatterplots of the observed values versus the posterior mean and 90% CI for the corresponding predictive distribution using $M_5$ for maximum (left), minimum (centre), and joint records (right), in $N_{64,month}(\mathbf{s})$ (top), $R_{55:64,month}(\mathbf{s})$ (centre), and $t \times \text{ERS}_{t,month}(D)$ (bottom), respectively. Red indicates CI's not containing the observed value.

## 3.3 Coregionalisation parameters

Table 4 shows the hyperparameters of the spatially varying coregionalisation matrix $\mathbf{A}(\mathbf{s})$. Two of the three processes show a strong positive relationship with distance to the coast. Figure 10 shows maps related to the elements in $\mathbf{A}(\mathbf{s})$. Different spatial patterns are observed, but the most notable is the strong relationship between $a_{11}(\mathbf{s})$ and consequently $\bar{a}(\mathbf{s})$ and the distance to the coast. Figure 11 shows the boxplots of the same spatial processes at the weather stations. The conclusions are similar but they offer an idea of uncertainty quantification for these processes.

## 3.4 Spatial correlation: isotropy vs. anisotropy

To illustrate the need to consider an anisotropic correlation, Figure 12 shows the spatial correlation pattern induced by models $M_3$ (isotropic) and $M_5$ (anisotropic) in all weather stations. The isotropic structure imposes that the correlation between the location and any point at the same distance is identical, which is not a realistic assumption due to the varied orography of the study region that can result in very different temperatures in locations at the same distance. The anisotropy is very clear, e.g., in Valencia, where it captures the existing higher correlation with locations close to the Mediterranean coast due to the effect of the coast. In another example, San Sebastián, the anisotropy allows to capture the effect of the sea, that yields stronger correlations along the Cantabrian coast, and the effect of the Ebro Valley that yields stronger correlation in the upper Ebro Basin; correlations with locations at the same distance in the northern Inner Plateau are much lower.

## 3.5 Calendar heatmaps

Figures 13 and 14 show calendar heatmaps that represent the daily evolution across time from 1999 to 2023 of the ERS for the three types of record. These plots allow us to observe the persistence of different heat waves, such as that of the summer of 2003 or those of 2022 and 2023.

Table 4: Posterior mean and 90% CI of the hyperparameters in the spatially varying coregionalisation matrix $\mathbf{A}(\mathbf{s})$ in $M_5$.

| Hyperparameter | Mean | 90% CI |
| --- | --- | --- |
| $\beta_{a_{11},0}$ (Intercept) | 0.06 | $(-0.03, 0.14)$ |
| $\beta_{a_{11},1} \log(1 + \text{dist}(\mathbf{s}))$ | **0.18** | $(0.16, 0.19)$ |
| $\beta_{a_{22},0}$ (Intercept) | $-0.16$ | $(-0.43, 0.12)$ |
| $\beta_{a_{22},1} \log(1 + \text{dist}(\mathbf{s}))$ | 0.06 | $(-0.01, 0.12)$ |
| $\beta_{a_{21},0}$ (Intercept) | **0.47** | $(0.31, 0.65)$ |
| $\beta_{a_{21},1} \log(1 + \text{dist}(\mathbf{s}))$ | **0.08** | $(0.04, 0.12)$ |
| $\sigma_{a_{11}}$ | 0.11 | $(0.09, 0.13)$ |
| $\sigma_{a_{22}}$ | 0.44 | $(0.35, 0.54)$ |
| $\sigma_{a_{21}}$ | 0.28 | $(0.23, 0.34)$ |

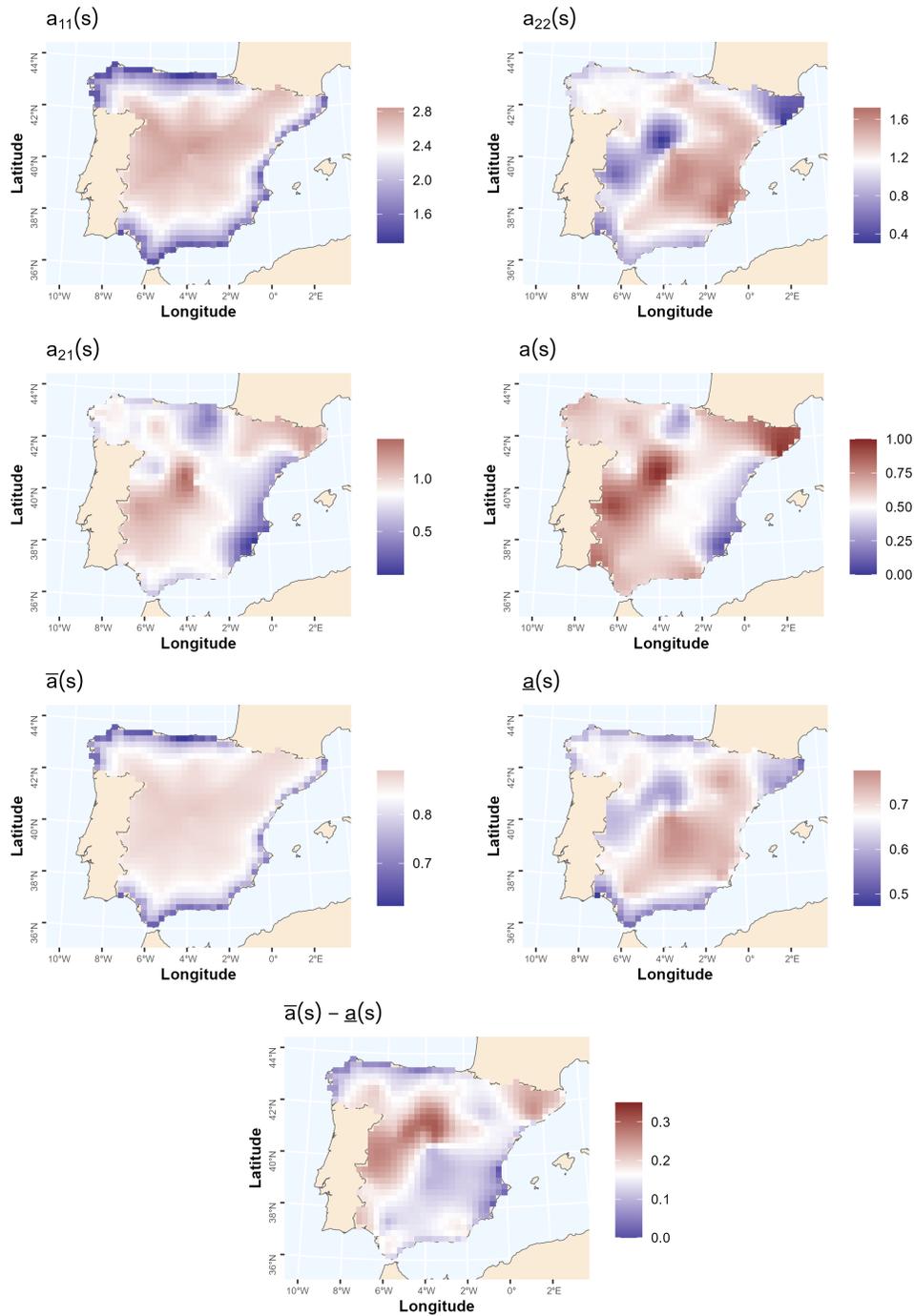

Figure 10: Posterior mean surfaces of the spatially varying coregionalisation coefficients and their products. Reference white is the posterior mean of the corresponding block average.

18

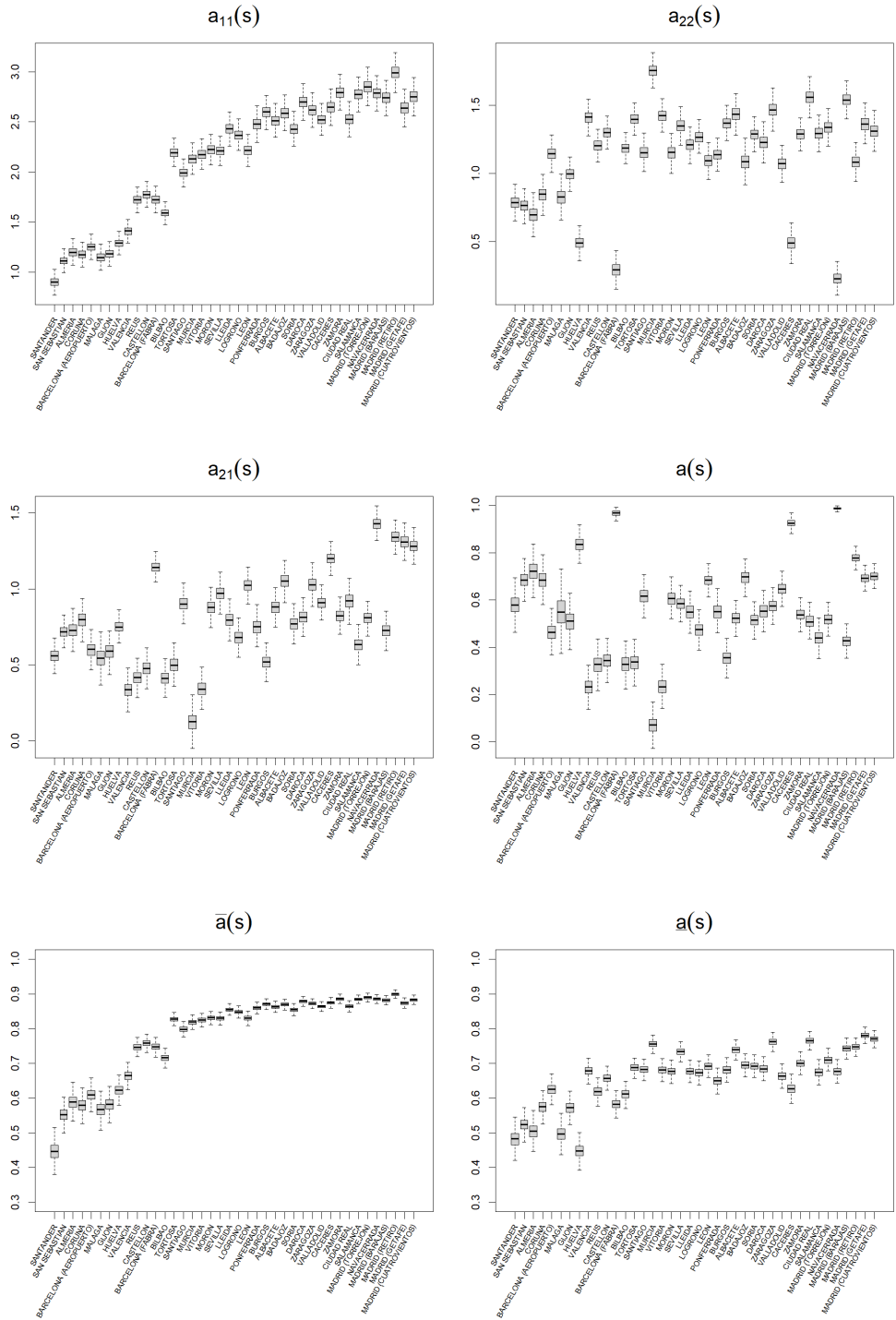

Figure 11: Boxplots of the posterior distribution of the spatially varying coregionalisation coefficients and their products. Locations are sorted by distance to the coast, from lowest to highest.

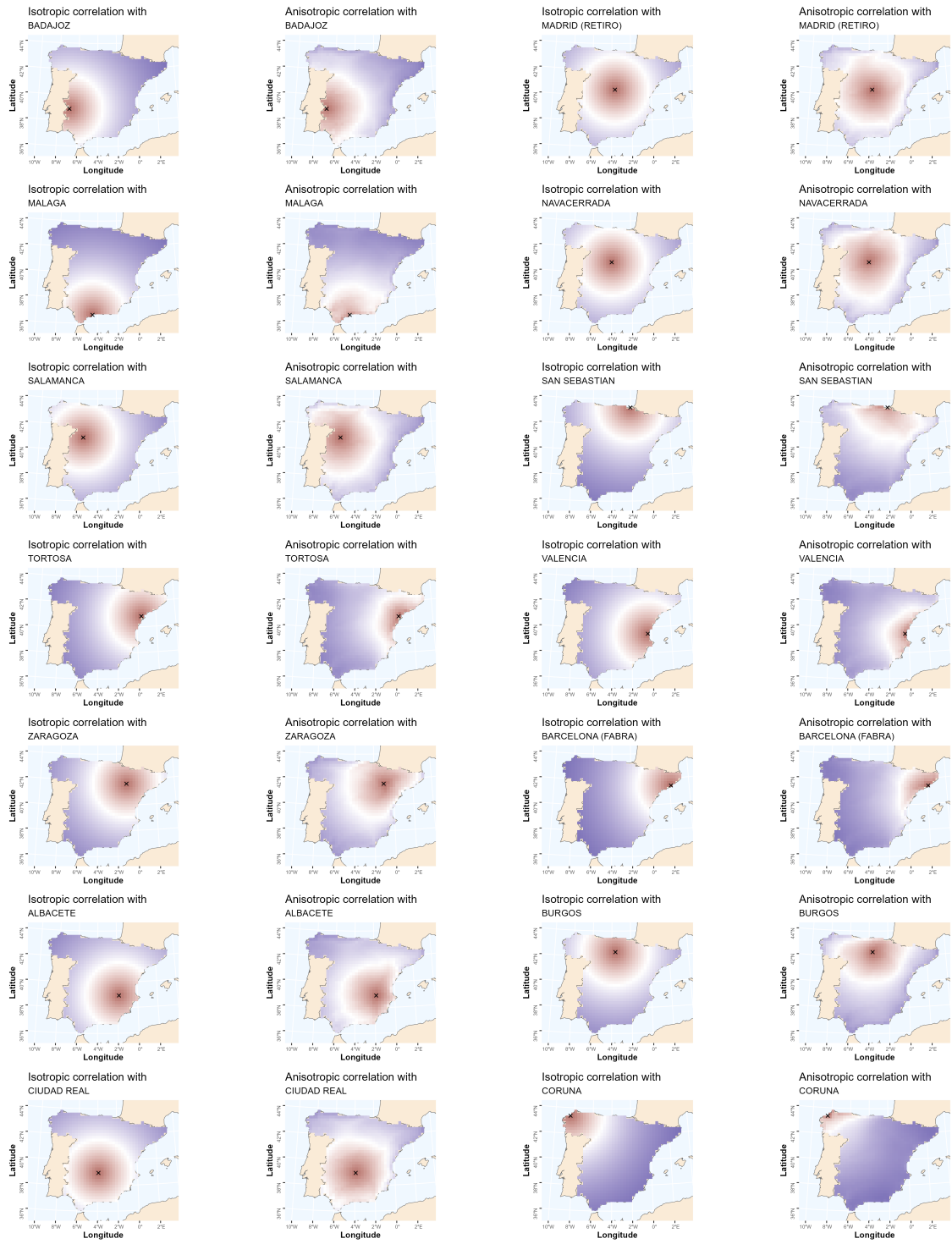

Figure 12: Maps of spatial isotropic ($M_3$) and anisotropic ($M_5$) correlation with respect to the weather stations over peninsular Spain. (1 of 3)

20

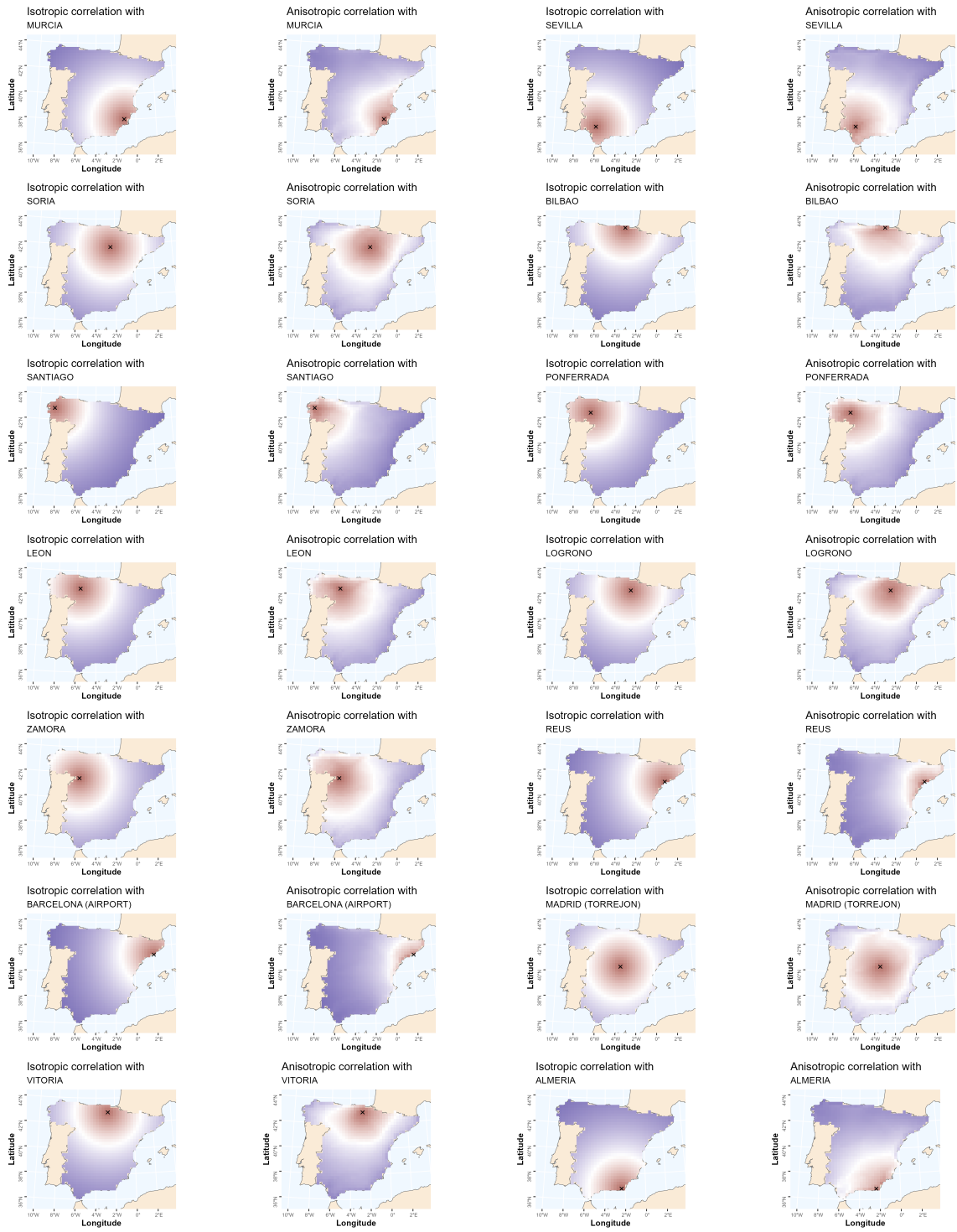

Figure 12: Maps of spatial isotropic ($M_3$) and anisotropic ($M_5$) correlation with respect to the weather stations over peninsular Spain. (2 of 3)

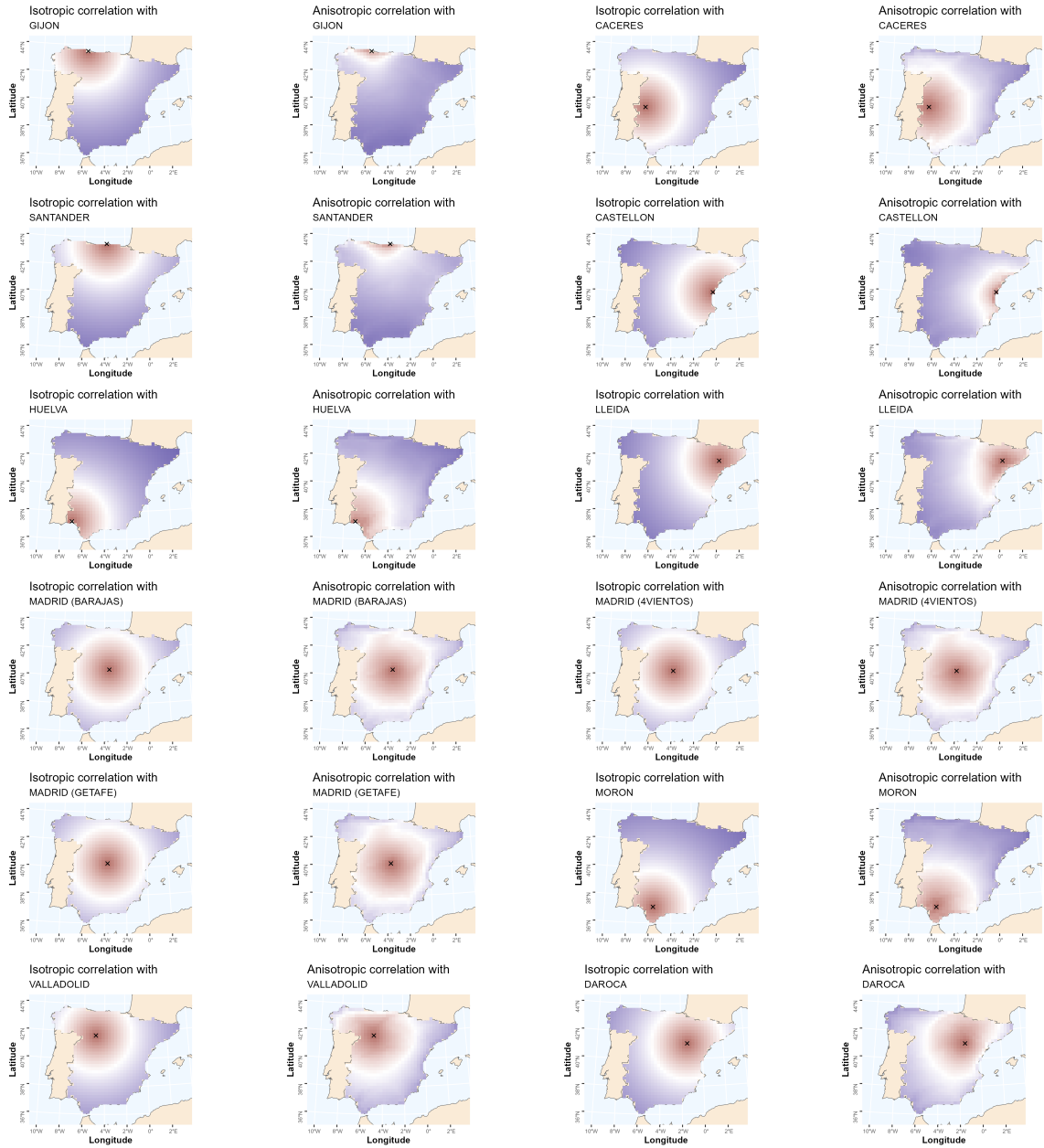

Figure 12: Maps of spatial isotropic ($M_3$) and anisotropic ($M_5$) correlation with respect to the weather stations over peninsular Spain. (3 of 3)

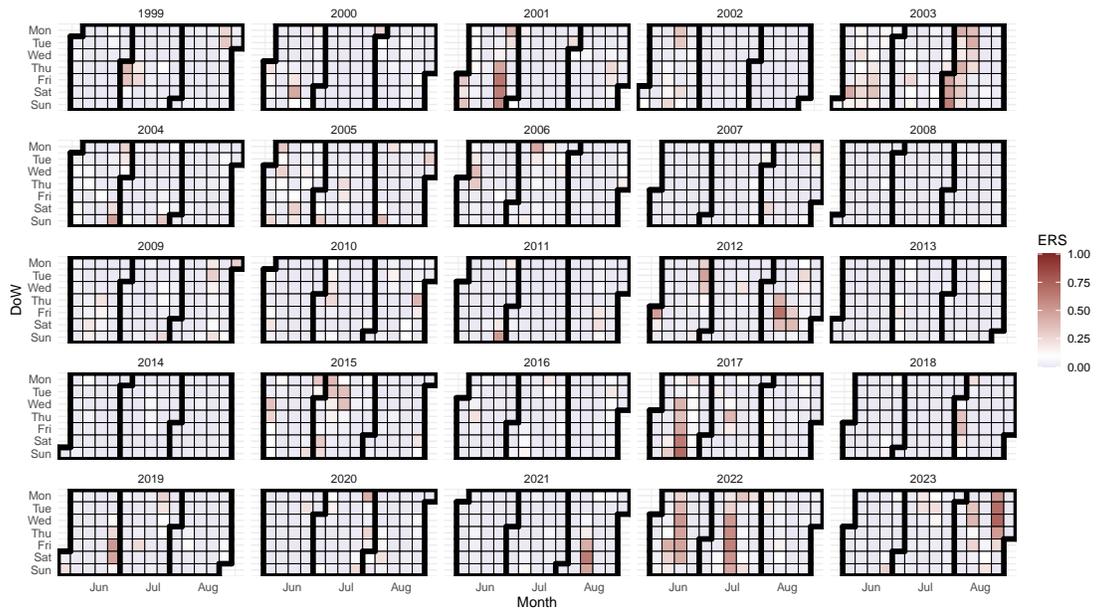

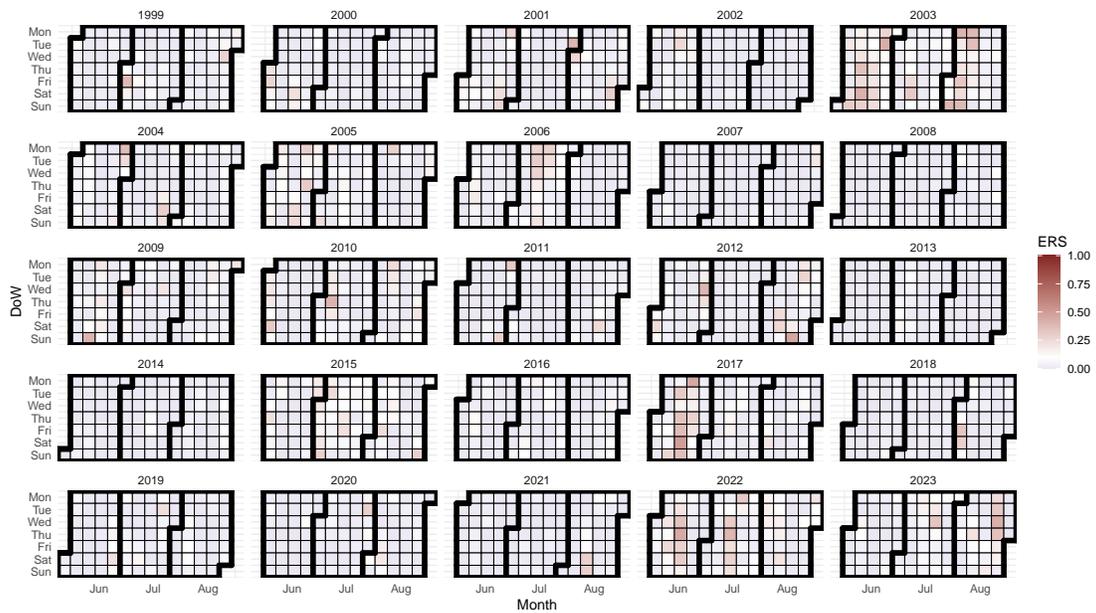

Figure 13: Calendar heatmaps for the posterior mean of $ERS_{t\ell}(D)$ for the maximum (top), and minimum (bottom) records for years from 1999 to 2023 by columns across days within the summer season (from top to bottom).

23

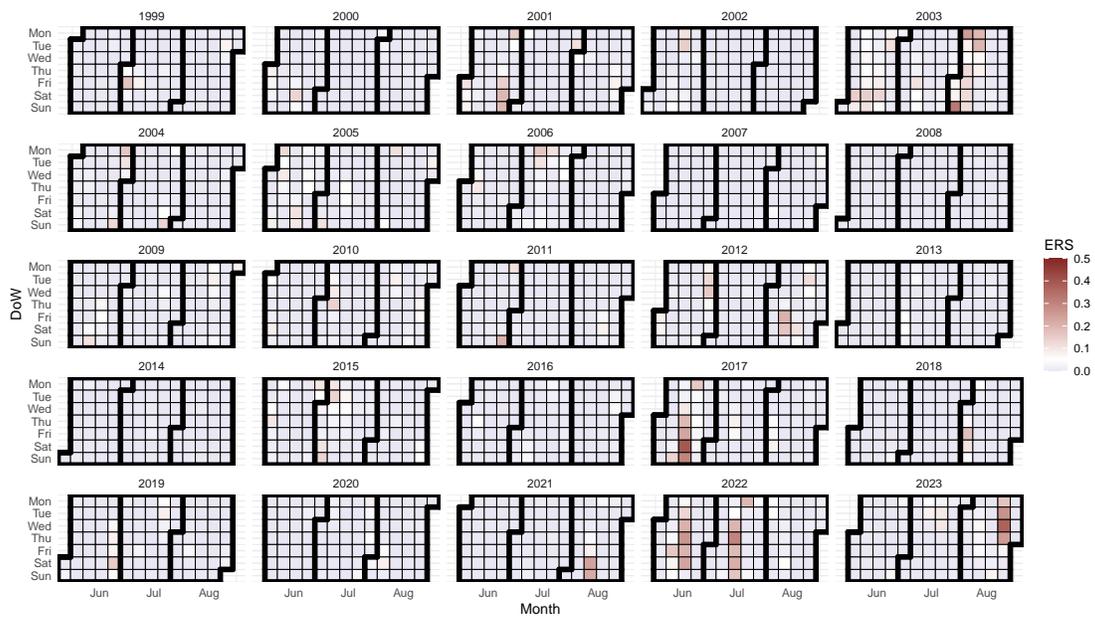

Figure 14: Calendar heatmaps for the posterior mean of $E\bar{R}S_{t\ell}(D)$ for the joint records for years from 1999 to 2023 by columns across days within the summer season (from top to bottom).

# References

Banerjee, S., Carlin, B. P., and Gelfand, A. E. (2014). *Hierarchical Modeling and Analysis for Spatial Data*. Chapman and Hall/CRC, New York, NY, 2 edition.

Castillo-Mateo, J., Cebrián, A. C., and Asín, J. (2023). Statistical analysis of extreme and record-breaking daily maximum temperatures in peninsular Spain during 1960–2021. *Atmospheric Research*, 293:106934.

Castillo-Mateo, J., Gelfand, A. E., Gracia-Tabuenca, Z., Asín, J., and Cebrián, A. C. (2024). Spatio-temporal modeling for record-breaking temperature events in Spain. *Journal of the American Statistical Association*.

Pebesma, E. J. (2004). Multivariable geostatistics in S: the gstat package. *Computers & Geosciences*, 30(7):683–691.

Plummer, M., Best, N., Cowles, K., and Vines, K. (2006). CODA: Convergence diagnosis and output analysis for MCMC. *R News*, 6(1):7–11.

Robert, C. P. (1995). Simulation of truncated normal variables. *Statistics and Computing*, 5(2):121–125.

Roberts, G. O. and Rosenthal, J. S. (2009). Examples of adaptive MCMC. *Journal of Computational and Graphical Statistics*, 18(2):349–367.

Robin, X., Turck, N., Hainard, A., Tiberti, N., Lisacek, F., Sanchez, J.-C., and Müller, M. (2011). pROC: an open-source package for R and S+ to analyze and compare ROC curves. *BMC Bioinformatics*, 12:77.

Spiegelhalter, D. J., Best, N. G., Carlin, B. P., and Van Der Linde, A. (2002). Bayesian measures of model complexity and fit. *Journal of the Royal Statistical Society: Series B (Statistical Methodology)*, 64(4):583–639.

WMO (2022). Centennial Observing Stations: State of Recognition Report – 2021. Technical Report WMO-No. 1296, WMO, Geneva, Switzerland.



\end{document}